\documentclass{article}
\usepackage{amssymb}
\usepackage[dvips]{graphicx}
\usepackage{amsmath}
\usepackage{amsfonts}

\input{tcilatex}

\begin{document}

\begin{quotation}
{\huge Gravity assisted solution}{\Huge \ }

{\huge of the mass gap problem }{\Huge \ }

{\huge for pureYang-Mills fields}

{\Huge \ \ \ \ \ \ \ \ \ \ \ \ \ \ \ \ \ \ \ \ \ \ \ \ \ \ \ \ \ \ \ \ \ \ \
\ \ \ \ \ \ \ \ \ \ \ \ \ \ \ \ \ \ \ \ \ \ \ \ \ \ \ \ \ \ \ \ \ \ \ \ \ \
\ \ \ }
\end{quotation}

\ \ \ \ \ \ \ \ \ \ \ \ \ \ \ \ \ \ \ \ \ \ \ \textbf{Arkady L.Kholodenko}

\ \ \ \ \ \ 

$\ \ \ \ \ $375 H.L.Hunter Laboratories, Clemson University, Clemson,

\ \ \ \ \ SC 29634-0973, USA. e-mail: string@clemson.edu

$\ \ \ \ \ \ \ \medskip $

In 1979 Louis Witten demonstrated that stationary axially symmetric Einstein
field equations and those for static axially symmetric self-dual SU(2) gauge
fields can both\ be reduced to the same (Ernst) equation. In this paper we
use this result as point of departure to prove the existence of the mass gap
for quantum source-free Yang-Mills (Y-M) fields. The proof is facilitated by
results of our recently published paper, JGP 59 (2009) 600-619. Since both
pure gravity, the Einstein-Maxwell and pure Y-M fields are described for
axially symmetric configurations by the Ernst equation classically, their
quantum\ descriptions are likely to be interrelated. Correctness of this
conjecture is successfully checked by reproducing (by different methods)
results of Korotkin and Nicolai, Nucl.Phys.B475 (1996) 397-439, on
dimensionally reduced quantum gravity. Consequently, numerous new results
supporting the Faddeev-Skyrme (F-S) -type models are obtained. We found that
the F-S-like model is best suited for description of electroweak
interactions while strong interactions require extension of Witten's results
to the SU(3) gauge group. Such an extension is nontrivial. It is linked with
the symmetry group SU(3)$\times $SU(2)$\times $U(1) of the Standard Model.
This result is quite rigid and should be taken into account in development
of \ all grand unified theories. Also, the alternative (to the F-S-like)
model emerges as by-product of such an extension. Both models are related to
each other via known symmetry transformation. Both models possess gap in
their excitation spectrum and are capable of producing knotted/linked
configurations of gauge/gravity fields. \ In addition, the paper discusses
relevance of \ the obtained results \ to heterotic strings \ and to
scattering processes involving topology change. It ends with discussion
about usefulness of this information for searches of Higgs boson.

\smallskip



\pagebreak

\ 

\bigskip

\bigskip

\section{Introduction}

\subsection{General remarks}

History of physics is full of situations when experimental observations lead
to deep mathematical results. Discovery of Yang-Mills (Y-M) fields in 1954
[1] falls out of this trend. Furthermore, if one believes that theory of
these fields makes sense, they should never be directly observed. \ To make
sure that these fields do exist, it is necessary to resort to all kinds of
indirect methods to probe them. Physically, the rationale for the Y-M fields
is explained already in the original Yang and Mills \ paper [1].
Mathematically, such a field is easy to understand. It is a non Abelian
extension of Maxwell's theory of electromagnetism. In 1956 Utiyama [2]
demonstrated that gravity, Y-M and electromagnetism can be obtained from
general principle of local gauge invariance of the underlying Lagrangian.
The explicit form of the Lagrangian is fixed then by assumptions about its
symmetry. For instance, by requiring invariance of such a Lagrangian with
respect to the Abelian U(1) group, the \ functional for the Maxwell field is
obtained, while\ doing the same operations but using the Lorentz group the
Einstein-Hilbert functional for gravitational field is recovered. By
employing the SU(2) non Abelian gauge group the original Y-M result [1] is
recovered.

Only Maxwell's electromagnetic field is reasonably well understood both at
the classical and quantum level. Due to their nonlinearity, the Y-M fields
are much harder to study even at the semi/classical level. In particular, no
classical solutions e.g. solitons (or lumps)\ with finite action are known
in Minkowski space-time. This result was proven by many authors, e.g. see
[3-4] and refrerences therein. The situation changes dramatically in
Euclidean space where the self-duality constraint allows to obtain
meaningful classical solutions [5,6]. These are helpful for development of
the theory of quantum Y-M fields. Such solutions are useful in the fields
other than quantum chromodynamics (QCD) since the self-duality equations are
believed to be at the heart of all exactly integrable systems [7]. Although
the self-duality equations\ originate from study of the Y-M functional, not
all solutions [6] of these equations are relevant to QCD. \ In this paper we
discuss the rationale behind the selection procedure. In QCD solutions of
self-duality equations, known as \textsl{instantons}, are describing
tunneling between different \ QCD vacua [8]. It should be noted though that
\ treatment of instantons in mathematics [9-11] and physics literature [8]
is different. This fact is important.\ It is important since one of the
major tasks of nonperturbative QCD lies in developing \ mathematically
correct \ and physically meanigful description of these vacua. According to
a point of view existing in physics literature the QCD has a countable
infinity of topologically different vacua. \ Supposedly, the Faddeev-Skyrme
(F-S) model \ is designed for description of these vacua. If this model can
be used for this purpose, then each vacuum state is expected to be
associated with a particular knot (or link) configuration. Under these
conditions the instantons are believed to be \ well localized objects
interpolating between different knotted/linked vacuum configurations
[12-16]. These configurations upon quantization are expected to posses a
tower of excited states. Whether or not such a tower has a gap in its
spectrum or \ the spectrum is gapless is the essence of the millennium prize
problem\footnote{%
E.g. see
\par
http://www.claymath.org/millennium/Yang-Mills\_Theory/}. Originally, the
above results were obtained and discussed only for SU(2) gauge fields [17].
They were extended to SU(N) case, $N\geq 2,$ only quite recently [18]%
\footnote{%
In this work, in accord with experimental evidence, we demonstrate that $%
N\leq 3.$}. Although such a description of \ QCD vacua is in accord with
general principles of instanton calculations [8], it is in\textsl{\ formal}
disagreement with results known in mathematics [9-11]. Indeed, it is well
known that complement of a particular knot in $S^{3}$ is 3-manifold. Since
instantons "live" in $\mathbf{R}^{4}$ (or any Riemannian 4-manifold allowing
an anti self -dual decomposition of the Y-M field (e.g. see Ref.[9], pages
38-39 \footnote{%
In physics literature, both anti and self dual instantons are allowed to
exist, e.g. see Ref.[19], page 481.}), this means that all knots \ in $%
\mathbf{R}^{4}($or $S^{4}$) are trivial and one should talk about knotted
spheres instead of knotted rings [20]. This known topological fact is in
apparent contradiction with results of [13-15]. In this work we shall
provide evidence that such a contradiction is only apparent and that,
indeed, knotted configurations in $S^{3}$ are consistent with the notion of
instantons as formulated in mathematics. This is achieved by using results
by Floer [21]. It should be noted, though, that known to us "proofs" \
[22-24] of the existence of the mass gap in pure Y-M theory done at the
physical level of rigor ignore instanton effects altogether. Among these
papers only Ref.[22] uses the F-S SU(2) model for such mass gap
calculations. It also should be noted that results\ of such calculation
sensitively depend upon the way the F-S model is quantized. For instance, in
the work by Faddeev and Niemi [25], done for the SU(2) gauge group, the
results of quantization produce gapless spectrum. To fix the problem the
same authors suggested to extend the original model in ad hock fashion.
Other authors, e.g. see Ref.[26], proposed different ad hoc solution of the
same problem.

The above results are formally destroyed by the effects of gravity. Indeed,
in 1988 Bartnik and McKinnon numerically demonstrated [27] that the combined
Y-M and gravity fields lead to a stable particle-like (solitonic) solutions
while neither source-free gravity nor pure Y-M fields are capable of
producing such solutions\footnote{%
More accurately, neither pure Y-M fields nor pure gravity have nontrivial
static globally regular (i.e.nonsingular, asymptotically flat) solitons.}.
Such situation has interesting cosmological ramifications\footnote{%
E.g. Einstein-Y-M hairy black holes} [28] causing disappearance of
singularities in spacetime as shown by \ Smoller et al [29]. In this work we
do not discuss implications of these remarkable results. Instead, in the
spirit of Floer's ideas [21], we argue that even without taking these
results into account, the effects of gravity on processes of high energy
physics are quite substantial.

\subsection{ Statements of the problems to be solved}

In this paper several problems are posed and solved. \ In particular, we
would like to investigate the physics and mathematics behind gravity-Y-M
correspondence discovered by Louis Witten [30] for SU(2) gauge fields. Is
this correspondence accidental? If it is not accidental, how it should be
related to commonly shared opinion that the Standard Model (SM) of particle
physics does not account for gravity? Can \ this correspondence be extended
to other gauge fields, e.g. SU(N), N\TEXTsymbol{>}2 ? If the answer is
"yes", will such correspondence be valid for all N's or just for few? \ In
the last case, what such a restriction means physically? How the noticed
correspondence is helping to solve the gap problem? \ What role the F-S
model is playing in this solution? \ Is this model instrumental in solving
the gap problem or are there other aspects of this problem which the F-S
model is unable to account? How this correspondence affects \ known
string-theoretic and loop quantum gravity (LQG) results ? \ What place the
topology-changing (scattering) processes occupy in this correspondence? Is
there any relevance of the results of this work to searches for Higgs boson?

\subsection{Organization of the rest of the paper and summary of obtained
results}

Sections 2,3 and 6,\textsl{\ \ }and Appendix A are devoted to detailed
investigation of gravity-Y-M correspondence. Section 4 is devoted to the
physics-style exposition of works by Andreas Floer [11, 21] on Y-M theory
with purpose of connecting his mathematical formalism for Y-M fields with
the F-S model. \ In the same section we also consider the Y-M fields
monopole and instanton solutions and their \ meaning and place within
Floer's theory. Our exposition is based on results of Sections 2 and 3.
Section 5 is entirely devoted to solution of the gap problem for pure Y-M
fields. \ Although \ the solution \ depends on results of previous sections,
numerous additional facts from statistical mechanics and nuclear physics are
being used. \ In Section 6 we discuss various implications/corollaries of
the obtained results, especially for the SM of particle physics. \ \ In
Section 7 we discuss possible directions for further research based on the
results presented in this paper. These include (but not limited to):
connections with the LQG, the role and place of the Higgs boson,
relationship between real space-time scattering processes of high energy
physics and processes of topology change associated with such scattering.
Based on the results of this paper, we argue that this task can be
accomplished with help of the formalism developed by G. Perelman for his
proof of the Poincare$^{\prime }$ and geometrization conjectures.

\ The major \ new results \ of this paper are summarized as follows.

1. In subsection 5.4.4, while solving the gap problem, we reproduced \ by
employing entirely different methods, the main results of the paper by
Korotkin and Nicolai [31] on quantizing dimensionally reduced gravity. From
these results it follows that for gravity and Y-M fields possessing the same
symmetry the nonperrturbative quantization proceeds essentially in the same
way.

2. In subsection 6.3 we demonstrated that gravity-Y-M correspondence
discovered by L.Witten for gauge group SU(2) can be extended \textsl{only}
to the SU(3) gauge group. This group contains SU(2)$\times $U(1) group as a
subgroup. This fact allowed us to come up with the anticipated (but never
proven!) conclusion about \ symmetry of the SM. It is given by SU(3)$\times $%
SU(2)$\times $U(1). The obtained result is very rigid. It is deeply rooted
into not widely known/appreciated (discussed in Appendix A) properties of \
the gravitational field. It is these properties which\ ultimately determine
the conditions of gravity-Y-M correspondence.

3. The latest papers Refs.[32-34] \ are aimed at reproduction of the
classification scheme of particles and fields in the SM within the framework
of LQG formalism. These results match perfectly with the results of our
paper because of the noticed and developed gravity-Y-M correspondence. In
view of this correspondence, the results of Refs.[32-34] can be reproduced
with help of \textsl{minimal} \ gravity model described in subsections 3.2,
3.4, and 7.2 . This \ minimal model has differential-geometric /topological
meaning in terms of the dynamics of the extended Ricci flow [35,36]. Such a
flow is the minimal extension of the Ricci flow \ now famous because of its
relevance in proving the Poincare$^{\prime }$ and geometrization
conjectures. \ 

4. The formalism developed in this paper explains why using pure gravity one
can talk about the particle/field content of the SM. Not only it is
compatible with just mentioned LQG results but also with those, coming from
noncommutative geometry [37], where it is demonstrated that use of pure
gravity (that is "minimal model") combined with 0- dimensional internal
space is sufficient for description of the SM.

\section{Emergence of the Ernst equation in pure gravity and Y-M fields}

\subsection{Some facts about the Ernst equation}

Study of static vacuum Einstein fields was initiated by Weyl in 1917.
Considerable progress made in later years \ is documented in Ref.[38]. To
develop formalism of this paper we need to discuss some facts about these
static fields. Following Wald [39], a spacetime is considered to be \textsl{%
stationary }if there is a one-parameter group of isometries $\sigma _{t\text{
}}$ whose orbits are time-like curves e.g. see [40]. With such group of
isometries is associated a time-like Killing vector $\xi ^{i}.$ Furthermore,
a spacetime is \textsl{axisymmetric} if there exists a one-parameter \ group
of isometries $\chi _{\phi \text{ }}$ whose orbits are closed spacelike
curves. Thus, a spacelike Killing vector field $\psi ^{i}$ has integral
curves \ which are closed. The spacetime is \textsl{stationary and
axisymmetric} if it possesses both of these symmetries, provided that $%
\sigma _{t\text{ }}\circ \chi _{\phi \text{ }}=\chi _{\phi \text{ }}\circ
\sigma _{t\text{ }}.$ If $\xi =(\frac{\partial }{\partial t})^{\text{ \ }}$%
and $\psi =(\frac{\partial }{\partial \phi })^{\text{ \ }}$ so that $[\xi
,\psi ]=0,$ one can choose coordinates as follows: $x^{0}=t,x^{1}=\phi
,x^{2}=\rho ,x^{3}=z.$ Under such identification, the metric tensor $g_{\mu
\nu }$ becomes a function of only $x^{2}$and $x^{3}.$ Explicitly,%
\begin{equation}
ds^{2}=-V(dt-wd\phi )^{2}+V^{-1}[\rho ^{2}d\phi ^{2}+e^{2\gamma }(d\rho
^{2}+dz^{2})],  \tag{2.1}
\end{equation}%
where functions $V$, $w$ and $\gamma $ depend on $\rho $ and $z$ only. In
the case when $V=1,w=\gamma =0,$ the metric can be presented as $%
ds^{2}=-\left( dt\right) ^{2}$ +$\left( d\tilde{s}\right) ^{2}$, where%
\begin{equation}
\left( d\tilde{s}\right) ^{2}=\rho ^{2}d\phi ^{2}+d\rho ^{2}+dz^{2} 
\tag{2.2}
\end{equation}%
is the standard flat 3 dimensional metric written in cylindrical
coordinates. The four-dimensional set of vacuum Einstein equations $R_{ij}=0$
with help of metric given by Eq.(2.2) acquires the following form%
\begin{equation}
\mathbf{\nabla \cdot \{}V^{-1}\mathbf{\nabla }V+\rho ^{-2}V^{2}w\mathbf{%
\nabla }w\}=0  \tag{2.3a}
\end{equation}%
and%
\begin{equation}
\mathbf{\nabla \cdot \{}\rho ^{-2}V^{2}\mathbf{\nabla }w\}=0.  \tag{2.3b}
\end{equation}%
In these equations $\mathbf{\nabla \cdot }$ and $\mathbf{\nabla }$ are
three-dimensional flat (that is with metric given by Eq.(2.2)) divergence
and gradient operators respectively. In addition to these two equations,
there are another two needed for determination of factor $\gamma $ in the
metric, Eq.(2.1). They require knowledge of $V$ and $w$ as an input.
Solutions of\ Eq.s(2.3) is described in great detail in the paper by Reina
and Trevers [41] with final result:

\begin{equation}
\left( \func{Re}\epsilon \right) \nabla ^{2}\epsilon =\mathbf{\nabla }%
\epsilon \cdot \mathbf{\nabla }\epsilon .  \tag{2.4}
\end{equation}%
This equation is known in literature as the Ernst equation. The complex
potential $\epsilon $ is defined in by $\epsilon =V+i\omega $ with $V$
defined as above and $\omega $ being an auxiliary potential whose explicit
form we do not need in this work. As it was recognized by Ernst [42,43] such
an equation can be also used for description of the combined
Einstein-Maxwell fields. \ We shall exploit this fact in Section 6. \textsl{%
In Appendix A and in} \textsl{Section 6 we provide proofs that knowledge of
static vacuum solutions} of the Ernst equation \textsl{is necessary and
sufficient} for restoration of static Einstein-Maxwell fields.\footnote{%
Surprisingly, upon changes of variables in these \ static solutions, the \
exact results for propagating gravitational waves can be obtained as well.}\
Fields other than Y-M should be also restorable \footnote{%
This is so because each of these fields is a source of gravitational field
which, in turn, can be eliminated locally. See Appendix A.}. To proceed, we
need to list several properties of the Ernst equation to be used below.
First, following [41] and using prolate spheroidal coordinates, the Ernst
equation reproduces the Schwarzschild metric, and with another choice of
coordinates it reproduces the Kerr and Taub-NUT metric. Thus, the Ernst
equation is the most general equation describing \ physically interesting
vacuum spacetimes compatible with the Cauchy formulation of general
relativity\ [39,40,44,45]. Such a formulation is convenient staring point
for quantization of gravitational field via superspace formalism [39]
leading to the Wheeler- De Witt equation, etc. \textsl{Since in this work we
advocate different \ approach to quantization of gravity, this topic is not
being discussed further. \ }Second, following Ref.[38], page 283, a \textsl{%
stationary} solution of Einstein's field equations is called \textsl{static }%
if the timelike Killing vector is orthogonal to the Cauchy surface. In such
a case from the Table 18.1. of the same reference \ it follows that the
Ernst potential $\epsilon $ is real. This observation allows us to simplify
Eq.(2.4) considerably. For the sake of \ notational comparison with Ref.[38]
we redefine the potential $\epsilon =V+i\Phi .$ In the static case we have $%
\epsilon \equiv -F\equiv -e^{2u}\footnote{%
The minus sign in front of $F$ is written in accord with \ conventions of
Chapter 30.2 of the 1st edition of Ref.[38].}.$ Using this result in
Eq.(2.4) produces%
\begin{equation}
\Delta _{\rho ,z}u=0,  \tag{2.5}
\end{equation}%
where $\Delta _{\rho ,z}$ is flat Laplacian written in cylindrical
coordinates defined by the metric, Eq.(2.2).

\subsection{ Isomorphism between the SU(2) self-dual gauge \ and vacuum
Einstein \ field equations}

This isomorphism was discovered by Louis Witten in 1979 [30]. His work was
inspired by \ earlier works of Ernst [42] and Yang [46]. \ To our knowledge,
since time when Ref.[30] was published such an isomorphism was \ left
undeveloped. In this paper we correct this omission in order to demonstrate
that when both fields are mathematically indistinguishable, their
quantization should proceed in the same way. The result analogous to that
discovered by Witten was obtained using different arguments\ a year later by
Forgacs, Horvath and Palla [47] and, in a simpler form, by Singleton [48].
These authors used essentially the paper by Manton, Ref.[49], in which it
was cleverly demonstrated that the 't Hooft-Polyakov monopole can be
obtained \textsl{without} \ actual use of the auxiliary Higgs field. Both
Refs.[47,48] and the original paper by Witten [30] use the axial symmetry of
either gravitational or Y-M fields essentially. Only in this case it can be
shown that the axisymmetric version of the self-duality equations obtained
by Manton can be rewritten in the form of the Ernst equation. In the light
of above information, following Ref.[5\textbf{],} we shall discuss briefly
contributions of Yang and Witten. For this purpose, we need to consider
first the following auxiliary system of \textsl{linear} equations%
\begin{equation}
\mathbf{\Psi }_{x}=\mathbf{X\Psi ;\Psi }_{t}=\mathbf{T\Psi }.  \tag{2.6}
\end{equation}%
Here $\mathbf{\Psi }_{x}=\frac{\partial }{\partial x}\mathbf{\Psi }$ and $%
\mathbf{\Psi }_{t}=\frac{\partial }{\partial t}\mathbf{\Psi }.$ In this
system $\mathbf{X}$ and $\mathbf{T}$ are square matrices of the same
dimension and such that 
\begin{equation}
\mathbf{X}_{t}-\mathbf{T}_{x}+[\mathbf{X},\mathbf{T}]=0  \tag{2.7}
\end{equation}%
This\ result easily follows from the compatibility condition: $\mathbf{\Psi }%
_{xt}=\mathbf{\Psi }_{tx}$. The matrices \textbf{X }and \textbf{T} can be
realized as 
\begin{equation}
\mathbf{X}=\left( 
\begin{array}{cc}
-i\zeta & q(x,t) \\ 
r(x,t) & i\zeta%
\end{array}%
\right) \text{ , \ }\mathbf{T}=\left( 
\begin{array}{cc}
A & B \\ 
C & -A%
\end{array}%
\right)  \tag{2.8}
\end{equation}%
with $\zeta $ being a spectral parameter and, $\ A,B$ and $C$ being some
Laurent polynomials in $\zeta .$ The above system can be extended to four
variables $x_{1},x_{2},t_{1},t_{2}$ in a simple minded fashion as follows%
\begin{equation}
(\frac{\partial }{\partial x_{1}}+\zeta \frac{\partial }{\partial x_{2}})%
\mathbf{\Psi =(X}_{1}\mathbf{+}i\mathbf{X}_{2}\mathbf{)\Psi ,}  \tag{2.9a}
\end{equation}%
\begin{equation}
(\frac{\partial }{\partial t_{1}}+\zeta \frac{\partial }{\partial t_{2}})%
\mathbf{\Psi =(T}_{1}\mathbf{+}i\mathbf{T}_{2}\mathbf{)\Psi .}  \tag{2.9b}
\end{equation}%
In the most general case, the matrices $\mathbf{X}_{1},\mathbf{X}_{2},%
\mathbf{T}_{1},\mathbf{T}_{2}$ \ are made of functions which "live" in 
\textbf{C}$^{4}.$ They are representatives of the Lie algebra $sl(n,\mathbf{C%
})$ of $n\times n$ trace-free matrices. The compatibility conditions for
this case are equivalent to the self-duality condition for the Y-M fields \
associated with algebra $sl(n,\mathbf{C}).$It is instructive to illustrate
these general statements explicitly.

In $\mathbf{R}^{4}$ the (anti)self-duality condition for the Y-M curvature
reads: $\ast F=(-1)F$ so that for the self-dual case we obtain: 
\begin{equation}
F_{01}=F_{23},F_{02}=F_{31},F_{03}=F_{12}.  \tag{2.10}
\end{equation}%
In the "light cone" coordinates $\sigma =\frac{1}{\sqrt{2}}%
(x_{1}+ix_{2}),\tau =\frac{1}{\sqrt{2}}(x_{0}+ix_{3})$ the Y-M field
one-form can be written as $A_{\mu }dx^{\mu }=A_{\sigma }d\sigma +A_{\tau
}d\tau +A_{\bar{\sigma}}d\bar{\sigma}+A_{\bar{\tau}}d\bar{\tau}$ with the
overbar labeling the complex conjugation. In such notations $A_{0}=\frac{1}{%
\sqrt{2}}(A_{\tau }+A_{\bar{\tau}}),$ $A_{1}=\frac{1}{\sqrt{2}}(A_{\sigma
}+A_{\bar{\sigma}}),$ $A_{2}=\frac{1}{\sqrt{2}}(A_{\sigma }-A_{\bar{\sigma}%
}),$ $A_{3}=\frac{1}{\sqrt{2}}(A_{\tau }-A_{\bar{\tau}}).$ In these
notations Eq.s (2.9) acquire the following form%
\begin{equation}
F_{\sigma \tau }=0,\text{ }F_{\bar{\sigma}\bar{\tau}}=0\text{ and }F_{\sigma 
\bar{\sigma}}+F_{\tau \bar{\tau}}=0.  \tag{2.11}
\end{equation}%
They can be obtained as compatibility condition for the isospectral linear
problem%
\begin{equation}
(\partial _{\sigma }+\zeta \partial _{\bar{\tau}})\mathbf{\Psi =(}A_{\sigma }%
\mathbf{+}\zeta A_{\bar{\tau}}\mathbf{)\Psi }\text{ and }(\partial _{\tau
}-\zeta \partial _{\bar{\sigma}})\mathbf{\Psi =(}A\mathbf{_{\tau }-\zeta }A%
\mathbf{_{\bar{\sigma}})\Psi ,}  \tag{2.12}
\end{equation}%
where the spectral parameter is $\zeta $ and $\mathbf{\Psi }$ is the local
section of the Y-M fiber bundle. The compatibility condition reads: $%
(\partial _{\sigma }-\zeta \partial _{\bar{\tau}})(\partial _{\sigma }+\zeta
\partial _{\bar{\tau}})\mathbf{\Psi }=(\partial _{\sigma }+\zeta \partial _{%
\bar{\tau}})(\partial _{\sigma }-\zeta \partial _{\bar{\tau}})\mathbf{\Psi ,}
$ thus leading to 
\begin{equation}
\lbrack F_{\sigma \tau }-\zeta (F_{\sigma \bar{\sigma}}+F_{\tau \bar{\tau}%
})+\zeta ^{2}F_{\bar{\sigma}\bar{\tau}}]\mathbf{\Psi }=0.  \tag{2.13}
\end{equation}%
This equation allows us to recover Eq.s(2.11). \ The first two equations of
Eq.s(2.11) can be used in order to represent the $A$ -fields as follows: $%
A_{\sigma }=\left( \partial _{\sigma }C\right) C^{-1},$ $A\mathbf{_{\tau }=}%
\left( \partial _{\tau }C\right) C^{-1},$ $A\mathbf{_{\bar{\sigma}}=}\left(
\partial _{\bar{\sigma}}D\right) D^{-1}$ and $A_{\bar{\tau}}=\left( \partial
_{\tau }D\right) D^{-1},$ where both $C$ and $D$ are some matrices in the
Lie group $G$, e.g. $G=SU(2)$. \ By introducing the matrix $M=C^{-1}D\in G$
the last of equations in Eq.(2.11) becomes%
\begin{equation}
\partial _{\bar{\sigma}}(M^{-1}\partial _{\sigma }M)+\partial _{\bar{\tau}%
}(M^{-1}\partial _{\tau }M)=0.  \tag{2.14a}
\end{equation}%
Thus, the self-duality conditions for the Y-M fields are equivalent to
Eq.(2.14a). For the future use, following Yang [46], we notice that in such
formalism the gauge transformations for Y-M fields are expressible through $%
D\rightarrow DE$ and $C\rightarrow CE$ so that $F_{\sigma \bar{\sigma}%
}\rightarrow E^{-1}F_{\sigma \bar{\sigma}}E$ and $F_{\tau \bar{\tau}%
}\rightarrow E^{-1}F_{\tau \bar{\tau}}E$ with the matrix $E=E(\sigma ,\bar{%
\sigma},\tau ,\bar{\tau})\in $ $SU(2)$ leaving self-duality Eq.s(2.10) (or
(2.13)) unchanged.

To connect Eq.(2.14a) with the Ernst equation, following L.Witten [30] it is
sufficient to assume that the matrix $M$ is a function of $\rho =\sqrt{%
x_{1}^{2}+x_{2}^{2}}$ and $z=x_{3}.$ In such a case it is useful to remember
that $\rho ^{2}=2\sigma \bar{\sigma}$ and $z=\frac{i}{\sqrt{2}}(\tau -\bar{%
\tau})$. With help of these facts Eq.(2.14a) can be rewritten as 
\begin{equation}
\partial _{\rho }(\rho M^{-1}\partial _{\rho }M)+\rho \partial
_{z}(M^{-1}\partial _{z}M)=0.  \tag{2.14b}
\end{equation}%
By assuming that the matrix $M$ is representable by the $SL(2,R)$-type
matrix, and writing it in the form 
\begin{equation}
M\mathbf{=}\frac{1}{V}\left( 
\begin{array}{cc}
1 & \Phi \\ 
\Phi & \Phi ^{2}+V^{2}%
\end{array}%
\right) ,  \tag{2.15}
\end{equation}%
Eq.(2.14b) is reduced to the pair of equations 
\begin{equation*}
V\nabla ^{2}V=\nabla V\cdot \nabla V-\nabla \Phi \cdot \nabla \Phi \text{
and }V\nabla ^{2}\Phi =2\nabla V\cdot \nabla \Phi .
\end{equation*}%
With help of the Ernst potential $\epsilon =V+i\Phi $ these \ two equations
can be brought to the canonical form of the Ernst equation, Eq.(2.4). Below,
we shall provide sufficient evidence that such a reduction of the Ernst
equation is compatible with analogous reduction in instanton/monopole
calculations for the Y-M fields.

\section{From analysis to synthesis}

\subsection{General remarks}

The results of previous section \ demonstrate that for axially symmetric
fields both pure gravity and pure self-dual Y-M fields are described by the
same (Ernst) equation. In this section we reformulate these results in terms
of the nonlinear sigma model with purpose of using such a reformulation
later in the text. To do so we need to recall some results from our recent
works, Ref.s[50,51]. In particular, \ we notice that under conformal
transformations $\hat{g}=e^{2u}g$ in $d$-dimensions the curvature scalar \ $%
R(g)$ changes as follows: 
\begin{equation}
\hat{R}(\hat{g})=e^{-2u}\{R(g)-2(d-1)\Delta _{g}u-(d-1)(d-2)\left\vert
\bigtriangledown _{g}u\right\vert ^{2}\}.  \tag{3,1}
\end{equation}%
Since this equation is Eq.(2.11) of \ our Ref.[50] we shall be interested
only in transformations for which $\hat{R}(\hat{g})$ is a constant. This is
possible only if the total volume of the system is conserved. Under this
constraint we need to consider Eq.(3.1) for $d=3$ in more detail.\ Without
loss of generality we can assume that initially $R(g)=0$. For this case we
shall write $g=g_{0}$ so that Eq.(3.1) acquires the form%
\begin{equation}
\hat{R}(\hat{g})=-2e^{-2u}[2\Delta _{g_{0}}u+\left\vert \bigtriangledown
_{g_{0}}u\right\vert ^{2}]  \tag{3.2}
\end{equation}%
in which $\Delta _{g_{0}}$ is the flat space Laplacian. Now we can formally
identify it with that in Eq.(2.5). \ Accordingly, we shall be interested in
such conformal transformations for which $\Delta _{g_{0}}u=0$ in Eq.(3.2).
If they exist, Eq.(3.2) can be rewritten as 
\begin{equation}
e^{2u}\hat{R}(\hat{g})=-2\left( \vec{\bigtriangledown}_{g_{0}}u\right) \cdot
\left( \vec{\bigtriangledown}_{g_{0}}u\right) .  \tag{3.3}
\end{equation}%
This allows us to interpret Eq.(3.3) and%
\begin{equation}
\Delta _{g_{0}}u=0  \tag{3.4}
\end{equation}%
as interdependent equations: solutions of Eq.(3.4) determine the scalar
curvature $\hat{R}(\hat{g})$ in Eq.(3.3)$.$ Clearly, under conditions at
which these results are obtained only those solutions of Eq.(3.4) should be
used which yield the constant scalar curvature $\hat{R}(\hat{g}).$ \
Eq.(3.3) \ contains information about the Ricci tensor. To recover this
information we notice that $\hat{g}_{ij}=-e^{2u}\delta _{ij}$. Therefore we
obtain: 
\begin{equation}
\hat{R}_{ij}(\hat{g})=2\nabla _{i}u\nabla _{j}u,  \tag{3.5}
\end{equation}%
in accord with Eq.(18.55) of Ref.[38] where this result was obtained by
employing entirely different arguments. From the same reference we find that
Eq.(3.4) comes as result of use of the contracted Bianci identities applied
to $\hat{R}_{ij}(\hat{g})\footnote{%
Ref.[38], page 283, bottom}.$ It is instructive to place the obtained
results into broader context. This is accomplished in the next subsection.

\subsection{Connection with the nonlinear sigma model}

Some time ago Neugebauer and Kramer (N-K), Ref.[38], obtained Eq.s(3.4) and
(3.5) using variational principle. In less general form this principle was
used previously by Ernst [42] \ resulting in now famous Ernst equation.
Neugebauer and Kramer proposed the Lagrangian \ and the associated with it
action functional $S_{N-K}$ producing upon minimization both Eq.s(3.4) and
(3.5). To describe these results, we also use some results by Gal'tsov [52].

The functional $S_{N-K}$ is given by 
\begin{equation}
\mathcal{S}_{N-K}=\frac{1}{2}\int\limits_{M}\sqrt{\hat{g}}[\hat{R}(\hat{g})-%
\hat{g}^{ij}G_{AB}(\mathbf{\varphi })\partial _{i}\varphi ^{A}\partial
_{j}\varphi ^{B}]d^{3}x,  \tag{3.6}
\end{equation}%
easily recognizable as three-dimensional nonlinear sigma model coupled to
3-d Euclidean gravity. The number of \ components for \ the auxiliary field $%
\mathbf{\varphi }$ \ as well as the metric tensor $G_{AB}(\mathbf{\varphi })$
of the target space is determined by the problem in question. In our case
upon variation of $S_{N-K}$ with respect to $\varphi _{i}^{A}$ and $\hat{g}%
_{ij}$ we should be able to re obtain Eq.s(3.4) and (3.5).\ To do so,\
following Ref.[53], \ we introduce the current%
\begin{equation}
J_{i}=M^{-1}\partial _{i}M.  \tag{3.7}
\end{equation}%
In view of results of subsection\textsl{\ }2.2, we have to identify the
matrix $M$ with that defined by Eq.(2.15) and, taking into account
Eq.(2.14a), the index $i$ should take two values: $\sigma $ and $\tau .$
With such definitions we can replace the functional $\mathcal{S}_{N-K}$ by \
\ \ \ \ \ \ \ \ \ \ \ \ \ \ \ \ \ 
\begin{equation}
\mathcal{S}=\frac{1}{2}\int\limits_{M}\sqrt{\hat{g}}[\hat{R}(\hat{g})-\hat{g}%
^{ik}\frac{1}{4}tr(J_{i\text{ }}J_{k})]d^{3}x.  \tag{3.8}
\end{equation}%
\ \ The actual calculations with such type of functionals \ can be made
using results of Ref.[53]. Thus, \ using this reference we obtain, 
\begin{equation}
\hat{R}_{ij}(\hat{g})=\frac{1}{4}tr(J_{i\text{ }}J_{j})  \tag{3.9}
\end{equation}%
and%
\begin{equation}
\partial _{i}J_{i\text{ }}=0.  \tag{3.10}
\end{equation}%
Evidently, by construction Eq.(3.10) coincides with Eq.(2.14a) and,
ultimately, with Eq.(3.4). It is also easy also to check that Eq.(3.9) \
does coincide with Eq.(3.5). For this purpose it is sufficient to notice
that 
\begin{equation}
tr(J_{i\text{ }}J_{j})=-tr(\partial _{i}M\partial _{j}M^{-1}).  \tag{3.11}
\end{equation}%
To check correctness of our calculations the entries of the matrix $M$,
Eq.(2.15), can be restricted to $V$ (that is we can put $\Phi =0).$ Since $%
V=-F\equiv -e^{2u}$ (e.g. see discussion prior to Eq.(2.5)), a simple
calculation indeed brings Eq.(3.9) back to Eq.(3.5) as required.

It is interesting and important to observe at this point that\textsl{\ the
equation} \textsl{of motion, Eq.(3.10), formally is not affected by effects
of gravity}. This conclusion requires some explanation. From subsection 2.2,
especially from Eq.s(2.14),(2.15), it should be clear that Eq.(3.10) is the
Ernst equation determining gravitational field. Hence, it is physically
wrong to expect that it is going to be affected by the effects of gravity.
Eq.s(3.9) and (3.10) are the same as Eq.s(3.5) and (3.4) whose meaning was
explained in the previous subsection. Clearly, the functional, Eq.(3.8), can
be used for coupling of \textsl{other} fields to gravity. This is indeed
demonstrated in Ref.[52]. This is done with purpose of connecting results
for the nonlinear sigma models with those for heterotic strings. We would
like to discuss this connection now since it will be used later in the text.

\subsection{Connection with heterotic string models}

The functional $\mathcal{S}$, Eq.(3.8), \ is related to that for the
heterotic string model, e.g. see Ref.[54]. For such a model the sigma
model-like functional is obtainable from 10 dimensional supersymmetric
string model by means of compactification scheme (ideologically similar to
that used in the Kaluza-Klein theory of gravity and electromagnetism) aimed
at bringing the \ effective dimensionality to physically acceptable values
(e.g. 2, 3 or 4). For dimensionality $D<10$ such compactified/reduced action
functional reads (e.g. see Ref.[54], Eq.(9.1.8)): 
\begin{eqnarray}
S_{D}^{heterotic} &=&\int d^{D}x\sqrt{-\det G}e^{-2\phi }[R+4\partial ^{\mu
}\phi \partial _{\mu }\phi -\frac{1}{12}\hat{H}^{\mu \nu \rho }\hat{H}_{\mu
\nu \rho }  \notag \\
&&-\frac{1}{4}(M^{-1})_{ij}F_{\mu \nu }^{i}F^{j\mu \nu }+\frac{1}{8}%
tr(\partial _{\mu }M\partial ^{\mu }M^{-1})].  \TCItag{3.12}
\end{eqnarray}%
The compactification procedure is by no means unique. There are many ways to
make a compactified action to look exactly like that given by Eq.(3.8) (e.g.
see [55\textbf{]}). Evidently, there should be a way to relate such actions
to each other since they all are having the same origin - 10 dimensional
heterotic string action. Because of this, we would like to make some
comments on action given by Eq.(3.12) by specializing to $D=3$ for reasons
explained in Refs[\textbf{68,69}] and to be clarified below, in Section 6.
Under such conditions if we require the dilaton $\phi $, the antisymmetric $%
H $-field (associated with string orientation) and the electromagnetic field 
$F $ to vanish$,$ the remaining action will coincide with that given by
Eq.(3.8). Because of this, the following steps can be made.

\ First, \ as explained in our work, Ref.[50], for closed 3-manifolds we
can/should drop the dilaton field $\phi $. Second, by properly selecting
string model we can ignore the antisymmetric field $H$. Third, by taking
into account results of Appendix A we can also drop the electromagnetic
field since it can be always restored from pure gravity. Thus, we end up
with the action functional $\mathcal{S}$, Eq.(3.8), which we shall call "%
\textsl{minimal}". In Section 6 we shall provide evidence that its
minimality is deeply rooted into gravity-Y-M correspondence which does not
leave much room for "improvements" abundant in physics literature. We shall
begin explaining this fact immediately below and will end our arguments in
Section 6.

\subsection{The extended Ricci flow}

Thus far use of the variational principle apparently had not brought us any
new results (at least at the classical level). Situation changes in the
light of recent work by List [35]. Following Ref.s[35,36\textbf{]}, it is
convenient to introduce Perelman-like entropy functional $\mathcal{F}(\hat{g}%
_{ij},u,f)$%
\begin{equation}
\mathcal{F}(\hat{g}_{ij},u,f)=\int\limits_{M}(\hat{R}(\hat{g})-2\left\vert
\nabla _{\hat{g}}u\right\vert ^{2}+\left\vert \nabla _{\hat{g}}f\right\vert
^{2})e^{-f}dv  \tag{3.13}
\end{equation}%
coinciding with Eq.(7.22b) of our work, Ref.[50], when $u=0$.$\footnote{%
It should be noted that there is an obvious typographical error in
Eq.(7.22b): the term $\left\vert \nabla _{h}f\right\vert ^{2}$ is typed as $%
\left\vert \nabla _{h}f\right\vert .$}$. Because of this observation, if
formally we make a replacement $\mathcal{R(}\hat{g};u)=\hat{R}(\hat{g}%
)-2\left\vert \nabla u\right\vert ^{2}$ in Eq.(3.13), \ we are able to
identify Eq.(3.13) with Perelman's entropy functional enabling us to follow
the same steps as were made in Perelman's papers aimed at proof of the
geometrization and Poincare$^{\prime }$ conjectures. Such a program was
indeed completed in the PhD thesis by List [36]. Minimization of entropy
functional $\mathcal{F}(\hat{g}_{ij},f)$ produces \ the \ following set of
equations%
\begin{equation}
\frac{\partial }{\partial t}g_{ij}=-2(\hat{R}_{ij}+\nabla _{i}\nabla _{j}f)%
\text{ }+4\nabla _{i}u\nabla _{j}u,  \tag{3.14a}
\end{equation}%
\begin{equation}
\frac{\partial }{\partial t}u=\Delta _{\hat{g}}u-\left( \nabla u\right)
\cdot \left( \nabla f\right) ,  \tag{3.14b}
\end{equation}%
and%
\begin{equation}
\frac{\partial }{\partial t}f=-\hat{R}-\Delta _{\hat{g}}f+2\left\vert \nabla
u\right\vert ^{2},  \tag{3.14c}
\end{equation}%
coinciding with Eq.s(7.28a), (7.28b) of our work, Ref.[50], when $u=0.$In
these equations $\left\vert \nabla _{\hat{g}}u\right\vert ^{2}$ $=\hat{g}%
^{ij}\nabla _{i}u\nabla _{j}u$ , etc. \ From the next section and results
below it follows that physically we should be interested in closed 3
manifolds. \ For such manifolds one can use Lemma 2.13, proven by List [36],
which can be formulated as follows:

Let $\hat{g},u,f$ be a solution of Eq.s(3.14) for $t\in \lbrack 0,T)$ on a
closed manifold $M$. Then the evolution of the entropy is given by%
\begin{equation}
\partial _{t}\mathcal{F}(\hat{g}_{ij},u,f)=2\int\limits_{M}[\left\vert 
\mathcal{R}_{ij}\mathcal{(}\hat{g};u)+\nabla _{i}\nabla _{j}f\right\vert
^{2}+2(\Delta _{\hat{g}}u-\left( \nabla u\right) \cdot \left( \nabla
f\right) )^{2}]e^{-f}dv\geq 0.  \tag{3.15}
\end{equation}%
Thus, the entropy is non decreasing with equality taking place if and only
if the solution of Eq.(3.14) is a gradient soliton. This happens when the
following two conditions hold%
\begin{equation}
\mathcal{R}_{ij}\mathcal{(}\hat{g};u)+\nabla _{i}\nabla _{j}f=0\text{ and }%
\Delta _{\hat{g}}u-\left( \nabla u\right) \cdot \left( \nabla f\right) =0. 
\tag{3.16}
\end{equation}%
For $u=0$ the result of Perelman, Eq.(7.30) of Ref [50], for steady gradient
soliton is reobtained, as required. Since for closed compact manifolds $%
f=const$ Eq.s(3.16) coincide with Eq.s(3.4) and (3.5) as anticipated. 
\textsl{Thus, existence of steady gradient solitons in the present context
is equivalent to existence of solutions of static Einstein's equations for
pure gravity.} This fact alone could be mathematically interesting but
requires some reinforcement to be of interest physically. We initiate this \
reinforcement process in the following subsection.

\subsection{Relationship between the F-S and the Ernst functionals}

The F-S functional was mentioned in the Itroduction. In this subsection we
would like to initiate study of its connection with the Ernst functional. We
begin with the following observation. In steps leading to Eq.(2.14b) (or
(3.10)) the Euclidean time coordinate $x_{0}$ was eventually dropped
implying that solutions of selfduality for Y-M equations, when substituted
back into Y-M action functional, will produce physically meaningless
(divergent) results. While in subsection 4.4 we discuss a variety of means
for removing of such apparent divergence, in this subsection we notice that
already Ernst [42] suggested the action functional whose minimization
produces the Ernst equation. He gave two equivalent forms for such a
functional, now bearing his name. These are either%
\begin{equation}
S_{E_{1}}[\epsilon ]=\int\limits_{M}dv\frac{\mathbf{\nabla }\epsilon \cdot 
\mathbf{\nabla }\epsilon ^{\ast }}{\left( \func{Re}\epsilon \right) ^{2}} 
\tag{3.17}
\end{equation}%
or%
\begin{equation}
S_{E_{2}}[\xi ]=\int\limits_{M}dv\frac{\mathbf{\nabla }\xi \cdot \mathbf{%
\nabla }\xi ^{\ast }}{\left( \xi \xi ^{\ast }-1\right) ^{2}}.  \tag{3.18}
\end{equation}%
Minimization of $S_{E_{1}}[\epsilon ]$ leads to Eq.(2.4) while functional,
Eq.(3.18), is obtained from $S_{E_{1}}[\epsilon ]$ by means of substitution: 
$\epsilon =(\xi -1)/(\xi +1).$ In both functionals $dv$ is 3-dimensional
Euclidean volume element so that apparently the manifold $M$ is just $E^{3}$
(or, with one point compactification, it is $S^{3}).$Evidently$,$both $%
S_{E_{1}}[\epsilon ]$ and $S_{E_{2}}[\xi ]$ are functionals for the
nonlinear sigma model. If we drop the curvature term in Eq.(3.6) such
truncated functional can be identified, for example, with $S_{E_{2}}[\xi ]$.
This explains why Eq.(3.10) is formally unaffected by gravity. In
mathematics literature the nonlinear sigma models are known as harmonic
maps. \ Since Reina [56] demonstrated that the functional $S_{E_{2}}[\xi ]$
describes the harmonic map from $S^{3}$ to \textbf{H}$^{2},$ it is not too
difficult to write analogous functional $S_{E_{3}}[\xi ]$ describing the
mapping from $S^{3}$ to $S^{2}.$ It is given by%
\begin{equation}
S_{E_{3}}[\xi ]=\int\limits_{M}dv\frac{\mathbf{\nabla }\xi \cdot \mathbf{%
\nabla }\xi ^{\ast }}{\left( \xi \xi ^{\ast }+1\right) ^{2}}  \tag{3.19}
\end{equation}%
and is part of the F-S model. If needed, both $S_{E_{2}}[\xi ]$ and $%
S_{E_{3}}[\xi ]$ can be supplemented by additional (topological) terms which
in the simplest case are winding numbers. Thus, we shall be dealing either
with the truncated F-S model, Eq.(3.19), or with its hyperbolic analog,
Eq.(3.18). \ The choice between these models is nontrivial and will
discussed in detail in Section 6\textsl{.} To facilitate this discussion, we
need to observe the following. In the static case, we argued, e.g. see
Eq.(2.5), that $\epsilon =-F=-e^{2u}.$ Substitution of this result back into 
$S_{E_{1}}[\epsilon ]$ produces (up to a constant) the following result:%
\begin{equation}
\tilde{S}_{E_{1}}[\epsilon ]=\int\limits_{M}dv\mathbf{\nabla }u\cdot \mathbf{%
\nabla }u  \tag{3.20}
\end{equation}%
leading to Eq.(2.5) as anticipated. At the same time, consider the following
H-E action functional%
\begin{equation}
S_{H-E}[\hat{g}]=\int\limits_{M}dv\sqrt{\hat{g}}\hat{R}(\hat{g}),  \tag{3.21}
\end{equation}%
and take into account Eq.(3.3) and the fact that $\hat{g}_{ij}=-e^{2u}\delta
_{ij\text{ }}$. Straightforward calculation leads us then to the result (up
to a constant):%
\begin{equation}
S_{H-E}[\hat{g}]=-\int\limits_{M}dv\mathbf{\nabla }u\cdot \mathbf{\nabla }u.
\tag{3.22}
\end{equation}%
The minus sign in front of the integral is important and will be explained
below. Before doing so, we notice that the Ernst functional (in whatever
form) is essentially equivalent to the H-E functional! Since in the original
paper by Ernst $M$ is $E^{3}$ (or $S^{3}),$ apparently, such a functional
should be zero. This is surely not the case in general but the explanation
is nontrivial.\ Suppose that minimization of the Ernst functional leads to
some knotted/linked structures\footnote{%
We shall postpone detailed discussion of this topic till Section 6\textsl{.}}%
. If such knots/links are hyperbolic then, by construction, complements of
these knots/links in $S^{3}$ are $\mathbf{H}^{3}$ modulo some discrete
group. This conclusion is in accord with properties of the Ernst equation
discovered by Reina and Trevers [41]. Following this reference,\ we
introduce the complex space $\mathbf{C}\times \mathbf{C}=\mathbf{C}^{2}$ so
that $\forall $ $\mathbf{z=(}u\mathbf{,}v\mathbf{)}\in \mathbf{C}^{2}$ the
scalar product $z_{\alpha }^{\ast }z^{\alpha }$ can be made with the metric $%
\pi _{\alpha \beta }=diag\{1,-1\}$. Furthermore, the Ernst Eq.(2.4) can be
rewritten with help of substitution $\epsilon =(u-v)/(u+v)$ as the set of
two equations 
\begin{equation}
z^{\alpha }z_{\alpha }^{\ast }\nabla ^{2}z^{\beta }=2z_{\alpha }^{\ast }%
\mathbf{\nabla }z^{\alpha }\cdot \mathbf{\nabla }z^{\beta }.  \tag{3.23}
\end{equation}%
Such a system of equations is invariant with respect to transformations from
unimodular group SU(1,1) which is equivalent to SL(2,\textbf{C}). But SL(2,%
\textbf{C}) is the group of isometries of hyperbolic space $\mathbf{H}^{3}$
as was discussed extensively in our work, Ref.[57]. Thus, minimization of
both the F-S and Ernst functionals should account for knotted/linked
structures. This conclusion is strengthened in the next subsection.

\subsection{Relationship between the Ernst and Chern-Simons functionals}

Even though we need to find this relationship anticipating results of the
next section, by doing so, some unexpected connections with previous
subsection \ are also going to be revealed. For this purpose, we notice that
\ for $u=0$ the functional $\mathcal{F}(\hat{g}_{ij},u,f)$ introduced
earlier is just Perelman's entropy functional. As such, it was discussed in
our work, Ref.[50]. Evidently, both $\mathcal{F}$ and Perelman's functional
can be used for study of \ topology of 3-manifolds. We believe, though, that
use of Perelman's functional is more advantageous as we would like to
explain now. For this purpose,\ it is convenient to introduce the Raleigh
quotient $\lambda _{g}$ via%
\begin{equation}
\lambda _{g}=\inf_{\varphi }\frac{\int\limits_{M}dV(4\left\vert \nabla
\varphi \right\vert ^{2}+R(g)\varphi ^{2})}{\int\limits_{M}dV\varphi ^{2}}, 
\tag{3.24}
\end{equation}%
e.g. see Eq.(7.24) of [50], to be compared against the Yamabe quotient $(p=%
\frac{2d}{d-2}$ and $\alpha =4\frac{d-1}{d-2})$ .%
\begin{equation*}
Y_{g}=\frac{\int d^{d}x\sqrt{\hat{g}}\hat{R}(\hat{g})}{\left( \int d^{d}x%
\sqrt{\hat{g}}\right) ^{\frac{2}{p}}}=\left( \frac{1}{\int\nolimits_{M}d^{d}x%
\sqrt{g}\varphi ^{p}}\right) ^{\frac{2}{p}}\int\nolimits_{M}d^{d}x\sqrt{g}%
\{\alpha \left( \nabla _{g}\varphi \right) ^{2}+R(g)\varphi ^{2}\}\equiv 
\frac{E[\varphi ]}{\left\Vert \varphi \right\Vert _{p}^{2}}
\end{equation*}%
also discussed in [50]. Because of similarity of these two quotients the
question arises: Can they be equal to each other? The affirmative answer to
this question is obtained in Ref.[58]. It can be formulated as

\textbf{Theorem} [58]. Suppose that $\gamma $ is a conformal class on $M$
which \textsl{does not} contain metric of \textsl{positive scalar} $\QTR{sl}{%
curvature.}$ Then 
\begin{equation}
Y_{\gamma }=\sup_{g\in \gamma }\lambda _{g}V(g)^{\frac{2}{d}}\equiv \bar{%
\lambda}(M),  \tag{3.25a}
\end{equation}%
where $\bar{\lambda}(M)$ is Perelman's $\bar{\lambda}$ invariant.
Equivalently, 
\begin{equation}
\lambda _{g}V(g)^{\frac{2}{d}}\leq Y_{\gamma },  \tag{3.25b}
\end{equation}%
where $V(g)=\int d^{d}x\sqrt{\hat{g}}$ \ is \ the volume.

The equality happens when $g$ is the Yamabe minimizer. It is metric of unit
volume for manifold $M$ of constant scalar curvature (which, according to
theorem above, should be negative so that $M$ is hyperbolic 3-manifold). \ \
Only for hyperbolic 3-manifolds whose \textsl{Yamabe invariant} $\mathcal{Y}%
^{-}(M)=\sup_{\gamma }Y_{\gamma }$ $\ $the gravitational Cauchy problem for
source-free gravitational field is well posed [45,46]. For $g$ which is
Yamabe minimizer we have $S_{H-E}[\hat{g}]\leq Y_{\gamma }.$ This result can
be further extended by noticing that $\mathcal{N}S_{H-E}[\hat{g}]=CS(\mathbf{%
A}),$ where $\mathcal{N}$ is some constant whose value depends upon the
explicit form of the gauge field \textbf{A, }and\textbf{\ }$CS(\mathbf{A})$
is the Chern-Simons invariant to be described in the next section.

To demonstrate that $\mathcal{N}S_{H-E}[\hat{g}]=CS(\mathbf{A})$ it is
sufficient to use some results from works by Chern and Simons [59] and by
Chern [60]. In [59] it was proven that: a) the Chern-Simons (C-S) functional 
$\ CS(\mathbf{A})$ (to be defined in next section) is a conformal invariant
of $M$ (Theorem 6.3. of [59]) and, b) that the critical points of $CS(%
\mathbf{A})$ correspond to 3-manifolds which are (at least locally)
conformally flat (Corollary 6.14 of [59]). Subsequently, these results were
reobtained by Chern, Ref.[60], in much simpler and more \ physically
suggestive way. \ In view of these facts, at least for Yamabe minimizers we
obtain, $CS(\mathbf{A})=\mathcal{N}S_{Y}[\varphi ],$ where $\mathcal{N}$ is
some constant (different for different gauge groups). That this is the case
should come as not too big of a surprise since for Lorentzian 2+1 gravity
Witten, Ref.[61], demonstrated the equivalence of the Hilbert-Einstein and
C-S functionals without reference to results of Chern and Simons just cited.
At the same time, the Euclidean/Hyperbolic 3d gravity was discussed \ only\
much more recently, for instance, in the paper by Gukov, Ref.[62]. To avoid
duplications we refer our readers to these papers for further details.

\section{Floer-style nonperturbative treatment of Y-M fields}

\subsection{Physical content of the Floer's theory}

Striking resemblance between results of \ nonperturbative treatment of
4-dimensional Y-M fields and two dimensional nonlinear sigma model at the
classical level is well documented in Ref.[63\textbf{]}. Zero curvature
equations, e.g. Eq.(2.7), can be obtained either by using the two-
dimensional nonlinear sigma model or three-dimensional C-S functional. \ As
discussed in previous section, the self-duality condition for Y-M fields
also leads (upon reduction)\ to zero curvature condition. Since the Ernst
equation describing static gravitational (and electrovacuum) fields is
obtainable \ both from conditions of self-duality for the Y-M field and from
minimization of 3-dimensional nonlinear sigma model,\ it follows that 3-d
gravitational nonlinear sigma model, Eq.(3.8), contains nonperturbative
information about Y-M fields. Furthermore, in view of results of Appendix A,
it also should contain information about the static electromagnetic fields,
for the combined gravitational and electromagnetic waves and, with minor
modifications, for the combined gravitational, electromagnetic and neutrino
fields. The nonperturbative treatment of Y-M fields is usually associated
either with the instanton or monopole calculations. This observation leads
to the conclusion that, at least in some cases, zero curvature equation
should carry all nonperurbative information about Y-M fields. This point of
view\ is advocated and developed by Floer [11,21\textbf{]}. \ Below,we \
shall discuss Floer's point of view now in the language used in physics
literature. For the sake of illustration, it is convenient to present our
arguments for Abelian Y-M (that is electromagnetic) fields first.

\ The action functional $S$ in this case is given by\footnote{%
Up to an unimportant scale factor.} 
\begin{equation}
S=\frac{1}{2}\int\limits_{0}^{t}dt\int\limits_{M}dv[\mathbf{E}^{2}-\mathbf{B}%
^{2}],  \tag{4.1}
\end{equation}%
where $\mathbf{B}=\mathbf{\nabla }\times \mathbf{A}$ and $\mathbf{E}=$ $-%
\mathbf{\nabla }\varphi -\frac{\partial }{\partial t}\mathbf{A}$ , $\varphi
\equiv \mathbf{A}_{0}.$ It is known that, at least for electromagnetic
waves, it is sufficient to put $\mathbf{A}_{0}=0$ (temporal gauge). In such
a case the above action can be rewritten as 
\begin{equation}
S[\mathbf{A}]=\frac{1}{2}\int\limits_{0}^{t}dt\int\limits_{M}dv[\mathbf{\dot{%
A}}^{2}-\mathbf{(\nabla \times A)}^{2}],  \tag{4.2}
\end{equation}%
where $\mathbf{\dot{A}=}\frac{\partial }{\partial t}\mathbf{A.}$ From the
condition $\delta S/\delta \mathbf{A}=0$ we obtain $\frac{\partial \mathbf{E}%
}{\partial t}=\mathbf{\nabla }\times \mathbf{B}$. The definition of $\mathbf{%
B}$\textbf{\ }guarantees \ the validity of the condition $\mathbf{\nabla }%
\cdot \mathbf{B}=0$ while from the definition of $\mathbf{E}$ we get another
Maxwell equation $\frac{\partial \mathbf{B}}{\partial t}=-\mathbf{\nabla }%
\times \mathbf{E}$.\ The question arises: will these results imply the
remaining Maxwell's equation $\mathbf{\nabla }\cdot \mathbf{E}=0$ essential
for correct formulation of the Cauchy problem? If such a constraint
satisfied at $t=0,$ naturally, it will be satisfied for $t>0$.
Unfortunately, for $t=0$ the existence of such a constraint does not follow
from the above equations and should be introduced as independent. This
causes decomposition of the field $\mathbf{A}$ as $\mathbf{A}=\mathbf{A}%
_{\parallel }+\mathbf{A}_{\perp }$.Taking into account that $\mathbf{E}=$ $-%
\frac{\partial }{\partial t}\mathbf{A,}$ we obtain as well $\mathbf{\nabla }%
\cdot (\mathbf{E}_{\parallel }$ +$\mathbf{E}_{\perp }).$ Then, by design $%
\mathbf{\nabla }\cdot \mathbf{E}_{\perp }=0,$ while $\mathbf{\nabla }\cdot 
\mathbf{E}_{\parallel }$ remains to be defined by the initial and boundary
data. Because of this, it is always possible to choose $\mathbf{A}%
_{\parallel }=0$ and to use only $\mathbf{A}_{\perp }$ for description of
the field propagation [64]. Hence, the action functional $S$ can be finally
rewritten as 
\begin{equation}
S[\mathbf{A}_{\perp }]=\frac{1}{2}\int\limits_{0}^{t}dt\int\limits_{M}dv[%
\mathbf{\dot{A}}_{\perp }^{2}-\mathbf{(\nabla \times A}_{\perp }\mathbf{)}%
^{2}].  \tag{4.3}
\end{equation}%
\ In such a form it can be used \ as action in the path integrals, e.g. see
Ref.[64], page 152, describing free electromagnetic field. Such path
integral can be evaluated both in Minkowski and Eucldean spaces by the
saddle point method. There is, however, a closely related method more
suitable for our purposes. It is described in the monograph by Donaldson,
Ref.[11]. Following this reference, we replace time variable $t$ by $-i\tau $
in the functional $S[\mathbf{A}_{\perp }]$ .Consider now this replacement in
some detail. We have\footnote{%
We shall assume (without loss of generality) that \textbf{\.{A}}$_{\perp }$%
\textbf{\ }is collinear with\textbf{\ A}$_{\perp }.$}%
\begin{eqnarray}
&&\frac{1}{2}\int\limits_{0}^{T}d\tau \int\limits_{M}dv[\mathbf{\dot{A}}%
_{\perp }^{2}+\mathbf{(\nabla \times A}_{\perp }\mathbf{)}^{2}]  \notag \\
&=&\frac{1}{2}\int\limits_{0}^{T}d\tau (\int\limits_{M}dv[[\mathbf{\dot{A}}%
_{\perp }+\mathbf{(\nabla \times A}_{\perp }\mathbf{)}]^{2}-\frac{\partial }{%
\partial \tau }\mathbf{(A}_{\bot }\mathbf{\cdot \nabla \times A}_{\perp }%
\mathbf{)]).}  \TCItag{4.4}
\end{eqnarray}%
Since variation of $\mathbf{A}_{\perp }$ is fixed at the ends of $\tau $
integral, the last term can be dropped so that we are left with the
condition 
\begin{equation}
\frac{\partial }{\partial \tau }\mathbf{A}_{\perp }\mathbf{=-B}_{\perp } 
\tag{4.5}
\end{equation}%
extremizing the Euclidean action $S_{E}[\mathbf{A}_{\perp }].$ The above
results are transferable to the non Abelian Y-M field by continuity and
complementarity. Since in the Abelian case fields $\mathbf{E}$ and $\mathbf{B%
}$ are dual to each other, by applying the curl operator to both sides of
Eq.(4.5) (and removing \ the subscript $\perp )$ we obtain the \ equivalent
form of self-duality equations in accord with those on page 33 of Ref.[6].
This calculation provides an independent check of Donaldson's method of \
computation. Since the (anti)self-duality condition in the Abelian case can
be written as $\mathbf{B}=\mp \mathbf{E}$ [\textbf{9}].and since $\mathbf{E}%
= $ $-\frac{\partial }{\partial \tau }\mathbf{A}$, we conclude that Eq.(4.5)
is the self-duality equation. This conclusion is immediately transferable to
the non Abelian Y-M case where the analog of Eq. (4.5) is%
\begin{equation}
\frac{\partial }{\partial \tau }\mathbf{A=\ast F(A(}\tau \mathbf{)),} 
\tag{4.6}
\end{equation}%
in accord with Floer. The symbol * denotes the Hodge star operation in 3
dimensions. Following Donaldson [11] this result should be understood as
follows. Introduce a connection matrix $\mathbf{A}=A_{0}d\tau
+\sum\limits_{i=1}^{3}A_{i}dx_{i}$ such that both $A_{0}$ and $A_{i}$ depend
upon all four variables $\tau ,x_{1},x_{2}$ and $x_{3}.$ In the temporal
gauge $A_{0}$ should be discarded so that $\tau $ becomes a parameter in the
remaining $\ A_{i}^{\prime }s.$ Evidently, it can be associated with the
spectral parameter (e.g. see previous section). The Hodge star operator in
Eq.(4.6) \ is needed to make this equation as an equation for one-forms The
obtained results fit nicely into Cauchy formulation of dynamics of both Y-M
and gravity. Indeed,\ under conditions \ analogous to that discussed in
[45,46] the space-time (4-manifold) is decomposable into direct product $%
M\times R$ \ (a trivial fiber bundle) in such a way that all differential
operations \ acting on 4-manifold are been projected down to 3-manifold $M$.
This is essential part of Floer's theory. Furthermore, since $\delta CS(%
\mathbf{A})/\delta \mathbf{A}=\mathbf{F}(\mathbf{A})$ the above Eq.(4.6) can
be equivalently rewritten as 
\begin{equation}
\frac{\partial }{\partial \tau }\mathbf{A=\ast \lbrack }\delta CS(\mathbf{A}%
)/\delta \mathbf{A}]  \tag{4.7}
\end{equation}%
so that the Chern-Simons functional is playing a role of a "Hamiltonian" in
Eq.s(4.7). From the theory of dynamical systems it follows then that the
dynamics is taking place between the points of equilibria defined by zero
curvature condition $\mathbf{F}(\mathbf{A})=0.$ At the same time, using our
work, Ref.[50], it is easily recognize Eq.(4.7) as an equation for the
gradient flow, e.g. see Eq.s(3.14). For the sake of space we shall not
discuss this topic any further. Interested readers are encouraged to consult
Ref.[65]. For supersymmetric Y-M fields participating in Seiberg-Witten
theory the gradient flow equations are discussed in detail in Ref.[66]

The mechanical system described by Eq.(4.7) should be eventually quantized.
Since the quantization procedure is outlined in Ref.[67], to avoid
duplications, we shall concentrate attention of our readers on aspects \ of
Floer's theory not covered in [67] but still relevant to this paper. To do
so, we follow Donaldson [11]. This is accomplished in several steps.

\textsl{First}, in the previous section we noticed that the axially
symmetric self-dual solution for Y-M fields does not depend on $x_{0}$ (or $%
\tau )$ variable. Therefore, if such solution is substituted back into Y-M
functional, it produces divergent result. Although the cure for this issue
is discussed in subsection 4.4, in this subsection we provide needed
background. For this purpose, following Ref.[68] \ we consider the Y-M
action $S[\mathbf{F}]$ for the pure Y-M field\footnote{%
Strictly following notations of Ref.[68] we do not indicate that in general
the integration should be made over some 4-manifold $\mathcal{M}$. In
physics literature, and in Eq.s(2.11), it is assumed that we are dealing
with \textbf{R}$^{4}$ (or $S^{4}$ upon compactification). In Floer's theory
it is essential that the 4-manifold is decomposable as $M\times R$ . This
decomposition should be treated with care as described in the Donaldson's
book [11]}%
\begin{equation}
S[\mathbf{F}]=-\frac{1}{8}\int_{\mathbf{R}^{4}}d^{4}xtr(F_{\mu \nu }F_{\mu
\nu }).  \tag{4.8}
\end{equation}%
The duality condition\footnote{%
With the convention that $\varepsilon _{1234}=-1.$}$\ast F_{\mu \nu }=\frac{1%
}{2}\varepsilon _{\mu \nu \alpha \beta }F_{\alpha \beta }$ allows \ us then
to rewrite this action as follows%
\begin{equation}
S[\mathbf{F}]=-\frac{1}{16}\int_{\mathbf{R}^{4}}d^{4}x[tr((F_{\mu \nu }\mp
\ast F_{\mu \nu })(F_{\mu \nu }\mp \ast F_{\mu \nu }))\pm 2tr(F_{\mu \nu
}\ast F_{\mu \nu })]  \tag{4.9}
\end{equation}%
since $tr(F_{\mu \nu }F_{\mu \nu })=tr(\ast F_{\mu \nu }\ast F_{\mu \nu }).$
The winding number $N$ \ for SU(2) gauge field is defined as\footnote{%
We follows notations of Ref.\textbf{[}68] in which\textbf{\ R}$^{4}$ is
actually standing for $S^{4}\in SU(2)$} 
\begin{equation}
N=-\frac{1}{8\pi ^{2}}\int_{\mathbf{R}^{4}}d^{4}xtr(F_{\mu \nu }\ast F_{\mu
\nu })\equiv -\frac{1}{8\pi ^{2}}\int_{\mathbf{R}^{4}}tr(F_{\mu \nu }\wedge
F_{\mu \nu })  \tag{4.10}
\end{equation}%
so that use of this definition in Eq.s(4.8),(4.9) produces%
\begin{equation}
S[\mathbf{F}]\geq \pi ^{2}\left\vert N\right\vert  \tag{4.11}
\end{equation}%
with the equality taking place when the (anti) self-duality condition (e.g.
see Eq.(2.10) ) holds. In such a case the saddle point action is becoming
just a winding number (up to a constant).

\textsl{Second}, if our space-time 4-manifold $\mathcal{M}$ can be
decomposed as $M\times \lbrack 0,1],$ the following identity \ can be used
[11] 
\begin{equation}
\int_{M\times \lbrack 0,1]}tr(F_{\mu \nu }\wedge F_{\mu \nu })=\int_{M}tr(%
\mathbf{A}\wedge d\mathbf{A+}\frac{2}{3}\mathbf{A}\wedge \mathbf{A}\wedge 
\mathbf{A})\doteqdot CS(\mathbf{A}).  \tag{4.12}
\end{equation}%
Here the symbol $\doteqdot $ means "up to a constant". The decomposition $%
M\times \lbrack 0,1]$ reflects the fact that the C-S functional is defined
up to mod \textbf{Z}. This ambiguity can be removed if we agree to consider
C-S functional as a quotient \textbf{R}/\textbf{Z}. Accordingly, this allows
us to replace $M\times R$ by $M\times \lbrack 0,1].$ Details can be found in
Ref.[11]. Thus, one way or another the winding number $N$ in Eq.(4.10) can
be replaced by the Chern-Simons functional.

\textsl{Third}, since the equation of motion for the C-S functional is zero
curvature condition $\mathbf{F}=0$, i.e.%
\begin{equation}
d\mathbf{A}+\mathbf{A}\wedge \mathbf{A}=0,  \tag{4.13}
\end{equation}%
implying that the connection $\mathbf{A}$ is flat, we can use this result in
Eq.(4.12) in order to rewrite it as (e.g. for SU(2)) 
\begin{equation}
\frac{1}{8\pi ^{2}}\int_{M}tr(\mathbf{A}\wedge d\mathbf{A+}\frac{2}{3}%
\mathbf{A}\wedge \mathbf{A}\wedge \mathbf{A})=-\frac{1}{24\pi ^{2}}%
\int_{M}tr(\mathbf{A}\wedge \mathbf{A}\wedge \mathbf{A).}  \tag{4.14}
\end{equation}%
For other groups the prefactor and the domain of integration will be
different in general.

\textsl{Fourth}, zero curvature Eq.(4.13) involves connections which are
functions of three arguments and a spectral/time parameter. \ In such
setting minimization of Y-M functional is not divergent in view of Eq.(4.11).

\ \textsl{Fifth}, the obtained result, Eq.(4.14), coincides with that known
for the winding number for SU(2) instantons \ in physics literature[8,19]
where it was obtained with help of entirely different arguments. \ It should
be noted though that in spite \ of apparent simplicity of these results,
actual calculations of C-S functionals (invariants) for different
3-manifolds are, in fact, very sophisticated [69,70]. In accord with Floer
and Ref.[67], we conclude that nonperturbatively the 4-dimensional pure Y-M
quantum field theory is a topological field theory of C-S type.

\textsl{Sixth}, the isomorphism noticed by Louis Witten acquires\ now
natural explanation. It becomes possible in view of results just presented,
on one hand, and \ the fact that $\mathcal{N}S_{H-E}[\hat{g}]=CS(\mathbf{A})$
(previous section), on another. For fields with axial symmetry, equations of
motion, Eq.(4.13), for gravity and Y-M fields coincide. \ 

\textsl{Seventh}, the instantons in Floer's theory are \textsl{not} the same
as considered in physics literature [8,19]. To understand this, we must take
into account that in Floer's theory manifolds under consideration are
4-manifolds $\mathcal{M}$ with tubular ends. Such manifolds are complete
Riemannian manifolds with finite number of tubular ends \ made of half tubes 
$(0,\infty )$ so that locally each such manifold looks like $%
U_{i}=L_{i}\times $ $(0,\infty )$ with $L_{i\text{ }}$ being a compact
3-manifold (called a "crossection of a tube") and $i$ numbering the tubes.
The closure of $\mathcal{M\setminus }\bigcup\nolimits_{i=1}^{n}U_{i}$ is a
compact manifold with boundary. If the crossection \ is $S^{3},$ then $U$ is
conformally equivalent to a punctured ball $B^{4}\setminus \{0\}.$ This
implies that a manifold $\mathcal{M}$ with tubular ends is conformally
equivalent to a punctured manifold $\mathcal{\tilde{M}}\setminus \{$ $%
p_{1},...,p_{n}\}$ where $\mathcal{\tilde{M}}$ is compact. The instanton
moduli problem for $\mathcal{M}$ is equivalent to that for the punctured
manifold [11]. Recall that the moduli space of instantons is defined as set
of solutions of anti self-dual equations modulo gauge equivalence.

Being armed with these definitions and taking into account that the (anti)
self-duality Eq.(4.7) we can interpret the instanton as a path connecting
one flat connection $\mathbf{F}=0$ at "time" $\tau =-\infty $ with another
flat connection at "time" $\tau =\infty $ [11]. It is permissible for the
path to begin at one flat connection, to wind around a tube (modulo gauge
equivalence) and to end up at the same flat connection, Ref.[11], page 22.
Evidently, this case involves 4-manifolds with just one tubular end.
Physically, each flat connection $\mathbf{F}=0$ represents the vacuum state
so that the instantons discussed in the Introduction should be connecting
different vacua. In this sense there is a difference between the
interpretation of instantons in mathematics and physics literature. As in
the case of standard quantum mechanics, only imposition of some additional
physical constraints permits us to select between all possible solutions
only those which are physically relevant. In the present context it is known
that all exactly integrable systems are described by the zero curvature
equation $\mathbf{F}=0$ [5,6\textbf{]}. It is also known that differences
between these equations are caused in part by differences in a way the
spectral parameter enters into these equations. Since for the Floer's
instantons $\mathbf{F}\neq 0,$ it means that the curvature $\mathbf{F}$
should be parametrized in such a way that the "time" parameter \ should
become a spectral parameter when $\mathbf{F}=0.$ In this work we do not
investigate this problem\footnote{%
See Ref.[7] for introduction into this topic.}. Instead, we shall focus our
attention on different vacua, that is \ on different (knot-like) solutions
of zero curvature equation $\mathbf{F}=0.\footnote{%
A complement of each knot in $S^{3}$ is 3-manifold. Floer's instantons are
in fact connecting various three-manifolds. These 3- manifolds (with tubular
ends) should be glued together to form $\mathcal{M}.$The gluing procedure is
extremely delicate mathematical operation [11]. It is above the level of
rigor of this paper. To imagine the connected sum of knots [20] is much
easier than the connected sum of 3-manifolds. This sum has physical meaning
discussed in Section 6.}$

\subsection{The Faddeev-Skyrme model and vacuum states of the Y-M functional}

In the light of results just presented, we would like to argue that the F-S
model \ is indeed capable of representing the vacuum states of pure Y-M
fields. For this purpose it is sufficient to recall the key results of the
paper by Auckly and Kapitansky [71]. These authors were able to rewrite the
Faddeev functional%
\begin{equation}
E[\mathbf{n}]=\int\limits_{S^{3}}dv\{\left\vert d\mathbf{n}\right\vert
^{2}+\left\vert d\mathbf{n}\wedge d\mathbf{n}\right\vert ^{2}\}  \tag{4.15}
\end{equation}%
in the equivalent form given by 
\begin{equation}
E_{\varphi }[a]=\int\limits_{S^{3}}dv\{\left\vert D_{\mathbf{a}}\mathbf{%
\varphi }\right\vert ^{2}+\left\vert D_{\mathbf{a}}\mathbf{\varphi }\wedge
D_{\mathbf{a}}\mathbf{\varphi }\right\vert ^{2}\}.  \tag{4.16}
\end{equation}%
In this expression, the covariant derivative $D_{a}\mathbf{\varphi }=d%
\mathbf{\varphi }+[\mathbf{a},\mathbf{\varphi }]$. Evidently, $E_{\varphi }[%
\mathbf{a}]$ acquires its minimum when $\mathbf{\varphi }=\mathbf{a}$ and
the connection becomes flat (that is covariant derivative becomes zero). \
Since this result is compatible with those discussed in previous subsection,
it implies that, indeed, Faddeev's model can be used for description of
vacuum states for pure Y-M fields. The only question remains: Is this model
the only model describing QCD vacuum? \ In view of Eq.s (3.18),(3.19) it
should be clear that this is not the case. The full explanation is given \
below, in Sections 5,6. \ In addition, the disadvantage of the F-S model as
such (that is without modifications) lies in \ the absence of gap \ upon its
quantization as was recognized already by Faddeev and Niemi in Ref.[25]. \ \
In\textsl{\ }Sections 5,6 we shall eliminate this deficiency in a way
different from that described in the Introduction (e.g. in Ref.s[25,26]). In
the meantime, we would like to find the place for monopoles in our
calculations.

\subsection{Monopoles and the Ernst equation}

\subsubsection{Monopoles versus instantons}

To introduce notations and for the sake of uninterrupted reading, we need to
describe briefly the alternative point of view at the results of previous
subsection. For this purpose, following Manton [49], we need to make a
comparison between the Lagrangians for SU(2) Y-M and the Y-M-Higgs fields
described respectively by%
\begin{equation}
\mathcal{L}_{Y-M}=-\frac{1}{4}tr(F_{\mu \nu }F^{\mu \nu })  \tag{4.17}
\end{equation}%
and%
\begin{equation}
\mathcal{L}_{Y-M-H}=-\frac{1}{4}tr(\mathbf{F}_{\mu \nu }\mathbf{F}^{\mu \nu
})-\frac{1}{2}tr\left( D_{\mu }\mathbf{\Phi \cdot }D^{\mu }\mathbf{\Phi }%
\right) -\frac{\lambda }{2}(1-\mathbf{\Phi }\cdot \mathbf{\Phi })^{2} 
\tag{4.18}
\end{equation}%
with covariant derivative for the Higgs field defined as $D_{\mu }\mathbf{%
\Phi =\partial }_{\mu }\mathbf{\Phi +[A}_{\mu },\mathbf{\Phi }]$ and \ with
connection $\mathbf{A}_{\mu }$ used to define the Y-M curvature tensor $%
\mathbf{F}_{\mu \nu }=\mathbf{\partial }_{\mu }\mathbf{A}_{\nu }-\mathbf{%
\partial }_{\nu }\mathbf{A}_{\mu }+\mathbf{[A}_{\mu },\mathbf{A}_{\nu }],$
provided that $\mathbf{\Phi =}\Phi ^{a}t^{a}$, $\mathbf{A}_{\mu }=A_{\mu
}^{a}t^{a},$ and $[t^{a},t^{b}]=-2\varepsilon _{abc}t^{c}.$ Now the
self-duality condition $\mathbf{F}=\ast \mathbf{F}$ can be equivalently
rewritten as $\mathbf{F}_{ij}=-\varepsilon _{ijk}\mathbf{F}_{k0}$ with
indices $i,j,k$ running over 1,2,3. Incidentally, in the temporal gauge this
result is equivalent to Floer's Eq.(4.6) Consider now the limit $\lambda
\rightarrow 0$ in Eq.(4.18). In the Minkowski spacetime the field equations
originating from the Y-M-Higgs Lagrangian can be solved by using the
Bogomolny ansatz equations $\mathbf{F}_{ij}=-\varepsilon _{ijk}D_{k}\mathbf{%
\Phi }$ \ in which $\mathbf{A}_{0}=0$ (temporal gauge). \ Instead of
imposing the temporal gauge condition, we can identify the Higgs field $%
\mathbf{\Phi }$ with $\mathbf{A}_{0}$ so that the Bogomolny equations read
now as follows:%
\begin{equation}
\mathbf{F}_{ij}=-\varepsilon _{ijk}D_{k}\mathbf{A}_{0}.  \tag{4.19}
\end{equation}%
Bogomolny demonstrated that the Prasad-Sommerfield monopole solution can be
obtained using Eq.(4.19). Thus, any static (that is time-independent)
solution of self-duality equations is leading to
Bogomolny-Prasad-Sommerfield (BPS) monopole solution of the Y-M fields,
provided that we interpret the component $\mathbf{A}_{0}$ as the Higgs
field. Suppose now that there is an axial symmetry. Forgacs, Horvath and
Palla [72] (FHP) demonstrated equivalence of the\ set of axially symmetric
Bogomolny Eq.s (4.19) to the Ernst equation. The static monopole solution is
time-independent self-dual gauge field. Because of this time independence,
its four-dimensional action is \textsl{infinite}\textbf{\ (}because of the
time translational invariance) while that for instantons \textsl{is finite}.
Furthermore, the boundary conditions for monopoles and instantons are
different. The infinity problem for monopoles can be cured somehow by
considering the monopole dynamics [68] but this topic at this moment "is
more art than science", e.g. read [68], page 309. For the same reason we
avoid in this section talking about dyons (pseudo particles having both
electric and magnetic charge). Hence, we would like to conclude our
discussion with description of more \ mathematically rigorous treatments. By
doing so we shall establish connections with results presented in previous
sections

\subsection{Calorons$.$}

Calorons are instantons on $R^{3}\times S^{1}.$ From this definition it
follows that, physically, these are just instantons at finite temperature%
\footnote{%
This explains the word "caloron".}. Calorons are related to both instantons
on \textbf{R}$^{4}$ (or $S^{4})$ and monopoles on \textbf{R}$^{3}($or $%
S^{3}).$ Heuristically, the large period calorons are instantons while the
small period calorons \ are monopoles [73,74]. \ These results do not
account yet for the fact that both the Y-M action and the self-duality
equations are conformally invariant. Atiyah [75]. noticed that the Euclidean
metric \ can be represented either as 
\begin{equation}
ds_{E}^{2}=\left( dx^{1}\right) ^{2}+\left( dx^{3}\right) ^{2}+\left(
dx^{3}\right) ^{2}+\left( dx^{4}\right) ^{2}  \tag{4.20a}
\end{equation}%
or as 
\begin{equation}
ds^{2}=\frac{r^{2}}{R^{2}}[R^{2}(\frac{\left( dx^{1}\right) ^{2}+\left(
dx^{2}\right) ^{2}+\left( dr\right) ^{2}}{r^{2}})+R^{2}d\varphi ^{2}] 
\tag{4.20b}
\end{equation}%
with $R$ being some constant. \ The above representation involves polar $%
r,\varphi $ coordinates in the $(x^{3},x^{4})$ plane thus implying some kind
of axial symmetry. Since self-duality equations are conformally invariant,
the prefactor $\frac{r^{2}}{R^{2}}$ can be dropped so that the Euclidean
space \textbf{R}$^{4}$ becomes conformally equivalent to the product \textbf{%
H}$^{3}\times S^{1}.$For such manifold the constant scalar curvature of the
hyperbolic 3-space \textbf{H}$^{3}$ is $-1/R^{2}.$ Furthermore, the
remaining term represents the metric on a circle of radius $R$. Following
Ref.[73], let $(x^{1},x^{2},x^{3})$ be coordinates for the hyperbolic ball
model of \textbf{H}$^{3}$ so that the radial coordinate be $\mathcal{R}=%
\sqrt{\left( x^{1}\right) ^{2}+\left( x^{2}\right) ^{2}+\left( x^{3}\right)
^{2}}$. Let $0\leq \mathcal{R}\leq R.$ Let $\tau $ \ be a coordinate on $%
S^{1}$ with period $\beta ,$ then the metric on \textbf{H}$^{3}\times S^{1}$
can be represented as%
\begin{equation}
ds_{H}^{2}=d\tau ^{2}+\Lambda ^{2}(d\mathcal{R}^{2}+\mathcal{R}^{2}d\Omega
^{2}),  \tag{4.21a}
\end{equation}%
where $\Lambda =(1-\mathcal{R}/R)^{-1}$ and $d\Omega ^{2}$ is the metric on
2-dimensional sphere. If we introduce an auxiliary coordinate $\mu
=(R/2)\arctan $h$(\mathcal{R}/R),$ and complex coordinate $z=\mu +i\tau ,$%
the above hyperbolic metric can be rewritten as 
\begin{equation}
ds_{H}^{2}=d\tau ^{2}+d\mu ^{2}+\Xi ^{2}d\Omega ^{2}  \tag{4.21b}
\end{equation}%
with $\Xi =(R/2)\sinh (2\mu /R).$ By analogy with transition from Eq.(4.20a)
to (4.20b) we can proceed as follows. Let $r=\sqrt{\left( y^{1}\right)
^{2}+\left( y^{2}\right) ^{2}+\left( y^{3}\right) ^{2}}$ with ($%
y^{1},y^{2},y^{3},y^{0})$ being coordinates on \textbf{R}$^{4}.$ By letting $%
t=y^{0}$ the Euclidean metric can be written \ as usual, i.e. 
\begin{equation}
ds_{E}^{2}=dt^{2}+dr^{2}+r^{2}d\Omega ^{2}  \tag{4.22a}
\end{equation}%
so that 
\begin{equation}
ds_{H}^{2}=\xi ^{2}ds_{E}^{2},  \tag{4.22b}
\end{equation}%
with $\xi =(R/2)[\cosh (2\mu /R)+\cos (2\tau /R)].$ This correspondence
between \textbf{R}$^{4}\smallsetminus \mathbf{R}^{2}$ and \textbf{H}$%
^{3}\times S^{1}$ is made with help of the mapping $w=tanh(z/R)$ (with $%
w=r+it$ and $\beta =\pi R).$ Let $\mathcal{M}=$\textbf{H}$^{3}\times S^{1}$
(or \textbf{H}$^{3}\times R)$ then, in view of conformal invariance, we can
rewrite Eq.(4.8) as\ 
\begin{equation}
S[\mathbf{F}]=-\frac{1}{8}\int_{\mathcal{M}}tr(F_{\mu \nu }F^{\mu \nu })\Xi
^{2}d\tau d\mu d\Omega .  \tag{4.23}
\end{equation}%
\ We have to rewrite the winding number, Eq.(4.10), accordingly. Since it is
a topological invariant, this means that the self-duality equations must be
adjusted accordingly. For instance, \ for the \textsl{hyperbolic calorons}
on \textbf{H}$^{3}\times S^{1}$ the self-duality equation reads%
\begin{equation}
F_{0i}=\frac{1}{2\Lambda }\varepsilon _{ijk}F_{jk}.  \tag{4.24}
\end{equation}%
The action $S[\mathbf{F}]$ now is finite with $tr(F_{\mu \nu }F^{\mu \nu
})\rightarrow 0$ when $\mu \rightarrow \infty .$ For hyperbolic instantons
we have finite action with $tr(F_{\mu \nu }F^{\mu \nu })\rightarrow 0$ when $%
\mu ^{2}+\tau ^{2}\rightarrow \infty .$

The results just described match nicely with the results by Witten [76] on
Euclidean SU(2) instantons invariant under the action of SO(3) Lie group.
His results\ will be discussed in detail in the next section. Notice, that
Euclidean metric, Eq.(4.22a), becomes that for \textbf{H}$^{2}\times S^{2}$
if we rewrite it as 
\begin{equation}
ds_{E}^{2}=r^{2}(\frac{dt^{2}+dr^{2}}{r^{2}}+d\Omega ^{2})  \tag{4.25a}
\end{equation}%
and, as before, we drop the conformal factor $r^{2}$ so that it becomes%
\begin{equation}
ds_{H}^{2}=\frac{dt^{2}+dr^{2}}{r^{2}}+d\Omega ^{2}.  \tag{4.25b}
\end{equation}%
Interestingly enough, that \ results by Witten initially developed for 
\textbf{H}$^{2}\times S^{2}$ can be also used without change for \textbf{H}$%
^{3}\times S^{1}$ and \textbf{H}$^{3}\times \mathbf{R}$ since the action of
SO(3) pulls back to these manifolds [73]. \textsl{This fact is of importance
since such an} \textsl{extension makes his results compatible with both
Floer's \ method of calculations for Y-M fields and with} \textsl{results of
Section 3.} Omitting all details, the action, Eq.(4.23), is reduced to that
known for two dimensional Abelian Ginzburg-Landau (G-L) model "living" on
the hyperbolic 2 manifold $X$ coordinatized by $\mu $ and $\tau $ \ with the
metric%
\begin{equation}
ds_{H}^{2}=\frac{d\mu ^{2}+d\tau ^{2}}{\Xi ^{2}}.  \tag{4.26}
\end{equation}%
Explicitly, such G-L action functional $S_{G-L}$ is given by [73]%
\begin{equation}
S_{G-L}=\frac{\pi }{2}\int\limits_{X}d\tau d\mu \lbrack \Xi ^{2}(\mathbf{%
\nabla }\times \mathbf{A})^{2}+2\left\vert (\mathbf{\nabla }+i\mathbf{A})%
\mathbf{\phi }\right\vert ^{2}+\Xi ^{-2}(1-\left\vert \mathbf{\phi }%
\right\vert ^{2})]  \tag{4.27}
\end{equation}%
with $\mathbf{A}$ and $\mathbf{\phi }$ being respectively the Abelian gauge
\ and the Higgs fields, $\mathbf{\phi }=\phi _{1}+i\phi _{2}$ so that $%
\left\vert \mathbf{\phi }\right\vert ^{2}=\phi _{1}^{2}+\phi _{2}^{2}.$ This
functional is obtained upon substitution of solution of the self-duality
equations into the Y-M action functional, Eq.(4.23). We refer our readers to
the original paper, Ref.[73] for details. \ In the limit $\beta \rightarrow
\infty $ the above functional coincides with that obtained by Witten [76].
The self-duality equations obtained by Witten describe instantons which lie
along the fixed axis while Fairlie, Corrigan, 't Hooft, and Wilczek [77]
developed an ansatz \ (CFtHW ansatz)\ for the self-duality equations
producing instantons at arbitrary locations. Manton [78] demonstrated that
Witten's and CFtHW multi-instanton solutions are gauge equivalent while
Harland [73,74\textbf{] } demonstrated how these instantons and monopoles
can be obtained from calorons in various limits. The obtained results \
provide needed background information for solution of the gap problem. This
solution is discussed in the next section.

\section{Solution of the gap problem}

\subsection{ Idea of \ the proof}

By cleverly using symmetry of the problem Witten [76] reduced the non
Abelian Y-M action \ functional to that for the Abelian G-L model "living"
in the hyperbolic plane. This is one of examples of\ the Abelian reduction
of QCD discussed in our paper, Ref.[79]. Vortices existing in the G-L model
could be visualized as made of some two-dimensional surfaces (closed
strings) living in the ambient space-time. These are known as Nambu-Gotto
strings. Their treatment by Polyakov [80] made them to exist in spaces of
higher dimensionality. In order for them to be useful for QCD, Polyakov
suggested to modify string action by adding an extra (rigidity) term into
string action functional. By doing so the problem was created of reproducing
Polyakov rigid string model from QCD action functional. The latest proposal
by Polyakov [81] involves consideration of spin chain models while that by
Kondo [82] involves the F-S model derived directly from QCD action
functional. As explained in\textbf{[}79\textbf{]}, in the case of scattering
processes of high energy physics one is confronted essentially with \ the
same combinatorial problems as were encountered at the birth of quantum
mechanics. In Ref.[83] we explained in detail why Heisenberg's
(combinatorial) method of developing quantum mechanical formalism is
superior to that by Schr\"{o}dinger. In Ref.[74] using these general results
we demonstrated how the combinatorial analysis of scattering data leads to
spin chain models as microscopic models describing excitation spectrum of
QCD. Thus, the mass gap problem can be considered as \ already solved in
principle. Nevertheless, in [\textbf{94}] such a solution is obtained
"externally", just based on the rules of combinatorics. \ As with quantum
mechanics, where atomic model is used to test Heisenberg's ideas, there is a
need to reproduce this combinatorial result "internally" by using
microscopic model of QCD. \ For this purpose, we shall use the G-L
functional, Eq.(4.27). By analogy with the flat case, we expect that it can
be rewritten in terms of interacting vortices. In the \ present case, in
view of Eq.(4.26), vortices "live" not in the Euclidean plane but in 3+1
Minkowski space-time. This is easy to understand if we recall the SO(3)$%
\rightleftarrows $SU(2) correspondence and take into account the analogous
correspondence between SU(1,1) and SO(2,1).

Within such a picture it is sufficient to look at evolution dynamics of the
individual vortex. Typically, it is well described by the dynamics of the
continuous Heisenberg spin chain model [84,85] in Euclidean space. \ In the
present case, this formalism should be extended to the Minkowski space and,
eventually, to hyperbolic space (that is to the \ case of Abelian model
discovered by Witten). Details of such an extension are summarized in
Appendix B. After that, the next task lies in connecting these results with
the Ernst equation. In the next subsection we initiate this study.

\subsection{Heisenberg spin chain model and the Ernst equation}

For the sake of space, this subsection is written under assumption that our
readers are familiar with the book "Hamiltonian methods in the theory of
solitons" [86] (or its equivalent) where all needed details can be found.
The continuous XXX Heisenberg spin chain is described with help of the spin
vector\footnote{%
In compliance with\ [86] we suppress the time-dependence.} $\vec{S}%
(x)=(S_{1}(x),S_{2}(x),S_{3}(x))$ restricted to \ live on the unit sphere $%
S^{2}:$ 
\begin{equation}
\vec{S}^{2}(x)=\sum\limits_{i=1}^{3}S_{i}^{2}(x)=1  \tag{5.1}
\end{equation}%
while obeying the equation of motion 
\begin{equation}
\frac{\partial }{\partial t}\vec{S}=\vec{S}\times \frac{\partial ^{2}}{%
\partial x^{2}}\vec{S}  \tag{5.2}
\end{equation}%
known as the Landau-Lifshitz (L-L) equation. By introducing matrices $U$($%
\lambda )$ and $V$($\lambda )$ via%
\begin{equation}
U(\lambda )=\frac{\lambda }{2i}S\text{, }V(\lambda )=\frac{i\lambda ^{2}}{2}%
S+\frac{\lambda }{2}S\frac{\partial }{\partial x}S,S=\vec{S}\cdot \vec{\sigma%
}  \tag{5.3}
\end{equation}%
so that $\sigma _{i}$ \ is one of Pauli's spin matrices and $\lambda $ is
the spectral parameter and requiring that $S^{2}=I,$ where $I$ is the unit
matrix, \ the zero curvature condition 
\begin{equation}
\frac{\partial }{\partial t}U-\frac{\partial }{\partial x}V+[U,V]=0 
\tag{5.4}
\end{equation}%
is obtained. \ With account of the constraint $S^{2}=I$ it can be converted
into equation 
\begin{equation}
\frac{\partial }{\partial t}S=\frac{1}{2i}[S,\frac{\partial ^{2}}{\partial
x^{2}}S]  \tag{5.5}
\end{equation}%
equivalent to Eq.(5.2). The correspondence between Eq.s(5.2) and (5.5) can
be made for \ $S(x,t)$ matrices of arbitrary dimension.

Having in mind Witten's result [76], we want now to extend these Euclidean
results to the case of noncompact Heisenberg spin chain model "living"
either in Minkowski or hyperbolic space. In doing so we follow, in part,
Ref.[\textbf{56}] and Appendix B. For this purpose we need to remind our
readers some facts about the Lie group SU(1,1). \ Since this group is
related to SO(2,1), very much like SU(2) is related to SO(3), we can proceed
\ by employing the noticed analogy. In particular, since $S=\vec{S}\cdot 
\vec{\sigma},$ we can preserve this relation by writing now $S=\vec{S}\cdot 
\vec{\tau}.$ Using this result we obtain, 
\begin{equation}
S=\left( 
\begin{array}{cc}
S^{z} & iS^{-} \\ 
iS^{+} & -iS^{z}%
\end{array}%
\right) \in su(1,1),\text{ }S^{\pm }=S^{x}\pm iS^{y},  \tag{5.6}
\end{equation}%
where the form of \ matrices generating su(1,1) Lie algebra is \ similar to
that for Pauli matrices. This time, however, $\det S=-1$ even though $%
S^{2}=I $ . Explicitly, $\left( S^{z}\right) ^{2}-\left( S^{x}\right)
^{2}-\left( S^{y}\right) ^{2}=1,$that is the motion is taking place on the
unit pseudosphere $S^{1,1}.$ Matrices $\tau _{i}$ \ generating $su(1,1)$ \
are \ fully characterized by the following two properties%
\begin{equation}
tr(\tau _{\alpha }\tau _{\beta })=2g_{\alpha \beta }\text{ , }[\tau _{\alpha
},\tau _{\beta }]=2if_{\alpha \beta \gamma }\tau _{\gamma }\text{; }%
g_{\alpha \beta }=diag(-1,-1,1);\alpha ,\beta ,\gamma =1,2,3  \tag{5.7}
\end{equation}%
with $f_{\alpha \beta \gamma }$ being structure constants for $su(1,1)$
algebra. An analog of the equation of motion, Eq.(5.5), now reads \ 
\begin{equation}
\frac{\partial }{\partial t}S^{\alpha }=\sum\limits_{\beta ,\gamma
}f^{\alpha \beta \gamma }S_{\beta }\frac{\partial ^{2}}{\partial x^{2}}%
S_{\gamma }.  \tag{5.8}
\end{equation}%
If we defines the Poisson brackets as $\{S^{\alpha }(x),S^{\beta
}(y)\}=-f^{\alpha \beta \gamma }S^{\gamma }(x)\delta (x-y),$ then the above
equation of motion can be rewritten in the Hamiltonian form%
\begin{equation}
\frac{\partial }{\partial t}S^{\alpha }=\{H,S^{\alpha }\},  \tag{5.9}
\end{equation}%
provided that the Hamiltonian \ $H$ is given by 
\begin{equation}
H=\frac{1}{2}\int\limits_{-\infty }^{\infty }dx\left( \nabla _{x}S^{\alpha
}\right) g_{\alpha \beta }\left( \nabla _{x}S^{\beta }\right) \equiv \frac{1%
}{4}tr\int\limits_{-\infty }^{\infty }dx\left( \nabla _{x}S\right) ^{2}. 
\tag{5.10}
\end{equation}%
Since now the motion takes place on pseudosphere $\check{S}^{2}$, it is
convenient to introduce the pseudospherical coordinates by analogy with
spherical, e.g.%
\begin{equation}
S^{x}(x,t)=\text{sinh }\theta (x,t)\cos \varphi (x,t),S^{y}(x,t)=\sinh
\theta (x,t)\sin \varphi (x,t),S^{z}(x,t)=\cosh \theta (x,t).  \tag{5.11}
\end{equation}%
Also, by analogy with spherical case we can use the stereographic projection
: from pseudosphere to hyperbolic plane. Recall [\textbf{102}], that in the
case of a sphere $S^{2}$ the inverse stereographic projection: from complex
plane $\mathbf{C}$ to $S^{2}$ is given by%
\begin{equation}
S^{+}=\frac{2z}{1+\left\vert z\right\vert ^{2}},S^{-}=\frac{2z^{\ast }}{%
1+\left\vert z\right\vert ^{2}},S^{z}=\frac{1-\left\vert z\right\vert ^{2}}{%
1+\left\vert z\right\vert ^{2}}.  \tag{5.12}
\end{equation}%
The mapping from \textbf{C} to \textbf{H}$^{2}$ is obtained \ with help of
Eq.(5.12) in a straightforward way as 
\begin{equation}
S^{+}=\frac{2\xi }{1-\left\vert \xi \right\vert ^{2}},S^{-}=\frac{2\xi
^{\ast }}{1-\left\vert \xi \right\vert ^{2}},S^{z}=\frac{1+\left\vert \xi
\right\vert ^{2}}{1-\left\vert \xi \right\vert ^{2}}.  \tag{5.13}
\end{equation}%
Using this correspondence the equations of motion, Eq.(5.10), \ rewritten in
terms of $\xi $ and $\xi ^{\ast }$variables (while keeping in mind that they
are parametrized by $x$ and $t)$ are given by%
\begin{equation}
i\frac{\partial }{\partial t}\xi +\frac{\partial ^{2}}{\partial x^{2}}\xi +2%
\frac{\xi ^{\ast }}{1-\left\vert \xi \right\vert ^{2}}\left( \frac{\partial 
}{\partial x}\xi \right) ^{2}=0.  \tag{5.14}
\end{equation}%
In the static ($t-$independent) case \ the above equation is reduced to%
\begin{equation}
(\left\vert \xi \right\vert ^{2}-1)\nabla _{x}^{2}\xi =2\xi ^{\ast }\left(
\nabla _{x}\xi \right) ^{2}  \tag{5.15}
\end{equation}%
easily recognizable as the Ernst equation. \ In his paper, Ref. [42], Ernst
used variational principle applied to the functional Eq.(3.18). \ From
Appendix B we know that both the L-L equation and its hyperbolic version
describe the motion of (could be knotted) vortex filament. Because of this,
the functional, Eq.(3.18), should undergo the same reduction as was made in
going from Eq.(B1.a) to (B1.b). Explicitly, this means that the functional,
Eq.(3.18), should be reduced in such a way that the Hamiltonian, Eq.(5.10),
should be replaced by 
\begin{equation}
H=-2\int\limits_{-\infty }^{\infty }dx\frac{\left\vert \nabla _{x}\xi
\right\vert ^{2}}{(1-\left\vert \xi \right\vert ^{2})^{2}},  \tag{5.16}
\end{equation}%
\ where the sign in front is chosen in accord with Ref.[87] and our
Eq.(3.22). The Hamiltonian equation of motion, Eq.(5.9), reproducing
Eq.(5.14) \ can be obtained if the Poisson bracket is defined as by $\{\xi
(x),\xi ^{\ast }(y)\}=(1-\left\vert \xi \right\vert ^{2})^{2}\delta (x-y).$%
The obtained results set up the stage for quantization. It will be discussed
in subsection 5.4. In the meantime, we need to connect results of Witten's
work, Ref.[76], with those we just obtained.

\subsection{From Abelian Higgs to Heisenberg spin chain model}

\subsubsection{The Abelian Higgs model}

The work by Witten [76] had been \ further analyzed in the paper by Forgacs
and Manton [88]. The major outcome of their \ work lies in demonstration of
uniqueness of the self-duality ansatz proposed by Witten. The self-duality
equations obtained in Witten's work are reduced to the system of three
coupled equations describing interaction between the Abelian Y-M and Higgs
fields%
\begin{equation}
\partial _{0}\varphi _{1}+A_{0}\varphi _{2}=\partial _{1}\varphi
_{2}-A_{1}\varphi _{1},  \tag{5.17a}
\end{equation}%
\begin{equation}
\partial _{1}\varphi _{1}+A_{1}\varphi _{2}=-(\partial _{0}\varphi
_{2}-A_{0}\varphi _{1}),  \tag{5.17b}
\end{equation}%
\begin{equation}
r^{2}(\partial _{0}A_{1}-\partial _{1}A_{0})=1-\varphi _{1}^{2}-\varphi
_{2}^{2}.  \tag{5.17c}
\end{equation}%
To analyze these equations, we recall that the original self-duality
equations for the Y-M fields are conformally invariant. We can take
advantage of this fact now by temporarily dropping the conformal factor $%
r^{2}$ in Eq.(5.17c). Then, the above equations become the Bogomolny
equations for the \ flat space Abelian Higgs model, e.g. for the model
described by the action functional, Eq.(4.24), with the conformal factor $%
\Xi =1$ [89]. Such obtained equations contain all information about the
Abelian Higgs model and, hence, they are equivalent to this model. It is of
importance for us to demonstrate this explicitly for both Euclidean and
hyperbolic spaces. For this purpose we introduce a covariant derivative $%
D_{\mu }=\partial _{\mu }-iA_{\mu }$, $\mu =0,1,$ and the complex field $%
\phi =\phi _{1}+i\phi _{2}$. Consider the Bogomolny equation following [68]:%
\begin{equation}
D_{0}\phi +iD_{1}\phi =0.  \tag{5.18}
\end{equation}%
Using the above definitions straightforward computation reproduces
Eq.s(5.17a,b). These equations can be used to obtain%
\begin{equation}
r^{2}\left( D_{0}-iD_{1}\right) (D_{0}+iD_{1})\phi =0  \tag{5.19}
\end{equation}%
implying%
\begin{equation}
r^{2}(D_{0}D_{0}+D_{1}D_{1})\phi =-ir^{2}[D_{0},D_{1}]\phi =-r^{2}\left(
\partial _{0}A_{1}-\partial _{1}A_{0}\right) \phi =-(1-\varphi
_{1}^{2}-\varphi _{2}^{2})\phi ,  \tag{5.20}
\end{equation}%
where the last equality was obtained with help of Eq.(5.17c). Evidently, the
equation%
\begin{equation}
(D_{0}D_{0}+D_{1}D_{1})\phi +\frac{1}{r^{2}}(1-\varphi _{1}^{2}-\varphi
_{2}^{2})\phi =0  \tag{5.21}
\end{equation}%
is one of the equations of "motion" for the G-L model on \textbf{H}$^{2}$,
e.g. see Ref.[89](Eq.(11.3) page 98). The second is the Ampere's equation 
\begin{equation}
\varepsilon _{\mu \nu }\partial _{\mu }(r^{2}B)=i(\phi \bar{D}_{\nu }\bar{%
\phi}-\bar{\phi}D_{\nu }\phi )  \tag{5.22}
\end{equation}%
with the "magnetic field" $B=\partial _{0}A_{1}-\partial _{1}A_{0}.$ Details
of derivation are given in Ref.[68], pages 198-199. Eq.(5.22) also coincides
with that given in the book by Taubs and Jaffe, Ref.[\textbf{105}]
(Eq.(11.3) page 98).

\textbf{Corollary 1}. \textsl{Since both equations can be obtained by
mimization of the functional,} \textsl{Eq.}(4.27)\textsl{, they are
equivalent to the Abelian Higgs model which, in turn, is the reduced form of
the Y-M functional for pure gauge fields.}

\ We continue with the discussion \ of Witten's treatment of Eq.s(5.17)
since we shall need his results later on. First, he selects physically
convenient gauge condition via $\partial _{\mu }A_{\mu }=0.$ This leads to
the choice: $A_{\mu }=\varepsilon _{\mu \nu }\partial _{\nu }\psi $ (for
some scalar $\psi )$. With such a choice for $A_{\mu }$ the first two of
Eq.s(5.17) can be rewritten as 
\begin{equation}
(\partial _{0}-\partial _{0}\psi )\varphi _{1}=(\partial _{1}-\partial
_{1}\psi )\varphi _{2},  \tag{5.23a}
\end{equation}%
\begin{equation}
(\partial _{1}-\partial _{1}\psi )\varphi _{1}=-(\partial _{0}-\partial
_{0}\psi )\varphi _{2}.  \tag{5.23b}
\end{equation}%
Let now $\varphi _{1}=e^{\psi }\chi _{1}$ and $\varphi _{2}=e^{\psi }\chi
_{2}.$ Then the above equations are reduced to the Cauchy-Riemann-type
equations: $\partial _{0}\chi _{1}=\partial _{1}\chi _{2}$ and $\partial
_{1}\chi _{1}=\partial _{0}\chi _{2}.$ Introduce the function $f=\chi
_{1}-i\chi _{2}$. Then, the last of Eq.s(5.17) acquires the form%
\begin{equation}
-r^{2}\nabla ^{2}\psi =1-ff^{\ast }e^{2\psi }.  \tag{5.24}
\end{equation}%
Notice that $-r^{2}\nabla ^{2}=-r^{2}(\frac{\partial ^{2}}{\partial t^{2}}+%
\frac{\partial ^{2}}{\partial r^{2}})$ is the hyperbolic Laplacian [90].
Eq.(5.24) is still gauge invariant in the sense that by changing $%
f\rightarrow fh$ and $\psi \rightarrow \psi -\frac{1}{2}\ln (hh^{\ast })$ in
this equation we observe that it preserves its original form. This is so
because $\nabla ^{2}\ln (hh^{\ast })=0$ for any analytic function which does
not have zeros. If $h$ does have zeros for $r>0$, then substitution of $\psi
\rightarrow \psi -\frac{1}{2}\ln (hh^{\ast })$ into Eq.(5.24) produces
isolated singularities at these zeros. After these remarks, Eq.(5.24) can be
simplified further. For this purpose, let $\psi =\ln r-\frac{1}{2}\ln
(ff^{\ast })+\rho $, provided that $\nabla ^{2}\ln (ff^{\ast })=0$ for any
analytic function $f$ which does not have zeros\footnote{%
In the case if it does, the treatment is also possible \ as explained by
Witten. Following his work, we shall temporarily ignore this option.}. Under
such conditions we end up with the Liouville equation%
\begin{equation}
\nabla ^{2}\rho =e^{2\rho }.  \tag{5.25}
\end{equation}%
It is of major importance for what follows.

\subsubsection{The Heisenberg spin chain model}

The results of Appendix B imply that the L-L Eq.(5.2) (or\ their hyperbolic
equivalent, Eq.(5.8)) could be interpreted in terms of equations \ for the
Serret-Frenet moving triad. Treatment along these lines \ suitable for
immediate applications is given in papers by Lee and Pashaev [91] and
Pashaev [92]. Below we superimpose their results with those of our work,
Ref.[84], to achieve our goals.

We begin with definitions. \ A collection of smooth vector fields \textbf{n}$%
_{\mu }$(x,t), $\mu =0-2$, forming an orthogonal basis is called the "moving
frame". If $x\in \mathcal{S}$ where $\mathcal{S}$ is some two dimensional
surface, then let \textbf{n}$_{1}($x,t$)$ and \textbf{n}$_{2}($x,t$)$ form
basis for the tangent plane to $\mathcal{S}$ $\forall x\in \mathcal{S}$.
Then, the Gauss map (that is the map from $\mathcal{S}$ to two dimensional
sphere $S^{2}$ or pseudosphere $S^{1,1})$ is given by \textbf{n}$_{2}$(x,t)$%
\equiv \mathbf{s}$. By design, it should obey Eq.(5.1). This observation
provides needed link between the spin and the moving frame vectors. Details
are given in [91,92] and Appendix B\textsl{. }It should be clear that \
since one can draw curves on surfaces both formalisms should involve the
same elements. The restriction for the curve to lie at the surface causes
additional complications in general but nonessential in the present case.

Next, we introduce the combinations $\mathbf{n}_{\pm }=\mathbf{n}_{0}\pm i%
\mathbf{n}_{1}$ possessing the following properties%
\begin{equation}
(\mathbf{n}_{+},\mathbf{n}_{+})=(\mathbf{n}_{-},\mathbf{n}_{-})=0\text{ , }(%
\mathbf{n}_{+}\text{,}\mathbf{n}_{-})=2/\kappa ^{2},  \tag{5.26}
\end{equation}%
where $\kappa ^{2}=1$ for $S^{2}$ and $\kappa ^{2}=-1$ for $S^{1,1}$ and 
\textbf{H}$^{2}.$ Furthermore, $(..,..)$ defines the scalar product (in
Euclidean or pseudo-Euclidean spaces). Also, 
\begin{equation}
\mathbf{n}_{+}\times \mathbf{s=}i\mathbf{n}_{+},\mathbf{n}_{-}\times \mathbf{%
s=-}i\mathbf{n}_{-},\mathbf{n}_{-}\times \mathbf{n}_{+}\mathbf{=}2i\kappa
^{2}\mathbf{s.}  \tag{5.27}
\end{equation}%
In addition, we shall use the vectors 
\begin{equation}
q_{\mu }=\frac{\kappa ^{2}}{2}(\partial _{\mu }\mathbf{s},\mathbf{n}_{+})%
\text{ and }\bar{q}_{\mu }=\frac{\kappa ^{2}}{2}(\partial _{\mu }\mathbf{s},%
\mathbf{n}_{-})  \tag{5.28}
\end{equation}%
in terms of which the equations of motion for the moving frame vectors look
as follows:%
\begin{equation}
D_{\mu }\mathbf{n}_{+}=-2\kappa ^{2}q_{\mu }\mathbf{s,}  \tag{5.29}
\end{equation}%
\begin{equation}
\partial _{\mu }\mathbf{s=}q_{\mu }\mathbf{n}_{-}+\bar{q}_{\mu }\mathbf{n}%
_{+},  \tag{5.30}
\end{equation}%
with covariant derivative $D_{\mu }=\partial _{\mu }-\frac{i}{2}V_{\mu }$
and $V_{\mu }=-2\kappa ^{2}(\mathbf{n}_{1},\partial _{\mu }\mathbf{n}_{0}).$
Consider now Eq.(5.30) for $\mu =1.$ Apply to it the operator $\partial _{1}$
and use the equations of motion and \ the definitions just introduced in
order to obtain 
\begin{equation}
\partial _{1}^{2}\mathbf{s=}\left( D_{1}q_{1}\right) \mathbf{n}_{-}+\left( 
\bar{D}_{1}\bar{q}_{1}\right) \mathbf{n}_{+}-\frac{4}{\kappa ^{2}}\left\vert
q_{1}\right\vert ^{2}\mathbf{s.}  \tag{5.31}
\end{equation}%
It can be shown that $q_{0}=iD_{1}q_{1}.$ In view of this, Eq.(5.30) for $%
\mu =0$ acquires the following form:%
\begin{equation}
\partial _{0}\mathbf{s=}iD_{1}q_{1}\mathbf{n}_{-}-i\bar{D}_{1}\bar{q}_{1}%
\mathbf{n}_{+}.  \tag{5.32}
\end{equation}%
This equation happens to be of major importance because of the following.
Multiply (from the left) Eq.(5.31) by $\mathbf{s}\times $ and use
Eq.s(5.27). Then (depending on signature of $\kappa ^{2})$ the obtained
result is equivalent to the L-L Eq.(5.2) \ or its pseudoeuclidean version,
Eq.(5.8). Furthermore, for this to happen the fields $V_{\mu }$ and $q_{\mu
} $ must be subject to the following constraint equations obtainable
directly from Eq.s (5.29)%
\begin{equation}
D_{\mu }q_{\nu }=D_{\nu }q_{\mu },  \tag{5.33a}
\end{equation}%
\begin{equation}
\lbrack D_{\mu },D_{\nu }]=-2\kappa ^{2}(\bar{q}_{\mu }q_{\nu }-\bar{q}_{\nu
}q_{\mu }).  \tag{5.33b}
\end{equation}%
We are going to demonstrate now that these equations are equivalent to
Eq.s(5.17) obtained by Witten.

We begin with the following observation. Let indices $\mu $ and $\nu $ be
respectively 1 and 0. Then, taking into account that $q_{0}=iD_{1}q_{1}$ we
can rewrite Eq.(5.33b) as 
\begin{equation}
F_{10}=B_{1}=-2\kappa ^{2}i(\bar{q}_{1}D_{1}q_{1}-q_{1}\bar{D}_{1}\bar{q}%
_{1}).  \tag{5.34}
\end{equation}%
Surely, by symmetry we could use as well: $q_{1}=-iD_{0}q_{0}$. This would
give us an equation similar to Eq.(5.34). Take now the case $\kappa ^{2}=-1$
(that is consider $S^{1,1}$) in these equations and compare them with the
Ampere's law, Eq.(5.22). We notice that these equations are not the same.
However, since the G-L model was originally designed for phenomenological
(thermodynamical) description of superconductivity (as explained in detail
in our work, Ref.[84]), we know that the underlying equations (obtainable
from the G-L functional) contain the London equation which reads\footnote{%
This is not the form of the London equation one can find in textbooks. But
in our work, Ref.[84], we demonstrated that Eq.(5.35) is equivalent to the
London equation.}%
\begin{equation}
\mathbf{\nabla }\times \mathbf{B}=C\mathbf{B}  \tag{5.35}
\end{equation}%
with $C$ being some constant (determined by physical considerations).
Evidently, in view of the London (5.35), Eq.s(5.22) and (5.34) become
equivalent. Consider now Eq.(5.33a). To understand better this equation, it
is useful to rewrite Eq.(5.18) as follows%
\begin{equation}
D_{0}\phi =-iD_{1}\phi \text{ or }D_{0}\phi _{1}=D_{1}\phi _{0},  \tag{5.36}
\end{equation}%
where $\phi _{1}=\phi $ and $\phi _{0}=-i\phi .$ Take into account now that $%
\phi =a+ib$ and identify $\phi _{1}$ with $q_{1}$ and $\phi _{0}$ with $%
q_{0}.$ Then, Eq.(5.33b) acquires the following form $(\kappa ^{2}=-1):$ 
\begin{equation}
(\partial _{0}V_{1}-\partial _{1}V_{0})=-i4(\bar{\phi}_{0}\phi _{1}-\phi _{0}%
\bar{\phi}_{1})=-4(a^{2}+b^{2}).  \tag{5.37}
\end{equation}%
Looking at Eq.(5.17c) we can make the following \ identifications: $%
V_{1}=A_{1},V_{0}=A_{0},\pm 2a=\varphi _{1},\pm 2b=\varphi _{2}.$ Then,
comparison between Eq.s(5.17c) and (5.37) indicates that we are still
missing a factor of $r^{2}$ in the l.h.s. and 1 in the r.h.s. Looking at
Witten's derivation of the Liouville Eq.(5.25), we realize that these two
factors are interdependent. By clever choice of the function $\psi $ they
can be made to disappear. This makes physical sense since locally the
underlying surface is almost flat. This observation makes Eq.s(5.37) and
(5.24) (or 5.17c) equivalent.

\textbf{Corollary 2}.\textsl{\ The L-L and\ 2 dimensional G-L models are \
essentially} \textsl{equivalent in the sense just described both in
Euclidean and in Minkowski spaces}.

\textbf{Corollary 3}.\ \textsl{The "hyperbolc" L-L Eq.}(5.14)\textsl{\ or
its Euclidean analog should be identified with Floer's Eq.}(4.6)\textsl{.}

These results play an important role in the rest of this work and, in
particular, in the next subsection.

\subsection{The proof (implementation)}

\subsubsection{General remarks}

In Ref.[79], we demonstrated how treatment of combinatorial data associated
with real scattering experiments leads to restoration of the underlying
quantum mechanical model reproducing the meson spectrum. It was established
that the underlying microscopic model is the Richardson-Gaudin (R-G) XXX
spin chain model originally designed for description of spectrum of
excitations in the Bardeen-Cooper-Schriefer (BCS) model of
superconductivity. \ Subsequently, the same model was used for description
of spectra of the atomic nuclei. Since the energy spectrum of the BCS model
has the famous gap between the ground and the first excited state, the
problem emerges :

\textsc{Can spectral properties of nonperturbative quantum Y-M field theory
be described by the R-G model}?

To answer this question affirmatively the "equivalence principle" discovered
by L.Witten is very helpful. Using it, we can proceed with quantization of \
pure Y-M fields by using results by Korotkin and Nicolai, Ref.[31], for
gravity. By comparing the main results of our paper, Ref.[79], done for QCD,
with those of Ref.[31], done for gravity, \ we found a complete agreement.
In particular, the Knizhnik-Zamolodchikov Eq.s(4.14),(4.15) and the R-G
Eq.(4.29) of Ref.[79] coincide \ respectively with Eq.s(4.27),(4.26) and
(4.50) of Ref.[31] even though methods of deriving of these equations are
entirely different! Both Ref.s [79] and [31] do not reveal the underlying
physics sufficiently deeply though. In the remainder of this section we
shall explain why this is indeed so and demonstrate ways this deficiency can
be corrected. Experimentally the challenge lies in designing scattering
experiments providing \ clean information about the spectrum of glueballs.
Thus far this task was accomplished only in lattice calculations done for
unphysically large number of colors, e.g. $N_{c}\rightarrow \infty .[$23$].$
When it comes to interpreting \textsl{real} experiments (always having only
three colors to consider\footnote{%
E.g. read Section 6\textbf{.}}), the situation is even less clear, e.g. see
Ref.[93]. Hence, the gap problem is full of challenges for both theory and
experiment. Fortunately, at least theoretically, the problem does admit
physically meaningful solution as \ we\ explained already. \ We continue
with ramifications in the next subsection.

\subsubsection{From Landau-Lifshitz to Gross-Pitaevskii equation via
Hashimoto map}

Since the F-S model \ is believed to be capable of describing QCD vacua and
\ is also capable of describing knotted/linked structures [17], two
questions arise: a) Is this the only model capable of describing QCD vacua?
b) To what extent it matters that the F-S model is also capable of
describing knots and links? The negative answer to the first question
follows from Corollary 3 implying that, in principle, both Euclidean and
hyperbolic versions of the L-L equation are capable of describing QCD vacua:
different vacua correspond to different steady-state solutions of the L-L
equations. The negative answer to the second question can be found in a
review, Ref.[85], by \ Annalisa Calini. From this reference it follows that,
besides the F-S model, knotted/linked structures can be also obtained by
using standard\ (that is Euclidean) L-L equation, e.g. see Eq.(B.4) of
Appendix B. This fact still does not explain why knots/links are of
importance to QCD. We address the above issues in more detail in Section 6.
In view of what is said above, wether or not the hyperbolic version of L-L
equation is capable of describing knotted structures is not immediately
important for us. Far more important is the connection between the
hyperbolic L-L and the Ernst equation. Only with this connection it is
possible to reproduce results by Korotkin and Nicolai [31].

Eq.(3.19) is just the F-S functional without winding number term. When the
commutation relations for su(1,1) introduced\ in subsection 5.2 \ are
replaced by those for su(2) this leads to the standard L-L equation (instead
of Eq.(5.14)). This replacement causes us to abandon \ the connection with
Ernst equation and, ultimately, with the results of Ref.[31]. In such a case
the gap problem should be investigated from scratch. In Ref.[25] Faddeev and
Niemi indicated that, unless some amendments to the F-S model are made, it
is gapless. At the same time from Appendix B it is known that the L-L
equation associated with the F-S model \ can be transformed into the NLSE\ \
with help of the Hashimoto map. \ Recently, Ding [94] and Ding and Inoguchi
[95] were able to find analogs of the Hashimoto map for \ the vortex
filaments in hyperbolic, de Sitter and anti de Sitter spaces. It is helpful
to describe their findings using terminology familiar from physics
literature [96].This leads us to the discussion of properties of the
Gross-Pitaevskii equation known in mathematics as the NLSE. In the system of
units in which $\hslash =1$ and $m=1/2$ this equation can be written as [86]%
\begin{equation}
i\psi _{t}=-\psi _{xx}+2\kappa \left( \left\vert \psi \right\vert
^{2}-c^{2}\right) \psi =0.  \tag{5.38}
\end{equation}%
Zakharov and Shabat [97,98] performed detailed investigation of this
equation for both positive and negative values of the coupling constant $%
\kappa .$ For $\kappa <0$ the above equation is used for description of
knots/links [85]. The standard Hashimoto map brings the L-L equation
associated with the truncated F-S model to the NLSE with $\kappa <0$ [94,
95]. \ From the same references it can be found that the Hashimoto-like map
brings the (hyperbolic) L-L-like equation to the NLSE for which $\kappa >0.$
Zakharov and Shabat studied in detail differences in treatments of the NLSE
for both negative and positive coupling constants. This difference is caused
by difference in underlying physics which in both cases can be explained in
terms of the properties of non ideal Bose gas [99,100]. The attentive reader
might have noticed at this point that Eq.(5.38) apparently contains no
information about the number of particles in such a gas. This parameter, in
fact, is hidden in the constant $c$ $($the chemical potential) or it can be
obtained selfconsistently with help of Eq.(5.38) (from which $c$ is removed
in a way described in Appendix B) as explained in Ref.[100]. With this
information at our disposal we are ready to make the next step.

\subsubsection{From non ideal Bose gas to Richardson-Gaudin equations}

Even though statistical mechanics of 1-d interacting Bose gas was considered
in detail by Lieb and Linger [101], solid state physics literature is full
of refinements of their results up to moment of this writing. These
refinements have been inspired by experimental and theoretical advancements
in the theory of Bose condensation [96]. Among this literature we selected
Ref.s[102,103] as the most relevant to our needs.

Following [102], the Hamiltonian \ for $N$ interacting bosons moving on the
circle of length $L$ is given by 
\begin{equation}
H=-\sum\limits_{i=1}^{N}\frac{\partial ^{2}}{\partial x_{i}^{2}}+2\check{c}%
\sum\limits_{1\leq i<j\leq N}\delta (x_{i}-x_{j})  \tag{5.39}
\end{equation}%
with constant $2\check{c}$ coinciding with $2\kappa $ in the system of units 
$\hslash =1$ and $m=1/2.$ The case $\check{c}>0$ (repulsive Bose gas)
corresponding to the L-L equation in the hyperbolic plane/space happens to
be of immediate relevance. \textsl{Only for this case it is} \textsl{%
possible to establish the connection with work by Korotkin and Nicolai} [31]!

We begin by noticing that in the standard BCS theory of superconductivity
electrons are paired into singlets (Cooper pairs) with zero centre of mass
momentum. The pairing interaction term in this theory accounts only for
pairs of attractive electrons with opposite spin and momenta so that the
degeneracy for each energy state is a doublet, with level degeneracy $\Omega
=2$ \footnote{%
We use here the same notations as in our work, Ref.[\textbf{94}].}. In the
interacting \textsl{repulsive} Bose gas model byRichardson [104] to mimic
this pairing he coupled two bosons with opposite momenta $\pm k_{j}$ into
one (pseudo) Cooper pair. An assembly of such formed pairs forms\textsl{\
repulsive} Bose gas \ which in the simplest case is described by the
Hamiltonian, Eq.(5.39). \textsl{Hence, the fermionic \ BCS-type model with
strong attractive pairing} \textsl{interaction can be mapped into bosonic
repulsive model proposed by Richardson.} Although the idea of such mapping\
looks very convincing, its actual implementation in Ref.[102] has some
flaws. Because of this, we shall use results of this reference selectively.
For this purpose, fist of all we need to make an explicit connection between
the repulsive Bose gas model described by Eq.(5.39) and the model proposed
by Richardson. In the weak coupling limit $\check{c}L\ll 1$ the Bethe ansatz
equations for the repulsive Bose gas model described by the Hamiltonian,
Eq.(5.39), acquire the following form:%
\begin{equation}
k_{i}=\frac{2\pi d_{i}}{L}+\frac{2\check{c}}{L}\sum\limits_{\substack{ j=1 
\\ \left( j\neq i\right) }}^{N}\frac{1}{k_{j}-k_{i}},i=1,...,N.  \tag{5.40}
\end{equation}%
Here $d_{i}=0,\pm 1,\pm 2,...$ denote the excited states for fixed $N$. To
link this result with Richardson's (repulsive boson) model, consider \ the
case of even number of bosons and make $N=2M$. Next, consider the ground
state of this model first. To the first order in $\check{c}$, it is clear
that we can write $k_{i}=\pm \sqrt{E_{i}}$. Specifically, let \ $k_{1,2}=\pm 
\sqrt{E_{1}},k_{3,4}=\pm \sqrt{E_{3}},...,k_{2M-1,2M}=\pm \sqrt{E_{M}}.$
Using these results in Eq.(5.40), with the accuracy just stated, the Bethe
ansatz equations after some algebra are converted into the following form: 
\begin{equation}
\frac{L}{2\check{c}}+\sum\limits_{\substack{ j=1  \\ \left( j\neq i\right) }}%
^{\tilde{M}}\frac{2}{E_{j}-E_{i}}=\frac{1}{2E_{i}},i=1,...,M;\text{ }\tilde{M%
}\leq M.  \tag{5.41}
\end{equation}%
To analyze these equations, we expect that our readers\ are familiar with
works of both Richardson-Sherman, Ref.[105], and Richardson, Ref.[104]. In
[105] diagonalization of the pairing force Hamiltonian describing the
BCS-type superconductivity was made. Such a Hamiltonian is given by 
\begin{equation}
H=\sum\limits_{f}2\varepsilon _{f}\hat{N}_{f}-g\sum\nolimits_{f}^{\prime
}\sum\nolimits_{f^{\prime }}^{\prime }b_{f}^{\dag }b_{f^{\prime }}, 
\tag{5.42}
\end{equation}%
where $\hat{N}_{f}=\frac{1}{2}(a_{f+}^{\dag }a_{f-}+a_{f-}^{\dag
}a_{f-}),b_{f}=a_{f-}a_{f+}$, with $a_{f\sigma }^{\dag }$ and $a_{f\sigma }$
being \textsl{fermion} creation and annihilation operators obeying usual
anticommutation relations $[a_{f\sigma },a_{f^{\prime }\sigma ^{\prime
}}^{\dag }]_{+}=\delta _{\sigma \sigma ^{\prime }}\delta _{ff^{\prime }}$,
where $\sigma =\pm $ denotes states conjugate under time reversal. The above
Hamiltonian is diagonalized along with the seniority operators (taking care
of the number of unpaired fermions at each level $f$) defined by 
\begin{equation}
\hat{\nu}_{f}=a_{f+}^{\dag }a_{f-}-a_{f-}^{\dag }a_{f-}.  \tag{5.43}
\end{equation}%
By construction, $[H,\hat{N}_{f}]=[H,\hat{\nu}_{f}]=0.$ The classification
of the energy levels is done in such a way that the eigenvalues $\nu _{f}$
of the operator $\hat{\nu}_{f}$ ($0$ and $\sigma )$ are appropriate for the
case when $g=0.$ This observation allows us to subdivide the Hamiltonian
into two parts:\ $H_{1},i.e.$that which does not contain Cooper pairs \ (for
which $\nu _{f}=\sigma )$ and $H_{2},i.e.$that which may contain such pairs
(for which $\nu _{f}=0).$ The matrix elements of $H_{2}$ are calculated with
help of the \textsl{bosonic-type} commutation relations%
\begin{equation}
\lbrack b_{f^{\prime }},\hat{N}_{f^{\prime }}]=\delta _{ff^{\prime }}b_{f}%
\text{ \ and }[b_{f}\text{ , }b_{f^{\prime }}^{\dag }]=\delta _{ff^{\prime
}}(1-2\hat{N}_{f^{\prime }}).  \tag{5.44}
\end{equation}%
These commutators are bosonic but nontraditional. In the traditional case we
have $[b_{f}$ , $b_{f^{\prime }}^{\dag }]=\delta _{ff^{\prime }}.$ \ We
refer our readers to Ref.[105] for details of how this commutator difficulty
is resolved. In the light of this resolution, in Ref.[104] Richardson
proposed to deal with the interacting bosons model from the beginning. 
\textsl{Supposedly, such bosonic model can be designed} \textsl{to reproduce
results of the fermionic pairing model of} Ref.[105]. \ An attempt to do
just this was made in Ref.[102]. In the repulsive boson model by Richardson
the "pairing" Hamiltonian is given by\footnote{%
To avoid ambiguities, our coupling constant $\frac{g}{2}$ is chosen exactly
the same as in [104].} 
\begin{equation}
H=\sum\limits_{l}2\varepsilon _{l}\hat{n}_{l}+\frac{g}{2}\sum\nolimits_{f}^{%
\prime }\sum\nolimits_{f^{\prime }}^{\prime }A_{f}^{\dag }A_{f^{\prime }}. 
\tag{5.45}
\end{equation}%
in which $\hat{n}_{l}$ and $A_{f^{\prime }}$ are bosonic analogs of $\hat{N}%
_{f}$ and $b_{f}.$ It is essential that the sign of the coupling constant $g$
is nonnegative (repulsive bosons). Upon diagonalization, the total energy $E$
is given by 
\begin{equation}
E=\sum\limits_{l=1}^{n}\varepsilon _{l}\nu _{l}+\sum\limits_{j=1}^{m}E_{j} 
\tag{5.46}
\end{equation}%
so that summation in the first sum takes place over the unpaired bosons
while in the second- over the paired bosons whose energies $E_{j}$ are
determined from the Richardson's equation (Eq.(2.29) of Ref. [104])\footnote{%
Since Gaudin's equation is obtained in the limit $\left\vert g\right\vert
\rightarrow \infty .$ from Eq.(5.47) the spin -like model described by this
equation is known as the Richardson-Gaudin (R-G) model.}%
\begin{equation}
\frac{1}{2g}+\sum\limits_{l=1}^{n}\frac{d_{l}}{2\varepsilon _{l}-E_{k}}%
+\sum\limits_{\substack{ i=1  \\ i\neq k}}^{m}\frac{2}{E_{i}-E_{k}}%
=0,k=1,...,m  \tag{5.47}
\end{equation}%
in which $n$ is the total number of single particle (unpaired) levels, $m$
is the total number of pairs, $d_{l}=\frac{1}{2}(2\nu _{l}+\Omega _{l}).$
From [104] it follows that for\ the bosonic model to mimic the BCS-type
pairing model the degeneracy factor $\Omega _{l}=1$ and $\nu _{l}=0.$ It
should be noted though that such an identification is not of much help in
comparing the repulsive bosonic model with the attractive BCS-type fermionic
model (contrary to claims made in Ref.[102]). This can be easily seen by
comparison between Eq.(5.47) (that is Eq.(2.29) of Ref.[104]) with such
chosen $\Omega _{l}$ and $\nu _{l}$ with Eq.(3.24) of Ref.[105]. By
replacing $g$ in Eq.(5.47) by $-g$ we still will not obtain the analog of
the key Eq.(3.24) of Ref.[105]! This fact has group-theoretic origin to be
explained in the next subsection. In the meantime, Eq.(5.47) still can be
used to connect it with Eq.(5.41) originating from \ different bosonic model
described by the Hamiltonian Eq.(5.39). To do so we follow the path
different from that suggested in Ref.[102]. Instead, following the original
Richardson's paper [104], \ we let $n=1$ in Eq.(5.47) then, without loosing
generality, we can put $\varepsilon _{1}=0$ so that Eq.(5.47) acquires the
following form%
\begin{equation}
\frac{1}{E_{k}}=\frac{1}{2g}+\sum\limits_{\substack{ i=1  \\ i\neq k}}^{M}%
\frac{2}{E_{i}-E_{k}},\text{ }k=1,...,M.  \tag{5.48}
\end{equation}%
The rationale for replacing $m$ by $M$ is given on page 1334 of [104].
Evidently, Eq.s (5.41) and (5.48) are identical. This observation allows us
to use the Richardson model instead of that described by Eq.(5.39). \ At
first sight such an identification looks a bit artificial. To convince our
readers that it does make sense, we would like to use the work by Dhar and
Shastry [106,107] on excitation spectrum of the ferromagnetic Heisenberg
spin chain. By analogy with Eq.(5.41) these authors derived a similar
equation obtained by reducing the Bethe ansatz equations for Heisenberg
ferromagnetic chain. It reads\footnote{%
The physical meaning of constants entering this equation is not important
for us. It is given in Ref.[106]..} 
\begin{equation}
\frac{1}{E_{l}}=\pi d-\frac{d}{n}\sum\limits_{\substack{ i=1  \\ i\neq l}}%
\frac{2}{E_{i}-E_{l}}.  \tag{5.49}
\end{equation}%
Even though Eq.s(5.48) and (5.49) look almost the same, they are not the
same! \ The crucial difference lies in the signs in front of the second term
in the r.h.s. of these equations. Because of this difference Heisenberg's
ferromagnetic spin chain model is mapped onto Bose gas model with \textsl{%
attractive} interaction in complete accord with what was said immediately
after Eq.(5.38). Regrettably, this result is still not the same as for the
BCS-type model \ investigated in Richardson-Sherman's paper, Ref.[105]. This
fact was recognized and discussed in some detail already by Richardson
[104]. For completeness, we mention that the problem of \ BCS-Bose-Einstein
condensation (BEC) crossover which follows exactly the qualitative picture \
\ just described was made quantitative only very recently in Ref.[108].
Fortunately, it is possible to by-pass this result as explained in the next
subsection.

\subsubsection{From Richardson-Gaudin to Korotkin-Nicolai equations}

In Ref.[109] bosonic and fermionic formalism for pairing models discussed in
the previous subsection was developed. This formalism happens to be the most
helpful for investigation of the gap problem. Indeed, define three operators 
$\hat{n}_{l}=\sum\nolimits_{m}a_{lm}^{\dag }a_{lm},$ $A_{l}^{\dag }=\left(
A_{l}^{{}}\right) ^{\dag }=\sum\nolimits_{m}a_{lm}^{\dag }a_{l\bar{m}}^{\dag
}$ . They can be used for construction of operators $K_{l}^{0}=\frac{1}{2}%
\hat{n}_{l}\pm \frac{1}{4}\Omega _{l}$ \ and $K_{l}^{+}=\frac{1}{2}%
A_{l}^{\dag }=\left( K_{l}^{-}\right) ^{\dag }$ such that they obey the
following commutator algebra%
\begin{equation}
\lbrack K_{l}^{0},K_{l}^{+}]=\delta _{ll}K_{l}^{+},\text{ }%
[K_{l}^{+},K_{l}^{-}]=\mp 2\delta _{ll}K_{l}^{0}.  \tag{5.50}
\end{equation}%
In this algebra as well as in the preceding expressions, the upper sign
corresponds to bosons and the lower to fermions. In Ref.[79], we discussed
such an algebra for the fermionic case only, e.g. see Eq.s (4.31) of [79].
These results can be extended now for the bosonic case. In fact, such an
extension is already developed in Ref.[109]. Unlike [79], where we used $%
sl(2,\mathbf{C})$ Lie algebra, only its compact version, that is $su(2)$,
was used in [109] for representing fermions. For bosonic case the
commutation relations, Eq.(5.50), are those for $su(1,1)$ Lie algebra.
Incidentally, in the paper by Korotkin and Nicolai, Ref.[31], exactly the
same Lie algebra was used. Furthermore, in the same paper it was argued that
it is permissible to replace $su(1,1)$ by $sl(2,\mathbf{R})$ Lie algebra
while constructing the K-Z-type equations, e.g. read p.428 of this
reference. Since in [79] the $sl(2,\mathbf{C})$ Lie algebra \ was used, that
is a complexified version of $sl(2,\mathbf{R}),$ this allows us to use many
results from our work. Thus, in this subsection we shall discuss only those
results of [109] which are absent in our Ref.[79]. In particular, following
this reference the set of Gaudin-like commuting Hamiltonians written in
terms of operators $K_{l}^{0},K_{l}^{+}$ and $K_{l}^{-}$ is given by 
\begin{equation}
H_{l}=K_{l}^{0}+2g\{\sum\limits_{l^{\prime }(\neq l)}\frac{X_{ll^{\prime }}}{%
2}(K_{l}^{+}K_{l^{\prime }}^{-}+K_{l}^{-}K_{l^{\prime }}^{+})\mp
Y_{ll^{\prime }}K_{l}^{0}K_{l^{\prime }}^{0}\}.  \tag{5.51}
\end{equation}%
Here $X_{ll^{\prime }}=Y_{ll^{\prime }}=(\varepsilon _{l}-\varepsilon
_{l^{\prime }})^{-1}.$ For $g\rightarrow \infty $ the first term can be
ignored and the remainder can be used in the K-Z-type equations. \ The
semiclassical treatment of these equations discussed in detail in [79] is
resulting in the following set of Bethe (or R-G) ansatz equations 
\begin{equation}
\sum\limits_{l=1}^{n}\frac{d_{l}}{2\varepsilon _{l}-E_{k}}\pm \sum\limits 
_{\substack{ i=1  \\ i\neq k}}^{m}\frac{2}{E_{i}-E_{k}}=0,\text{ }k=1,...,m 
\tag{5.52}
\end{equation}%
to be compared with Eq.(5.47). Unlike Eq.(5.47), in the present case $d_{l}=$
$\frac{1}{2}(2\nu _{l}\pm \Omega _{l}).$ The bosonic version of Eq.(5.52)
corresponding to $su(1,1)$ Lie algebra coincides with Eq.(4.50) of Korotkin
and Nicolai paper, Ref.[31], provided that the following identifications are
made: $d_{l}\rightleftarrows s_{l}$, $2\varepsilon _{l}\rightleftarrows
\gamma _{j}$. Unlike Ref.[31], where Eq.(5.52) was obtained by standard
mathematical protocol, in this work it is obtained based on the underlying
physics. Because of this, it is appropriate to extend our physics-stype
analysis by considering the case of finite $g^{\prime }s.$ Then, Eq.(5.52)
should be replaced by%
\begin{equation}
\frac{1}{2g}\pm \sum\limits_{l=1}^{n}\frac{d_{l}}{2\varepsilon _{l}-E_{k}}%
\pm \sum\limits_{\substack{ i=1  \\ i\neq k}}^{m}\frac{2}{E_{i}-E_{k}}%
=0,k=1,...,m.  \tag{5.53}
\end{equation}%
In Ref.[31] the gap problem was discussed in detail for the fermionic case
when the coupling constant $g$ is negative (BCS pairing Hamiltonian), e.g.
see Eq.s (4.43)-(4.45) of Ref.[31]. In the present case we are dealing with
the bosonic case for which the coupling constant is positive. Hence the gap
problem should be re analyzed. For this purpose, it is convenient to
consider both positive and negative coupling constants \ in parallel for
reasons which will become apparent upon reading.

\subsubsection{Emergence of the gap and \ the gap dilemma}

Eq.s(5.53) cannot be solved without some physical input. \ Initially, such
an input \ was coming from nuclear physics (e.g. read [110-112] for general
information on nuclear physics). \ \ Indeed, Richardson's papers [104,105]
were written having applications to nuclear physics in mind. Given this, the
question arises about the place of the R-G model among other models
describing nuclear spectra and nuclear properties. We need an answer to this
question to finish proof of the gap's existence in QCD.

Looking at the Gaudin-like Hamiltonian, Eq.(5.51), and comparing it with the
Hamiltonian, Eq.(6), in Ref.[113]\footnote{%
Published in 1961!} it is easy to notice that they are almost the same! \
Because of this, it becomes possible to transfer the methodology of
Ref.[113] to the present case. Thus, it makes sense to recall briefly
circumstances at which the gap emerges in nuclear physics.

As is well known, the nuclei are made of protons and neutrons. One can talk
about the number $\mathcal{N}$ of nucleons, the number Z of protons and the
number N of neutrons in a given nucleus. Nuclear and atomic properties
happen to be interrelated. For instance, in analogy with atomic physics one
can think of some effective nuclear potential in which nucleons can move
"independently". This assumption leads to the \textsl{shell model} of
nuclei. Use of Pauli principle guides fillings of shells \ the same way as
it guides these fillings in atomic physics. This leads to emergence of magic
numbers 2, 8, 20, 28, 50, 82 and 126 for either protons or neutrons for the
totally filled shells. Accordingly, the most stable are the doubly magic
(for both protons and neutrons) nuclei. It is of interest to know what kinds
of excitations are possible in such shell models? The simplest of these is
when some nucleon is moving from the closed shell to the empty shell thus
forming \textsl{a hole}. When the number of nucleons increases, the question
about the validity of the shell model emerges, again in analogy with atomic
physics. As in atomic physics, one can think about the Hartree-Fock (H-F)
and other many-body computational schemes, including that developed by
Richardson-Sherman and Gaudin. For our purposes, it is sufficient to use
only the Tamm-Dankoff (T-D) approximation \ to the H-F equations described,
for example, in Ref.[112]. The essence of this approximation lies in
restricting the particle-hole interactions to nucleons lying in the same
shell.\textsl{\ The T-D approximation is obtainable from the} is \textsl{R-G
Eq.s}(5.53)\textsl{\ when the last term (effectively taking care of Pauli
principle) in these equations is dropped}. The T-D approximation was
successfully applied for description of the giant nuclear dipole resonance
[110-112]. At the classical level the physics of this resonance was
explained in the paper by Goldhaber and Teller [114]. The resonance is
caused by two nuclear vibrational modes: one, when protons and neutrons move
in the opposite directions and another- when they move in the same
direction. Upon quantization of such classical model and taking into account
the isotopic spin of nucleons, the truncated Eq.s(5.53) are obtained in
which both signs for the coupling constant are allowed since the nucleon
system is expected to be in two isospin states : $T=1$ and $T=0\footnote{%
This can be easily understood based on the fact that isospin for both
particles and holes is equal to 1/2 [110-112].}$. Details of these
calculations are given in Ref.[112], page 221. Solutions of \ the T-D
equations can be obtained graphically in complete analogy with that
described in our work, Ref.[79]. These graphical solutions reflect the
particle-hole duality built into the T-D approximation. Because of this
duality, the magnitude of the gap in both cases should be the same. To
demonstrate this, the \textsl{seniority} scheme described in [110-112] is
helpful. The seniority operator was defined by Eq.(5.43). It determines the
number of unpaired particles in the nuclear system. Since it commutes with
the Hamiltonian, the many-body states can be classified with help of its
eigenvalues $\nu _{f}.$ Suppose at first that all single particle energies $%
\varepsilon _{f}$ are the same (that is $\varepsilon _{f}=\varepsilon )$ so
that all seniority eigenvalues $\nu _{f}$ are $\nu .$ Let then $\mathcal{N}$
be the total number of nucleons. Thus, the state for which $\nu =0$ contains
only pairs, analogously, the state $\nu =1$ contains just one unpaired
nucleon, $\nu =2$ has 2 unpaired nucleons and $\mathcal{N}$ should be even
and so on. So, states $\nu =0,\nu =2,\nu =4,...$ can exist only in even
nuclei. For such nuclei the gap is nonzero. To see this, we follow
Refs.[110-112] which we would like now to superimpose with the results of
the Richardson-Sherman paper, Ref.[105]. Specifically, on page 231 of this
reference one can find the following result for the ground state ($\nu =0)$
energy 
\begin{equation}
E_{\nu =0}(N)=2N\varepsilon -gN(\Omega -N+1)  \tag{5.54}
\end{equation}%
where $N$ is the number of pairs. To connect this result with that in
Refs.[110-112], let $N=\mathcal{N}$/2 and consider the difference 
\begin{equation}
\mathcal{E}_{\nu =0}(N\text{)}=E_{\nu =0}(\mathcal{N}/2)-\mathcal{N}%
\varepsilon =-\frac{g}{4}\mathcal{N}(2\Omega -\mathcal{N}+2).  \tag{5.55}
\end{equation}%
The obtained result coincides with Eq.(11.14) of Ref.[112] as required. To
obtain \ states of seniority $\nu =2n$ we use Eq.(3.2) of \ Ref.[105]. It
reads 
\begin{equation}
E_{\nu =2n}(N)=2N\varepsilon -g\left( N-n\right) (\Omega -N-n+1),\text{ }%
n=0,...,N.  \tag{5.56}
\end{equation}%
Repeating the same steps as in\ $\nu =0$ case we obtain, 
\begin{equation}
\mathcal{E}_{\nu }(\mathcal{N}\text{)}=-\frac{g}{4}\left( \mathcal{N}\text{-}%
\nu \right) (2\Omega -\mathcal{N}\text{-}\nu +2).  \tag{5.57}
\end{equation}%
Finally, consider the difference%
\begin{equation}
\mathcal{E}_{\nu }(\mathcal{N}\text{)}-\mathcal{E}_{\nu =0}(\mathcal{N}\text{%
)=}\frac{g}{4}\nu (2\Omega -\nu +2).  \tag{5.58}
\end{equation}%
This result is in accord with Eq.(11.22) of Ref.[112]. Since the obtained
difference is $\mathcal{N}$-independent it can be used both ways: a) for
calculations in the thermodynamic limit $\mathcal{N}\rightarrow \infty $ and
b) for making accurate calculations in the opposite limit of very small
number of nucleons. In the simplest case we should consider only one shell
and the first excited state of seniority 2 for this shell. Initially (the
ground state) we have just one pair while finally (the first excited state)
we have two independent particles occupying single particle levels.

Looking at Eq.s(5.53) and letting there $m=1$(one pair) we recognize that
the second sum in this set of equations disappears. Thus, by design, we are
left with the T-D approximation. Using Eq.(5.58) for $\nu =2$ we obtain the
\ following value of the gap $\Delta :$%
\begin{equation}
\Delta =\mathcal{E}_{2}(\mathcal{N}\text{)}-\mathcal{E}_{0}(\mathcal{N}\text{%
)}=g\Omega .  \tag{5.59a}
\end{equation}%
Notice, that since $\Omega $ is the degeneracy, there could be no more than $%
\mathcal{N}=\Omega $ particles at the single particle level. Thus, in
general we should have $\mathcal{N}\leq \Omega .$ Because of the
particle-hole duality, it is permissible to look also at the situation for
which $\mathcal{N}\geq \Omega .$This is equivalent to changing the sign in
front of the coupling constant. Repeating again all steps leads to the final
result for the gap%
\begin{equation}
\Delta =\mathcal{E}_{2}(\mathcal{N}\text{)}-\mathcal{E}_{0}(\mathcal{N}\text{%
)=}\left\vert \text{g}\right\vert \Omega .  \tag{5.59b}
\end{equation}%
It is demonstrated in Ref.s [110-112] that in the limit $\mathcal{N}%
\rightarrow \infty ,$ when the continuum approximation (replacing summation
by integration) can be used leading to a more familiar BCS-type equation for
the gap, the result just obtained survives. \ Indeed, in Ref.[103] the
BCS-type result is obtained in the continuum approximation for the \textsl{%
attractive} Bose gas. In view of the results just obtained, it should be
clear that such a result should hold for both attractive and repulsive Bose
gases. This conclusion is in accord with accurate recent Bethe ansatz
calculations done in Ref.[115] for \ systems of finite size. Thus, we just
arrived at the\ issue which we shall call \textsl{the gap dilemma}. While
the results obtained above strongly\textsc{\ }favor\textsc{\ }use of the
repulsive Bose gas model\textsc{\ }\textit{(\textsl{not linked with the F-S
model}), }the results\textsc{\ }obtained in this subsection indicate that,
after all, the F-S model\textit{\ (}\textsl{\ linked with the}\textit{\ 
\textsl{attractive Bose gas model }) }can also be used for description of
the ground and excited states\textsc{\ }of pure Y-M fields\textit{. }\textsc{%
The essence of the dilemma lies in deciding which of these results should
actually be used}.

While the answer is provided in the next section, we are not yet done with
the gap discussion. This is so because the seniority model is applicable
only to the case when all single-particle levels have the same energy. This
is too simplistic. We would like now to discuss more realistic case

Before doing so, few comments are appropriate. In particular, with all
successes of \ nuclear physics models, these models are much less convincing
than those in atomic physics. Indeed, all nuclei are made of hadrons which
are made of quarks and gluons. Thus the excitations in nuclei are in fact \
the excitations of quark-gluon plasma. This observation qualitatively
explains why the R-G equations work well both in nuclear and particle
physics. \ Some attempts to look at the processes in nuclear physics from
the standpoint of hadron physics can be found in Refs.[116,117]. \ 

Now we can return to the discussion of \ the T-D equations. Fortunately,
detailed analytical study of these equations was recently made in Ref.[118].
The same authors extended these results to the case of two pairs in [119].
Since the results obtained in [119] are in qualitative agreement with those
obtained in Ref.[118], we shall focus attention of our readers only on
results of Ref.[118]. Thus, we need to find some kind of analytic solution
of the following T-D equation%
\begin{equation}
\sum\limits_{i=1}^{L}\frac{\Omega _{i}}{2\varepsilon _{i}-E_{{}}}=\frac{1}{g}%
.  \tag{5.60}
\end{equation}%
For different $\varepsilon _{i}^{\prime }s$ normally it should have \ $L$
eigenvalues $E_{\mu }$ $(1\leq \mu \leq L).$ Since we are interested in
finding the gap, the above equation is written for just one nucleon pair.
Thus the seniority $\nu =0.$ It is of interest to check first what happens
when all $\varepsilon _{i}^{\prime }s$ coalesce. In such a case we obtain, 
\begin{equation}
\frac{\Omega _{{}}}{2\bar{\varepsilon}-E}=\frac{1}{g},  \tag{5.61}
\end{equation}%
where $\Omega =\sum\nolimits_{i}\Omega _{i}$ and $\varepsilon _{i}=\bar{%
\varepsilon}$ $\forall i=1,...,L.$ Eq.(5.61) can be\ equivalently rewritten
as%
\begin{equation}
E_{0}=2\bar{\varepsilon}-\Omega g.  \tag{5.62}
\end{equation}%
\ This result for the ground state is in agreement with Eq.(5.54) for $N=1$.
The first excited state is made of one broken pair so that the pairing
disappears and the energy $E_{\nu =2}=2\bar{\varepsilon}$. From here, the
value of the gap is obtained as $E_{\nu =2}-$ $E_{0}=$ $g\Omega $ in
agreement with Eq.(5.59). If now we make all energy levels different, then
one can see that solutions to Eq.(5.60) are subdivided into those lying in
between the single particle levels (\textsl{trapped solutions}) and those \
which lie outside these levels (\textsl{collectivized solutions}). For $%
\left\vert g\right\vert $ sufficiently large the solution, Eq.(5.61), is the
leading term (in the sense described below) representing the collectivized
solution. Since the trapped solutions represent corrections to energies of
single particle states, they do not contribute directly to the value of the
gap. They do contribute to this value indirectly. Indeed, following
Ref.[118] we rewrite Eq.(5.60) as 
\begin{equation}
\sum\limits_{i=1}^{L}\frac{\Omega _{i}}{2\varepsilon _{i}-E_{{}}}=\frac{1}{2%
\bar{\varepsilon}-E}\sum\limits_{i}\frac{\Omega _{i}}{1+2\frac{\varepsilon
_{i}-\bar{\varepsilon}}{2\bar{\varepsilon}-E}}=\frac{1}{g}  \tag{5.63}
\end{equation}%
and expand the denominator of Eq.(5.63) in a power series. As result, the
following expansion 
\begin{equation}
\frac{E-2\bar{\varepsilon}}{g\Omega }=-1-\alpha ^{2}+\gamma \alpha
^{3}+O(\alpha ^{4})  \tag{5.64}
\end{equation}%
is obtained in which $\bar{\varepsilon}=\frac{1}{\Omega }\sum\limits_{i}%
\Omega _{i}\varepsilon _{i}$, $\alpha =\frac{2\sigma }{g\Omega },\sigma =%
\sqrt{\frac{1}{\Omega }\sum\limits_{i}\Omega _{i}(\varepsilon _{i}-\bar{%
\varepsilon})^{2}}$ and $\gamma $ is related to the higher order moments (
details are in Ref.s[118,119]). Using these results, the gap is obtained in
the same way as before.

The quality of computations in Ref.[118] was tested for 3-dimensional
harmonic oscillator (by adjusting dimensionality of this oscillator\ it can
be thought of \ as "closed string model" representing \ both the shell model
for atomic nucleus and the gluonic ring for the Y-M fields) for which $%
\varepsilon _{i}=(i+3/2)$ \ (in the system of units in which $\hbar \omega
=1)$ and $\Omega _{i}=(i+1)(i+2)/2.$ \ For this 3-dimensional \ oscillator
corrections to the collectivized energy, Eq.(5.64), become negligible
already for $\left\vert g\right\vert \geq 0.2,$ provided that $L\geq 8.$
Obtained results allow us to close this section at this point. These results
are of no help in solving the gap dilemma though. This \ task is
accomplished in the next section.

\section{Resolution of the gap dilemma}

\subsection{Motivation}

In the previous section we provided evidence linking the gap problem for Y-M
fields with the problem about the excitation spectrum of the repulsive Bose
gas. \ The gap equation, Eq.(5.59), is also used in nuclear physics where it
is known to produce the same value for the gap for both signs of the
coupling constant $g$. Since both options are realizable in Nature in the
case of nuclear physics, the question arises about such possibility in the
present case. In the case of nuclear physics experimental realization (giant
nuclear dipole resonance) of both options for the coupling constant is
experimentally testable. Thus, in the present case we have to find some
alternative physical evidence. If, indeed, such evidence could be found,
this would allow us to bring back into play the well studied F-S model which
microscopically is essentially equivalent to the XXX 1d Heisenberg
ferromagnet as results of Appendix B and subsections 3.5 and 5.2 indicate. \
The next subsection supplies us with the alternative physical evidence.

\subsection{Some facts about harmonic maps and their uses in general
relativity}

Suppose we are interested in a map from $m-$dimensional Riemannian manifold
\ $\mathcal{M}$ with coordinates $x^{a}$ and metric $\gamma _{ab}(\mathbf{x}%
) $ to $n$-dimensional Riemannian manifold \ $\mathcal{N}$ with coordinates $%
\varphi ^{A\text{ }}$ and metric $G_{AB}(\varphi )$. A map $\mathcal{M}%
\rightarrow \mathcal{N}$ is called \textsl{harmonic }if $\varphi ^{A\text{ }%
}(x^{a})$ satisfies the Euler-Lagrange (E-L) equations originating from
minimization of the following Lagrangian%
\begin{equation}
\mathcal{L}=\sqrt{\gamma }G_{AB}(\mathbf{\varphi })\gamma ^{ab}(\mathbf{x}%
)\varphi _{,a}^{A}\varphi _{,b}^{B}  \tag{6.1}
\end{equation}%
in which $\gamma =\det (\gamma _{ab}).$ \ Since such defined Lagrangian is
part of the Lagrangian given by Eq.(3.6), the E-L equations for Eq.(6.1), in
fact, coincide with Eq.s(3.10). \ In the most general form they can be
written as [38]\footnote{%
We use the 1st edition of Ref.\textbf{[}38] for writing this equation. This
means that we have to define $\Gamma _{BC}^{A}$ as $\Gamma _{BC}^{A}=\frac{1%
}{2}G^{ad}\{\frac{\partial }{\partial \varphi ^{c}}G_{bd}+\frac{\partial }{%
\partial \varphi ^{b}}G_{cd}-\frac{\partial }{\partial \varphi ^{d}}%
G_{bc}\}. $} 
\begin{equation}
\varphi _{,a}^{A;a}+\Gamma _{BC}^{A}\varphi _{,a}^{B}\varphi ^{C_{,a}}=0. 
\tag{6.2}
\end{equation}%
In such a form we can look at transformations $\varphi ^{A^{\prime
}}=\varphi ^{A^{\prime }}(\varphi ^{B})$ keeping $\mathcal{L}$
form-invariant. To find such transformations, following Neugebauer and
Kramer [38], we introduce the auxiliary Riemannian space defined by the
metric%
\begin{equation}
dS^{2}=G_{AB}(\mathbf{\varphi })d\varphi ^{A}d\varphi ^{B}.  \tag{6.3}
\end{equation}%
Use of the above metric allows us to investigate the invariance of $\mathcal{%
L}$ with help of standard methods of Riemannian geometry. In the present
case, this means that one should study Killing's equations in spaces with
metric $G_{AB}.$ Specifically, let us consider the Lagrangian for
source-free Einstein-Maxwell fields admitting at least one non-null Killing
vector $\xi .$ To design such a Lagrangian we begin with the Ernst equation,
Eq.(2.4), for pure gravity and replace the Ernst potential $\epsilon
=-F+i\omega \footnote{%
Recall, that $-F=V$ according to notations introduced in connection with
Eq.(2.4).}$ by two complex potentials $\mathcal{E}$ and $\Phi $. Then, by
symmetry, the equations for stationary Einstein-Maxwell fields can be
written as follows [38] 
\begin{equation}
F\mathcal{E}_{,a}^{;a}+\gamma ^{ab}\mathcal{E}_{,a}(\mathcal{E}_{,b}+2\Phi
_{,b}\bar{\Phi})=0,F\Phi _{,a}^{;a}+\gamma ^{ab}\Phi _{,a}(\mathcal{E}%
_{,b}+2\Phi _{,b}\bar{\Phi})=0.  \tag{6.4}
\end{equation}%
These equations are obtained by minimization of the Lagrangian 
\begin{equation}
\mathcal{L}=\sqrt{\gamma }[\hat{R}_{ab}+2F^{-1}\gamma ^{ab}\Phi _{,a}\Phi
_{,b}+\frac{1}{2}F^{-2}\gamma ^{ab}(\mathcal{E}_{,a}+2\bar{\Phi}\Phi _{,a})(%
\mathcal{E}_{,b}+2\bar{\Phi}\Phi _{,b})],  \tag{6.5}
\end{equation}%
i.e. from equations $\frac{\delta \mathcal{L}}{\delta \gamma ^{ab}}=0,\frac{%
\delta \mathcal{L}}{\delta \Phi }=0$ and\ $\frac{\delta \mathcal{L}}{\delta 
\mathcal{E}}=0.$Taking these results into account, the auxiliary metric,
Eq.(6.3), can now be written as%
\begin{equation}
dS^{2}=2F^{-1}d\Phi d\bar{\Phi}+\frac{1}{2}F^{-2}\left\vert d\mathcal{E}+2%
\bar{\Phi}d\Phi \right\vert ^{2}.  \tag{6.6}
\end{equation}%
The analysis done by Neugebauer and Kramer [38] shows that there are eight
independent Killing vectors leading to the following finite transformations :%
\begin{equation}
\begin{array}{cc}
\mathcal{E}^{\prime }=\alpha \bar{\alpha}\mathcal{E}, & \Phi ^{\prime
}=\alpha \Phi ; \\ 
\mathcal{E}^{\prime }=\mathcal{E}+ib, & \Phi ^{\prime }=\Phi ; \\ 
\mathcal{E}^{\prime }=\mathcal{E}(1+ic\mathcal{E})^{-1}, & \Phi ^{\prime
}=(1+ic\mathcal{E})^{-1}; \\ 
\mathcal{E}^{\prime }=\mathcal{E}-2\bar{\beta}\Phi -\beta \bar{\beta}, & 
\Phi ^{\prime }=\Phi +\beta ; \\ 
\mathcal{E}^{\prime }=\mathcal{E}(1-2\bar{\gamma}\Phi -\gamma \bar{\gamma}%
\mathcal{E})^{-1}, & \Phi ^{\prime }=(\Phi +\gamma \mathcal{E})(1-2\bar{%
\gamma}\Phi -\gamma \bar{\gamma}\mathcal{E})^{-1}.%
\end{array}
\tag{6.7}
\end{equation}%
Complex parameters $\alpha ,\beta ,\gamma $ as well as real parameters $b$
and $c$ are connected with these eight symmetries. Evidently, solutions $%
\mathcal{E}^{\prime }$,$\Phi ^{\prime }$ are also solutions of Eq.s(6.4),
provided that $\gamma ^{ab}$ stays the same. Therefore if, say, we choose
some vacuum solution as a "seed", we would obtain, say, the electrovacuum
solution in accord with Appendix A. Incidentally, the electrovacuum
solutions obtained by Bonnor (Appendix A) cannot be obtained with help of
transformations given by Eq.s(6.7). They are considered separately below.
These observations allow us to reduce the Lagrangian $\mathcal{L}$ to the
absolute minimum without loss of information. In 1973 Kinnersley [38] found
that the group of symmetry transformations for the Einstein-Maxwell
equations with non null Killing vector is the group SU(2,1) which has eight
independent generators. In view of the above mentioned reduction of $%
\mathcal{L}$ it is sufficient to replace the metric in Eq.(6.6) by a
collection of much simpler metric related to each other by transformations
Eq.(6.7). All the possibilities are described in the Table 34.1 of Ref.[38].
For our needs we focus only on three of these (much simpler/reduced) metric
listed in this table. These are 
\begin{equation}
\text{ }dS^{2}=\frac{2d\xi d\bar{\xi}}{(1-\xi \bar{\xi})^{2}},\mathcal{E}=%
\frac{1-\xi }{1+\xi },  \tag{6.8}
\end{equation}%
\begin{equation}
dS^{2}=\frac{2d\Phi d\bar{\Phi}}{(1-\Phi \bar{\Phi})^{2}},  \tag{6.9}
\end{equation}%
and%
\begin{equation}
dS^{2}=\frac{-2d\Phi d\bar{\Phi}}{(1+\Phi \bar{\Phi})^{2}}.  \tag{6.10}
\end{equation}

\bigskip

The first \ and the second \ of these metric correspond to the vacuum state,
respectively, with $\Phi =0$ and $\mathcal{E}=-1$, of pure gravity
associated with the subgroup SU(1,1) of SU(2,1). The third metric,
Eq.(6.10), corresponds to a subgroup SU(2). It is related to the
electrostatic fields ($\mathcal{E}=1$) \ such that the space-time becomes
asymptotically flat for $\mathcal{E}\mathcal{\rightarrow }0.$It is important
that the metric, Eq.(6.10), is related to the vacuum metric,
Eq.s(6.8),(6.9), via transformations either listed in Eq.(6.7) or related to
these transformations. In particular, the related transformations can be
obtained as follows. Using Ref.[38], it is convenient to make the parameters 
$b$ and $c$ in Eq.s(6.7) complex and to consider all eight complex
parameters as independent of their complex conjugates. Under such
conditions\ the metric given by Eq.(6.10) \ is related to that given by
Eq.(6.8) by the simplest complex transformation: $\Phi ^{\prime }=i\xi $ \
and $\bar{\Phi}^{\prime }=i\bar{\xi}$ . These transformations indicate that,
starting with real vacuum solution for pure gravity as a seed, the above
transformations are capable of reproducing some electrovacuum solutions.
Additional details are discussed below.

These results can be interpreted as follows. \ While the Ernst functional,
Eq.(3.18), is representing pure axially symmetric gravity, the F-S-type
functional, Eq.(3.19), should describe some special case of electrovacuum
(Maxwell-Einstein) gravity. In view of results of Appendix C, it is possible
to use these transformations in reverse (see below), that is to obtain the
results for pure gravity from those for electrovacuum. This peculiar
"duality" property of gravitational fields provides physically motivated
resolution of the gap dilemma and, in addition, it allows us to obtain many
new results.

\subsection{Resolution of the gap dilemma and SU(3)$\times $SU(2)$\times $%
U(1) symmetry of the Standard Model}

The original F-S-type model thus far is limited only to SU(2) gauge theory.
SU(2) gauge theory is known to be used for description of electroweak
interactions where, in fact, one has to use the gauge group SU(2)$\times $%
SU(1) [19\textbf{]}. The hadron physics (that is QCD) requires us to use the
gauge group SU(3). This is caused by the fact that quark model of hadrons
uses flavors (e.g. u,d,s,c,b, t) labeling quarks of different masses. Each
of these quarks can be in three different colors (r,g,b) standing for "red",
"green" and "blue". Presence of different colors leads to fractional charges
for quarks. \ Far from the target\ the scattering products are always
colorless. The gauge group SU(3) is\ used for description of \ these colors.
Although theoretically the number of colors can be greater than three, this
number is strictly three experimentally [19]. The results of this work allow
us to reproduce this number of colors. For this purpose we have to be able
to provide the answer to the following \ \textsl{fundamental question}:

\textsc{Can equivalence between gravity and Y-M fields (for SU(2) gauge
group) discovered by Louis Witten be extended to the group SU(3)?}

Very fortunately, this can be done! \ For the sake of space, we shall be
brief whenever details can be found in literature, e.g. see Refs.[120-122].

To proceed, first, we have to go back to Eq.s(2.14),(2.15) and to modify
these equations in such a way that instead of the Ernst Eq.(2.4) for the
vacuum \ (gravity) field we should be able to obtain Eq.s (6.4) for
electrovacuum. In the limit $\Phi =0$ the obtained set of equations should
be reducible to Eq.(2.4). As it was noticed by G\"{u}rses and Xanthopoulos
[120], in general, this task cannot be accomplished. Indeed, these authors
demonstrated that the self-duality Eq.s(2.14) for SU(2) and for SU(3) Lie
groups look exactly the same for axially symmetric fields. Nevertheless, in
the last case, upon explicit computation instead of the vacuum Ernst
Eq.(2.4) one gets an electovacuum equations (e.g. see Eq.s(6.4)) which,
following Ernst [43], can be explicitly written as%
\begin{equation}
\left( \func{Re}\mathcal{E}+\left\vert \Phi \right\vert ^{2}\right) \nabla
^{2}\mathcal{E}=(\mathbf{\nabla }\mathcal{E}+2\bar{\Phi}\mathbf{\nabla }\Phi
)\cdot \mathbf{\nabla }\mathcal{E},  \tag{6.11a}
\end{equation}%
\begin{equation}
\left( \func{Re}\mathcal{E}+\left\vert \Phi \right\vert ^{2}\right) \nabla
^{2}\Phi =(\mathbf{\nabla }\mathcal{E}+2\bar{\Phi}\mathbf{\nabla }\Phi
)\cdot \mathbf{\nabla }\Phi .  \tag{6.11b}
\end{equation}%
These equations are obtained if, instead of the matrix $M$ given by
Eq.(2.15), one uses%
\begin{equation}
M=f^{-1}%
\begin{bmatrix}
1 & \sqrt{2}\Phi & -\frac{i}{2}(\mathcal{E}-\mathcal{\bar{E}}-2\Phi \bar{\Phi%
}) \\ 
\sqrt{2}\bar{\Phi} & -\frac{i}{2}(\mathcal{E}+\mathcal{\bar{E}}-2\Phi \bar{%
\Phi}) & -i\sqrt{2}\bar{\Phi}\mathcal{E} \\ 
\frac{i}{2}(\mathcal{\bar{E}}-\mathcal{E}-2\Phi \bar{\Phi}) & i\sqrt{2}%
\mathcal{\bar{E}}\Phi & \mathcal{E}\text{ }\mathcal{\bar{E}}%
\end{bmatrix}
\tag{6.12}
\end{equation}%
in which, instead of the one complex potential $\epsilon =-F+i\omega $ used
for solution of \ the vacuum Ernst Eq.(2.4), two complex potentials $%
\mathcal{E}$ and $\Phi $ are being used. In this expression the overbars
denote the complex conjugation and $f=$ $-\frac{1}{2}(\epsilon +\bar{\epsilon%
}+2\Phi \bar{\Phi}).$ Since the Einstein-Maxwell Eq.s(6.4) (or (6.11)) are
invariant with respect to transformations given by Eq.s(6.7), there should
be a matrix $A$ \ with constant coefficients such that the $M^{\prime
}=AMA^{\dag }$ will have primed potentials $\mathcal{E}$ and $\Phi $ taken
from those listed in the set Eq.(6.7). Authors of [120] found explicit form
of such $A$-matrices. However, when instead of matrix $M$ we substitute the
matrix $M^{\prime }$ into self-duality Eq.s(2.14), the combination $%
M^{\prime -1}\partial M^{\prime }$ looses this information. As result, we
are left with the following situation: while on the gravity side the matrix $%
M^{\prime }=AMA^{\dag }$ does allow us to obtain new and physically
meaningful solutions from the old ones, on the Y-M side all this information
is lost. Thus, the one-to-one correspondence discovered by L.Witten for
SU(2) is $\QTR{sl}{apparently}$ lost for SU(3). Very fortunately, this
happens only apparently! This is so because the Neugebauer- Kramer (N-K)
transformations described by Eq.s(6.7) do not exhaust all possible
transformations which can be applied to the matrix $M$, Eq.(6.12). Among
those which are not accounted by N-K transformations are those by Bonnor [38%
\textbf{,}123] whose work is mentioned in Appendix A. These are given by 
\begin{equation}
\mathcal{E}=\epsilon \bar{\epsilon};\Phi =\frac{1}{2}(\epsilon -\bar{\epsilon%
})=i\omega ,  \tag{6.13}
\end{equation}%
where $\epsilon =-F+i\omega $ is solution of the Ernst Eq.(2.4). In view of
the results of Appendix A one can be sure that the $\ $potentials $\mathcal{E%
}$ and $\Phi $ satisfy Eq.s(6.11). This means that one can use these
(Bonnor's) potentials in the matrix $M$ to reproduce Eq.s(6.11). This time,
there is one-to one correspondence between the self-duality Y-M and the
Einstein-Maxwell equations. Even though this is true, the question
immediately arises about relevance of such solutions to the solution of the
gap problem discussed in Section 5. In Section 5 the Ernst Eq.(2.4) was used
essentially for this purpose while Eq.s(6.11) are seemingly different from
Eq.(2.4). Again, fortunately, the difference is only apparent.

From the definition of Bonnor transformations, Eq.(6.13), it follows that
the potential $\mathcal{E}$ is real. Also, from the same definition it
follows that $\left\vert \Phi \right\vert ^{2}=\omega ^{2}.$Introduce now
new potential $Z=\mathcal{E}+\omega ^{2}$. For it, we obtain%
\begin{equation}
\mathbf{\nabla }Z=\mathbf{\nabla }(\mathcal{E}+\omega ^{2})=\mathbf{\nabla }%
\mathcal{E+}2\omega \mathbf{\nabla }\omega =\mathbf{\nabla }\mathcal{E}+2%
\bar{\Phi}\mathbf{\nabla }\Phi .  \tag{6.14}
\end{equation}%
Using this result, Eq.s(6.11) can be rewritten as follows%
\begin{equation}
\left( Z\nabla ^{2}\mathcal{-}\mathbf{\nabla }Z\cdot \mathbf{\nabla }\right)
\left( 
\begin{array}{c}
\mathcal{E} \\ 
\omega%
\end{array}%
\right) =0.  \tag{6.15}
\end{equation}%
Furthermore, consider the related equation%
\begin{equation}
\left( Z\nabla ^{2}\mathcal{-}\mathbf{\nabla }Z\cdot \mathbf{\nabla }\right)
\omega ^{2}=0.  \tag{6.16}
\end{equation}%
Evidently, if it can be solved, then equation $\left( Z\nabla ^{2}\mathcal{-}%
\mathbf{\nabla }Z\cdot \mathbf{\nabla }\right) \omega =0$ can be solved as
well. This being the case, the system of Eq.s(6.15) will be solved if the
Ernst-type vacuum equation 
\begin{equation}
Z\nabla ^{2}\mathcal{=}\mathbf{\nabla }Z\cdot \mathbf{\nabla }Z  \tag{6.17}
\end{equation}%
of the same type as Eq.(2.4) is solved. The obtained result is opposite to
that derived by Bonnor, described in Appendix A\textsl{\ }$\QTR{sl}{(}$see
also works Hauser and Ernst [124] and by Ivanov [125]). This means that, at
least in some cases (having physical significance) the self-dual Y-M fields
for both SU(2) and SU(3) gauge groups are obtainable as solutions of the
Ernst Eq.(2.4). This means that all results of Section 5 obtained for SU(2)
go through for the gauge group SU(3).

With these results at our disposal we would like to discuss their
applications to the Standard Model [19\textbf{, }126]. From Ref.[120] it is
known that the matrix $M\in $ SU(3) has subgroups which belong to SU(2). In
particular, one of such subgroups is obtained if we let $\Phi =0$ in
Eq.(6.12). Then, in view of Eq.(6.17), it is permissible to replace $%
\mathcal{E}$ by $\epsilon $ of Eq.(2.4). Thus, the obtained matrix $M$ is
decomposable as $M=M_{1}+M_{2},$ where the matrix $M_{1}$ is given by%
\begin{equation}
M_{1}=f^{-1}%
\begin{bmatrix}
1 & 0 & \omega \\ 
0 & 0 & 0 \\ 
\omega & 0 & \mathcal{\epsilon }\bar{\epsilon}%
\end{bmatrix}%
.  \tag{6.18}
\end{equation}%
in agreement with the matrix $M$ defined by Eq.(2.15) since in this case $f=$
$-\frac{1}{2}(\mathcal{\epsilon }+\mathcal{\bar{\epsilon}})=F.$ At the same
time, the matrix $M_{2}$ is given by 
\begin{equation}
M_{2}=%
\begin{bmatrix}
0 & 0 & 0 \\ 
0 & 1 & 0 \\ 
0 & 0 & 0%
\end{bmatrix}%
.  \tag{6.19}
\end{equation}%
\ Using elementary operations with matrices we can represent matrix $M$ in
the form 
\begin{equation}
\tilde{M}=%
\begin{bmatrix}
0 & 0 & 1 \\ 
a & b & 0 \\ 
b & c & 0%
\end{bmatrix}
\tag{6.20}
\end{equation}%
where $a=$ $1/F$, $b=$ $\omega /F$ and $c=(F^{2}+\omega ^{2})/F.$ \ Such a
form of the matrix $\tilde{M}$ is typical for the \ semidirect product of
groups (when group elements are represented by matrices). In general case
one should replace $\tilde{M}$ by 
\begin{equation*}
\tilde{M}=%
\begin{bmatrix}
0 & 0 & 1 \\ 
a & b & \alpha _{1} \\ 
b & c & \alpha _{2}%
\end{bmatrix}%
\end{equation*}%
\ Since the 2$\times $2 submatrix belongs to SU(2) (because its determinant
is 1) normally describing a rotation in 3d space (in view of SU(2)$%
\rightleftarrows $SO(3) correspondence), the parameters $\alpha _{1}$ and $%
\alpha _{2}$ are responsible for translation. In this, more general case,
the matrix $M$ describes the Galilean transformations, that is a combination
of translations and rotations. If the translational motion is one
dimensional it can be compactified to a circle \ in which case we obtain the
centralizer of SU(3) as SU(2)$\times $U(1). At the level of Lie algebra
su(3) this result was obtained in Ref.[127], pages 232 and 267. Its physical
interpretation discussed in this reference is essentially the same as ours.
The obtained centralizer is the symmetry group of the Weinberg-Salam model
(part of the standard model describing electroweak interactions).

All these arguments were meant only to demonstrate that the F-S-type model,
Eq.(3.19), should be used for description of electroweak interactions. For
description of strong interactions, in accord with Ref.[120], we claim that
the matrix $M$ given by Eq.(6.12) in which $\mathcal{E}$ and $\Phi $ are
taken from Bonnor's Eq.s(6.13) is \textsl{intrinsically} of SU(3) type. That
is,\textsl{\ it cannot be obtained} \textsl{from the matrix} $M$ (in which $%
\Phi =0)$ by applications of the N-K transformations, i.e. there are no
transformations of the type $M^{\prime }(\Phi )=AM(\Phi =0)A^{\dag }.$
Therefore, this type of SU(3) matrix should be associated with QCD part of
the SM. Hence we have to use the Ernst functional, Eq.(3.18), instead of the
F-S-type, Eq.(3.19). These results provide resolution of the gap dilemma. 
\textsl{\ Evidently,\ this resolution is equivalent to the} \textsl{%
statement that the symmetry group of the SM is} SU(3)$\times $SU(2)$\times $%
U(1). This result should be taken into account in designing all possible
grand unified theories (GUT). \ In the next subsection we shall discuss the
rigidity of this result.

\subsubsection{Remarkable rigidity of symmetries of the Standard Model and
the extended Ricci flow}

In addition to Bonnor's transformations there are many other transformations
from vacuum to electrovacuum. In particular, in Appendix A we mentioned
transformations discovered by Herlt. By looking at Eq.s(A.5)-(A.7)
describing these transformations and comparing them with those by Bonnor,
Eq.(6.13), \ it is an easy exercise to check that all arguments leading from
Eq.s(6.11) to (6.17) go through unchanged. \ By using \ superposition of N-K
transformations and those either by Bonnor or by Herlt it is possible to
generate a countable infinity of vacuum-to electrovacuum transformations
such that they could be brought back to the vacuum Ernst solution,
Eq.(6.17), using results of previous subsection. This property of Einstein
and Einstein -Maxwell equations we shall call "rigidity". \ In view of
results of previous subsection, this rigidity explains the remarkable
empirical rigidity of symmetries of the SM. Indeed, suppose that the color
subgroup SU(3) can be replaced by SU(N), N\TEXTsymbol{>}3. In such a case it
is appropriate again to pose a question : Can self-dual Y-M fields-gravity
correspondence discovered by L.Witten for SU(2) be extended for SU(N), N%
\TEXTsymbol{>}3? In Ref.[128] G\"{u}rses demonstrated that, indeed, this is
possible but under nonphysical conditions. Indeed, this correspondence
requires for SU(n+1) self-dual Y-M fields to be in correspondence with the
set of n-1 Einstein-Maxwell fields. \ Since $n=1$ and $n=2$ cases have been
already described, we need only \ to worry about $n>2$. In such a case we
shall have many-to-one correspondence between the replicas of electrovacuum
and vacuum Einstein fields which, while permissible mathematically, is not
permissible physically since the Bonnor-type transformations require
one-to-one correspondence between the vacuum and electrovacuum fields.
Herrera-Aguillar and Kechkin, Ref.[129], found a way of transforming the
compactified fields of heterotic string (e.g. see Eq.(3.12)) into
Einstein--multi-Maxwell fields of exactly the same type as discussed in the
paper by G\"{u}rses [128]. While in the paper by G\"{u}rses these replicas
of Maxwell's fields needed to be postulated, in [129] their stringy origin
was found explicitly. \ From here, it follows that results obtained in this
subsection make the minimal functional, Eq.(3.8), and the associated with it
Perelman-like functional, Eq.(3.13), universal. \ \ The universality of the
associated \ with it Ricci flow, Eq.s (3.14), has physical significance to
be discussed below.$\ \ \ \ \ \ \ \ \ \ \ $

\section{Discussion}

\subsection{Connections with loop quantum gravity}

A large portion of this paper was spent on justification, extension and
exploitation of the remarkable correspondence between gravity and self-dual
Y-M fields noticed by Louis Witten. \ Such correspondence is \ achievable
only non perturbatively. In a different form it was emphasized in the paper
by Mason and Newman [130] inspired by work by Ashtekar, Jacobson and Smolin
[131]. It is not too difficult to notice that, in fact, papers [130,131] are
compatible with Witten's result since reobtaining of Nahm's equations in the
context of gravity is the main result of Ref.[131]. In this context the Nahm
equations are just equations for moving triad on some 3-manifold. \ Since
the connection of Nam's equations with monopoles can be found in Ref.[68]
and \ with instantons in Ref.[132] the link with Witten's results can be
established, in principle. Since the authors of [131] are the main
proponents of loop quantum gravity (LQG) such refinements might be helpful
for developments in the field of LQG. We shall continue our discussion of
LQG in \ the next subsection.

\subsection{Topology changing processes, the extended Ricci flow and the
Higgs boson}

According to the existing opinion the SM does not account for effects of
gravity. At the same time, in the Introduction we mentioned that in recent
works by Smolin and collaborators [32-34] it was shown that "topological
features of certain quantum gravity theories\footnote{%
That is LQG.} can be interpreted as particles, matching known fermions and
bosons of the first generation in the Standard Model". Similar results were
also independently obtained in works by Finkelstein, e.g. see Ref.[133] and
references therein. In particular, Finkelstein recognized that all quantum
numbers describing basic building blocks(=particles) of the SM can be neatly
organized with help of numbers used for description of knots. More
precisely, with projections of these knots onto some plane. It happens, that
for description of all particles of the electroweak portion of the SM the
numbers describing trefoil knot are sufficient. The task of
topological/knotty description of the entire SM was accomplished to some
extent in Ref.[33]. This reference as well as Ref.s[32-34] \ in addition are
capable of describing particle dynamics/transformations. \ All these works
share one common feature: calculations \textsl{do not} require Higgs boson.
This fact is consistent with results discussed in subsection 4.3.1.

The question arises: Is this feature a serious deficiency of these
topological methods or are these methods\ so superior to other, that the
Higgs boson should be looked upon as an artifact of the\ previously existing
perturbative methods used in SM calculations? To answer this \ self-imposed
question requires several steps.

First, we recall that according to the existing opinion the SM does not
account for effects of gravity. In such a case all the above results should
have nothing in common with the SM which is not true.

Second, the results obtained in this paper indicate that knots/links/braids
mentioned above have not only virtual (combinatorial/topological) but also
differential-geometric description (Appendix B). Because of this,
topological description should be looked upon as complementary to that
obtainable with help of the F-S-type models.

Third, it is known that knot/link- describing Faddeev model can be converted
into Skyrme model [134]. It is also known that the Skyrme-type models 
\textsl{do not account for quarks explicitly}, Ref.[68], page 349. This is
not a serious drawback as we shall explain momentarily.

Fourth, much more important for us is the fact that the Skyrme model can be
used both in nuclear [135] and high energy [136\textbf{]} physics where it
is used for description of both QCD (\textsl{nicely describing} \textsl{the
entire known hadron spectra}) and electroweak interactions.

To account for quarks one has to go back to the Faddeev-type models capable
of describing knots/links and to make a connection between these \textsl{%
physical} knots/links and \textsl{topological/combinatorial} knots/ links
discussed in Refs[32-34\textbf{,}133]. This is still insufficient! It is
insufficient because Floer's Eq.(4.7) connects different vacua each is\
being described by the zero curvature condition Eq.(4.13). It is always
possible to look at such a condition as describing some knot/link
differential geometrically. With each knot, say in S$^{3},$ some 3-manifold
is associated. Furthermore such a manifold should be hyperbolic (subsection
3.6), that is either associated with hyperbolic-type knot/links [20,137] in S%
$^{3}$ or with knots/links \ "living" in hyperboloid embedded in the
Minkowski spacetime. Such a restriction is absent in Ref.s[32-34,133]. At
the same time the Y-M functional, Eq.(4.12), is defined for \ a particular
3-manifold whose construction is quite sophisticated. Eq.(4.7) describes
processes of topology change by connecting different vacua. Such changes
formally are not compatible with the fact that we are dealing with one and
the same 3-manifold M$\times $[0,1]. From the mathematical standpoint [11]
no harm is made if one considers just this 3-manifold, e.g. read Ref.[11],
page 22, bottom. Since particle dynamics is encoded in dynamics of
transformations between knots/links, it causes us to consider transitions
between different 3-manifolds. These 3-manifolds should be carefully glued
together as described in Ref.[11]. In this picture particle dynamics\
involving particle scattering/transformation is synonymous with processes
involving topology change. These are carried out naturally by instantons.
Such processes can be equivalently \ and more physically described in terms
of the \ properties of \ the (extended) Ricci flow (subsection 3.4)
following ideas of Perelman's proof of the Poincare$^{\prime }$ conjecture.
Indeed, experimentally there is only finite number of stable particles.
Without an exception, \ the end products of all scattering processes involve
only stable particles. This observation matches perfectly with the
irreversibility of Ricci flow processes involving changes in topology: from
more complex-to less complex 3-manifolds. Such Ricci flow model upon
development could provide mathematical justification to otherwise rather
vague statements \ by Finkelstein that "more complicated knots ( particles)
can therefore dynamically decay to trefoils (stable particles)", Ref.[133],
page 10, bottom.

\subsection{\protect\smallskip Elementary particles as black holes}

In the paper [138] by Reina and Treves and also in [139] by Ernst it was
found that for asymptotically flat Einstein-Maxwell fields generated from
the vacuum fields by means of transformations of the type described above,
in Section 6, the gyromagnetic factor $g=2$. For the sake of space, we refer
our readers to a recent review by Pfister and King [140] for definitions of $%
g$ \ and many historical facts and developments. In [140] it was noticed
that such value of $g$ is typical for most of stable particles of the SM. In
view of the quantum gravity-Y-M correspondence promoted in this paper, the
interpretation of elementary particles as black holes makes sense,
especially in view of \ the following excerpt \ from Ref.[38], page 526,
"There is one-to-one correspondence between stationary vacuum fields with
sources characterized by masses and angular momenta and stationary
Einstein-Maxwell fields with purely electromagnetic sources, i.e. charges
and currents."

\bigskip

\textbf{Appendix A}

\smallskip \medskip \textbf{Peculiar interrelationship between
gravitational, electromagnetic and other fields}

Unification of gravity and electromagnetism was initiated by Nordstr\"{o}m
in 1913- before general relativity was formulated by Einstein. Almost
immediately after \ Einstein's formulation, Kaluza, in 1921, and Klein, in
1926, proposed unification of electromagnetism and gravity by embedding
Einstein's 4-dimensional theory into 5 dimensional space in which 5th
dimension is a circle. These results and their generalizations (up to 1987)
can be found in the collection of papers compiled by Applequist, Chodos and
Freund [141]. Regrettably, this collection does not contain alternative
theories of unification. Since such alternative theories are much less
known/popular to/with string and gravity theoreticians, here we provide a
brief \ representative sketch of these alternative theories.

The 1st unified Einstein-Maxwell theory in 4-dimenssional space-time was
proposed and solved by Rainich in 1925. It was discussed in great detail by
Misner and Wheeler [142]. After Rainich there appeared many other works on
exact solutions of Einstein-Maxwell fields [38]. The most striking outcome
of these, more recent, works is the fact that multitude of exact solutions
of the \textsl{combined} Einstein-Maxwell equations can be obtained from
solutions of the \textsl{vacuum} Einstein equations.

In 1961 Bonnor [123]obtained the following remarkable result (e.g. read his
Theorem 1). Suppose solutions of the vacuum \ Einstein equations are known.
Using these solutions, it is possible to obtain a certain class of solutions
of Einstein-Maxwell equations.

In Section 6 we obtained the reverse result: Einstein's solutions for pure
gravity were obtained from solutions of the Einstein-Maxwell equations.
Without doing extra work, the electrovacuum solution \ obtained by Bonnor
can be converted into that describing propagation of the combined
cylindrical gravitational and electromagnetic waves. With some additional
efforts one can use the obtained results as an input for results describing
the combined gravitational, electromagnetic \ and neutrino wave propagation
[143-144\textbf{].}

The results \ by Bonnor comprise only a small portion of results connecting
static gravity fields with electromagnetic and neutrino fields. The next
example belongs to Herlt [38\textbf{,}145]. It provides a flavor of how this
could be achieved. \ We begin with Eq.(2.5). When written explicitly, this
equation reads%
\begin{equation}
\left( \partial _{\rho }^{2}+\frac{1}{\rho }\partial _{\rho }+\partial
_{z}^{2}\right) u=0.  \tag{A.1}
\end{equation}%
This type of solution is the result of use of the matrix $M$, Eq.(2.15), in
Eq.(2.14b). Nakamura [146] demonstrated that there is another matrix $Q$
given by 
\begin{equation}
Q=\left( 
\begin{array}{cc}
f & f\omega \\ 
f\omega & f^{2}\omega ^{2}-\rho ^{2}f^{-1}%
\end{array}%
\right)  \tag{A.2}
\end{equation}%
and the associated with it analog of Eq.(2.14b)%
\begin{equation}
\partial _{\rho }(\rho \partial _{\rho }Q\cdot Q^{-1})+\partial _{z}(\rho
\partial _{z}Q\cdot Q^{-1})=0  \tag{A.3}
\end{equation}%
leading to the equation analogous to Eq.(A.1), that is%
\begin{equation}
\left( \partial _{\rho }^{2}-\frac{1}{\rho }\partial _{\rho }+\partial
_{z}^{2}\right) \tilde{u}=0.  \tag{A.4}
\end{equation}%
Nakamura demonstrated that the solution $\tilde{u}$ is obtainable from
solution of Eq.(A.1). and vice versa. Thus, instead of the Ernst Eq.(2.4) we
can use Eq.(A.4). This fact plays crucial role in Hertl's work. \ In it, he
uses Eq.(A.4) to obtain $u$ in Eq.(A.1) as follows%
\begin{equation}
\exp (2u)=\left( \tilde{u}^{-1}+G\right) ^{2}  \tag{A.5}
\end{equation}%
with $G$ given by%
\begin{equation}
G=\tilde{u}_{,\rho }[\rho (u_{,\rho }^{2}+u_{,z}^{2})-\tilde{u}\tilde{u}%
_{,\rho }]^{-1}.  \tag{A.6}
\end{equation}%
These results allow him to introduce a potential $\chi $ via%
\begin{equation}
\chi =\tilde{u}^{-1}-G.  \tag{A.7}
\end{equation}%
Using the original work of Ernst [43] as well as Ref.[38], we find that
solution of the static axially symmetric coupled Einstein-Maxwell equations
is given in terms of complex potentials $\mathcal{\epsilon }$ and $\Phi .$
In particular, in purely electrostatic case one has $\mathcal{\epsilon =\bar{%
\epsilon}=}e^{2u}-\chi $ and $\Phi =\bar{\Phi}=\chi $ while the
magnetostatic case is obtained from the electrostatic by requiring -$\Phi =%
\bar{\Phi}=\psi $ and $\mathcal{\epsilon =\bar{\epsilon}=}e^{2u}-\psi $ . In
this case $\psi $ is just relabeled $\chi .$ Ref.[38] contains many other
examples of the coupled Einstein-Maxwell equations obtained from the vacuum
solutions of Einstein equations.

The above results should be looked upon from the standpoint of fundamental
problem of the energy-momentum conservation in general relativity requiring
introduction (in the simplest case) of the Landau-Lifshitz (L-L)
energy-momentum pseudotensor. The description of more complicated
pseudotensors (incorporating that by L-L) can be found in the monograph by
Ortin [147]. To this one should add the problem about the positivity of mass
in general relativity. The difficulties with these concepts stem from the
very basic observation, lying at the heart of general relativity, that at
any given point\ of space-time gravity field can be eliminated by moving in
the appropriately chosen accelerating frame (the equivalence principle).
This fact leaves unexplained the origin \ of the tidal forces requiring
observation of motion of at least two test particles separated by some
nonzero distance. The explanation of this phenomenon within general
relativity framework is nontrivial.It can be found in [148]. In turn, it
leads to speculations about the limiting procedure leading to elimination of
gravity at a given point\footnote{%
The abundance of available energy-momentum pseudotensors is result of these
speculations.}. Apparently, this problem is still not solved
rigorously[147]. \ An outstandig collection of rigorous results on general
relativity can be found in the recent monograph by Choquet-Bruhat [149]
while \ [150] discusses peculiar relationship between the Newtonian and
Einsteinian gravities at the scale of \ our Solar system.

Conversely, one can think of other fields at the point/domain where gravity
is absent as subtle manifestations of gravity. Interestingly enough, such an
idea was originally put forward \ by Rainich already in 1925 ! Recent status
of \ these ideas is given in paper by Ivanov [125]. From such a standpoint,
the functional given by Eq.(3.13) (that is the Perelman-like entropy
functional) is sufficient for description of all fields with integer spin.
With minor modifications (e.g. involving either the Newman-Penrose formalism
[143,144] or supersymmeric formalism used in calculation of Seiberg-Witten
invariants [66]), it can be used for description of all known fields in
nature.\ 

\smallskip

\textbf{Appendix B}

\textbf{Some facts about integrable dynamics of knotted vortex filaments}

\smallskip

B.1 \textsl{Connection with the Landau-Lifshitz equation}

Following Ref.[85], we discuss motion of a vortex filament in the
incompressible fluid. Some historical facts relating this problem to string
theory are given in our \ recent work, Ref.[84]. \ Let $\mathbf{u}$ be a
velocity field in the fluid such that div$\mathbf{u}=0$. Therefore, we can
write $\mathbf{u}=\mathbf{\nabla }\times \mathbf{A}$ . Next, we define the
vorticity $\mathbf{w}=\mathbf{\nabla }\times \mathbf{u}$ so that eventually,%
\begin{equation}
u=-\frac{1}{4\pi }\int d^{3}x\frac{(\mathbf{x}-\mathbf{x}^{\prime })\times 
\mathbf{w}(\mathbf{x}^{\prime })}{\left\Vert \mathbf{x}-\mathbf{x}^{\prime
}\right\Vert ^{3}}.  \tag{B.1a}
\end{equation}%
This expression can be simplified by assuming that there is a \textsl{line}
vortex which is modelled by a tube with a cross-sectional area $dA$ \ and
such that the vorticity \textbf{w} is everywhere tangent to the line vortex
and has a constant magnitude w. Let then $\Gamma =\int wdA$ so that 
\begin{equation}
u=-\frac{\Gamma }{4\pi }\oint \frac{(\mathbf{x}-\mathbf{x}^{\prime })\times d%
\mathbf{\gamma }}{\left\Vert \mathbf{x}-\mathbf{x}^{\prime }\right\Vert ^{3}}
\tag{B.1b}
\end{equation}%
with $d\mathbf{\gamma }$ being an infinitesimal line segment along the
vortex. Such a model of a vortex resembles very much model used for
description of dynamics of ring polymers [84]. Because of this, it is
convenient to make the following identification : $\mathbf{u}(\mathbf{\gamma 
}(s,t))=\frac{\partial \mathbf{\gamma }}{\partial t}(s,t)$, with $s$ being a
position along the vortex contour and $t$-time. This allows us to write 
\begin{equation}
\frac{\partial \mathbf{\gamma }}{\partial t}(s^{\prime },t)=-\frac{\Gamma }{%
4\pi }\oint \frac{(\mathbf{\gamma (}s^{\prime }\mathbf{,}t\mathbf{)}-\mathbf{%
\gamma (}s\mathbf{,}t\mathbf{)})}{\left\Vert \mathbf{\gamma (}s^{\prime }%
\mathbf{,}t\mathbf{)}-\mathbf{\gamma (}s\mathbf{,}t\mathbf{)}\right\Vert ^{3}%
}\times \frac{\partial \mathbf{\gamma }}{\partial s}ds  \tag{B.1c}
\end{equation}%
and to make a Taylor series expansion in order to rewrite Eq.(B1c) as 
\begin{equation}
\frac{\partial \mathbf{\gamma }}{\partial t}=\frac{\Gamma }{4\pi }[\frac{%
\partial \mathbf{\gamma }}{\partial s^{\prime }}\times \frac{\partial ^{2}%
\mathbf{\gamma }}{\partial s^{\prime 2}}\int \frac{ds}{\left\vert
s-s^{\prime }\right\vert }+...].  \tag{B.1d}
\end{equation}%
In this expression only the leading order result is written explicitly. By
introducing a cut off $\varepsilon $ such that $\left\vert s-s^{\prime
}\right\vert \geq \varepsilon $ and by rescaling time: $t\rightarrow \frac{%
\Gamma }{4\pi }t\ln (\varepsilon ^{-1})$ one finally arrives at the basic
vortex filament equation%
\begin{equation}
\frac{\partial \mathbf{\gamma }}{\partial t}=\frac{\partial \mathbf{\gamma }%
}{\partial s^{\prime }}\times \frac{\partial ^{2}\mathbf{\gamma }}{\partial
s^{\prime 2}}.  \tag{B.2}
\end{equation}%
Introduce now the Serret-Frenet frame made of vectors $\mathbf{B}$,$\mathbf{T%
}$ and $\mathbf{N}$ so that $\mathbf{B}=\mathbf{T}\times \mathbf{N,\check{%
\kappa}N=}\frac{d\mathbf{T}}{ds},\mathbf{T}=\frac{\partial \mathbf{\gamma }}{%
\partial s},$ where $\check{\kappa}$ is a curvature of $\mathbf{\gamma }$.
Then, Eq.(B.2) can be equivalently rewritten as 
\begin{equation}
\frac{\partial \mathbf{\gamma }}{\partial t}=\check{\kappa}\mathbf{B} 
\tag{B.3}
\end{equation}%
or, as 
\begin{equation}
\frac{\partial \mathbf{T}}{\partial t}=\mathbf{T}\times \mathbf{T}_{xx}. 
\tag{B.4}
\end{equation}%
In the last equation the replacement $s\leftrightharpoons x$ was made so
that the obtained equation coincides with the Landau-Lifshitz (L-L) equation
describing dynamics of 1d Heisenberg ferromagnets [86].

\medskip

B.2 \textsl{Hashimoto map and the Gross-Pitaevskii equation}

Hashimoto [85] found ingenious way to transform the L-L equation into the
nonlinear Scr\"{o}dinger equation (NLSE) which is also widely known in
condensed matter physics literature as the Gross-Pitaevskii (G-P) equation
[96]. \ Because\ of its is uses in nonlinear optics and in condensed matter
physics for description of the Bose-Einstein condensation (BEC) theory of
this equation is well developed. Some facts from this theory are discussed
in the main text. Here we provide a sketch of how Hashimoto arrived at his
result.

Let $\mathbf{T}$,$\mathbf{U}$ and $\mathbf{V}$ be another triad such that 
\begin{equation}
U=\cos (\int\limits^{x}\tau ds)\mathbf{N}-\sin (\int\limits^{x}\tau ds)%
\mathbf{B,}\text{ }\mathbf{V}=\sin (\int\limits^{x}\tau ds)\mathbf{N}+\cos
(\int\limits^{x}\tau ds)\mathbf{B}  \tag{B.5}
\end{equation}%
in which $\tau $ is the torsion of the curve. Introduce new curvatures $%
\kappa _{1}$ and $\kappa _{2}$ \ in such a way that%
\begin{equation*}
\kappa _{1}=\check{\kappa}\cos (\int\limits^{x}\tau ds)\text{ and }\kappa
_{2}=\check{\kappa}\sin (\int\limits^{x}\tau ds),
\end{equation*}%
then, it can be shown that 
\begin{equation}
\frac{\partial \mathbf{\gamma }}{\partial t}=-\kappa _{2}\mathbf{U}+\kappa
_{1}\mathbf{V}  \tag{B.6a}
\end{equation}%
and%
\begin{equation}
\frac{\partial ^{2}\mathbf{\gamma }}{\partial x^{2}}=\kappa _{1}\mathbf{U}%
+\kappa _{2}\mathbf{V.}  \tag{B.6b}
\end{equation}%
Using these equations and taking into account that $\mathbf{U}_{t}\mathbf{V}%
=-\mathbf{UV}_{t}$ after some algebra one obtains the following equation 
\begin{equation}
i\psi _{t}+\psi _{xx}+[\frac{1}{2}\left\vert \psi \right\vert ^{2}-A(t)]\psi
=0  \tag{B.7}
\end{equation}%
in which $\psi =\kappa _{1}+i\kappa _{2}$ and $A(t)$ is some arbitrary
x-independent function. By replacing $\psi $ with $\psi \exp
(-i\int\limits^{t}dt^{\prime }A(t^{\prime }))$ in this equation we arrive at
the canonical form of the NLSE which is also known as focussing cubic NLSE. 
\begin{equation}
i\psi _{t}+\psi _{xx}+\frac{1}{2}\left\vert \psi \right\vert ^{2}\psi =0 
\tag{B.8}
\end{equation}%
It can be shown that its solution allows us to restore the shape of the
curve/filament $\mathbf{\gamma }(s,t).$ The G-P equation can be identified
with Eq.(B.7) if we make $A(t)$ time-independent. In its canonical form it
is written as (in the system of units in which $\hslash =1$, $m=1/2$) [86]%
\begin{equation}
i\psi _{t}=-\psi _{xx}+2\kappa \left( \left\vert \psi \right\vert
^{2}-c^{2}\right) \psi =0.  \tag{B.9}
\end{equation}%
In general, the \ sign of the coupling constant $\kappa $ can be both
positive and negative. In view of Eq.(B.8), when motion of the vortex
filament takes place in Euclidean space, the sign of $\kappa $ is negative.
This is important if one is interested in dynamic of \textsl{knotted} vortex
filaments [85]. For purposes of this work it is \ also of interest to study
motion of \ vortex\ filaments in the Minkowski and related (hyperbolic, \ de
Sitter ) spaces. This should be done with some caution since the transition
from Eq.(B.1a) to (B.2) is specific for Euclidean space. Thus, study can be
made at the level of Eq.s (B.3) and (B.4). Fortunately, such study was
performed quite recently [94,95]. The summary of results obtained in these
papers can be made with help of the following definitions. Introduce a
vector $n=\{n_{1},n_{2},n_{3}\}$ so that the unit sphere $S^{2}$ is defined
by%
\begin{equation}
S^{2}:n_{1}^{2}+n_{2}^{2}+n_{3}^{2}=1.  \tag{B.10}
\end{equation}%
Respectively, the de Sitter space $S^{1,1}$ \ (or unit pseudo sphere in 
\textbf{R}$^{2,1}$ ) is defined by 
\begin{equation}
S^{1,1}:n_{1}^{2}+n_{2}^{2}-n_{3}^{2}=1,  \tag{B.11}
\end{equation}%
while the hyperbolic space \textbf{H}$^{2}($or hyperboloid embedded in 
\textbf{R}$^{2,1}$) is defined by%
\begin{equation}
\mathbf{H}^{2}:n_{1}^{2}+n_{2}^{2}-n_{3}^{2}=-1,n_{3}>0.  \tag{B.12}
\end{equation}%
Using these definitions, it was proven in [94,95] that: a) for both de
Sitter and \textbf{H}$^{2}$ spaces there are analogs of the L-L equation (
e.g. those discussed in the main text, in subsection 5.2.); b) the Hasimoto
map can be extended for these spaces so that the respective L-L equations
are transformed into the same NLSE (or G-P) equation in which $\kappa $ is 
\textsl{positive}.

\smallskip \pagebreak

\bigskip

\textbf{References}

\bigskip

[1] \ \ \ C.N Yang, R.Mills, Conservation of isotopic spin and isotopic gauge

\ \ \ \ \ \ \ invariance, Phys.Rev. 96 (1954) 191-195. \ 

[2] \ \ \ R.Utiyama, Invariant theoretical interpretation of interaction,

\ \ \ \ \ \ \ Phys.Rev. 101 (1956) 1597-1607.

[3] \ \ \ S.Coleman, There are no classical glueballs, Comm.Math.Phys.

\ \ \ \ \ \ \ 55 (1977) 113-116.

[4] \ \ \ H.Heseng, On the classical lump of Yang-Mills fields,

\ \ \ \ \ \ \ Lett.Math.Phys. 22 (1991) 267-275.

[5] \ \ \ M.Ablowitz, S.Chakabarty, R.Halburd, Integrable systems

\ \ \ \ \ \ \ and reductions of the self-dual Yang-Mills equations,

\ \ \ \ \ \ \ J.Math.Phys. 44 (2003) 3147-3173.

[6] \ \ \ L.Mason, N.Woodhouse, Integrability, Self-Duality,

\ \ \ \ \ \ \ and Twistor Theory, Clarendon Press, Oxford, 1996.

[7] \ \ \ N.Nekrasov, S.Shatashvili, Quantization of integrable systems and

\ \ \ \ \ \ \ \ four dimensional gauge theories, arXiv:0908.4052.

[8] \ \ \ \ T. Shafer, E. Shuryak, Instantons in QCD,

\ \ \ \ \ \ \ \ Rev.Mod.Phys. 70 (1998) 323-425.

[9] \ \ \ \ S.Donaldson, P.Kronheimer, The Geometry of Four Manifolds,

\ \ \ \ \ \ \ \ Clarendon Press, Oxford, 1990.

[10] \ \ D.Freed, K.Uhlenbeck, Instantons and Four Manifolds, 2nd Edition,

\ \ \ \ \ \ \ \ Springer-Verlag, New York, 1991.

[11] \ \ \ S.Donaldson, Floer Homology Groups in Yang-Mills Theory,

\ \ \ \ \ \ \ \ Cambridge University Press, Cambridge, 2002.

[12] \ \ E.Langmann, A.Niemi, Towards a string representation

\ \ \ \ \ \ \ \ of infrared SU(2) Yang-Mills theory, Phys.Lett.B 463\ (1999)
252-256.

[13] \ \ \ P.van Baal, A.Wipf, Classical gauge vacua as knots,

\ \ \ \ \ \ \ \ \ Phys.Lett.B 515 (2001) 181-184.

[14] \ \ \ T. Tsurumaru, I.Tsutsui, A.Fujii, Instantons, monopoles and the

\ \ \ \ \ \ \ \ \ flux quantization in the Faddeev-Niemi decomposition,

\ \ \ \ \ \ \ \ \ Nucl.Phys. B 589 (2000) 659-668.

[15] \ \ \ O.Jahn, Instantons and monopoles in general Abelian gauges,

\ \ \ \ \ \ \ \ \ J.Phys. A 33 (2000) 2997-3019.

[16] \ \ \ Y.Cho, Khot topology of classical QCD vacuum,

\ \ \ \ \ \ \ \ \ Phys.Lett.B 644 (2007) 208-211.

[17] \ \ \ L.Faddeev, Knots as possible excitations of the quantum Yang-Mills

\ \ \ \ \ \ \ \ \ fields, arXiv: 0805.1624.

[18] \ \ \ K-I.Kondo, T.Shinohara, T.Murakami, Reformulating SU(N)

\ \ \ \ \ \ \ \ \ Yang-Mills theory based on change of variables, arXiv:
0803.0176.

[19] \ \ \ T.Cheng, L.Li, Gauge Theory of Elementary Particle Physics,

\ \ \ \ \ \ \ \ \ Oxford U.Press, Oxford, 1984.

[20] \ \ \ C.Adams, The Knot Book, W.H.Freeman and Co., New York, 1994.

[21] \ \ \ \ A.Floer, An instanton-invariant for 3-manifolds,

\ \ \ \ \ \ \ \ \ Comm.Math.Phys. 118 (1988) 215-240.

[22] \ \ \ \ K-I Kondo, A.Ono, A.Shibata,T.Shinobara,T.Murakami,

\ \ \ \ \ \ \ \ \ Glueball mass from\ quantized knot solitons and
gauge-invariant

\ \ \ \ \ \ \ \ \ gluon mass, J.Phys. A 39 (2006) 13767-13782.

[23] \ \ \ \ L. Freidel, R.Leigh, D. Minic, A.Yelnikov On the spectrum of
pure

\ \ \ \ \ \ \ \ \ Yang-Mills theory, arxiv: 0801.1113.

[24] \ \ \ H.Meyer, Glueball Regge trajectories, Ph D, Oxford University.,

\ \ \ \ \ \ \ \ \ 2004; arxiv: hep/lat/0508002.

[25] \ \ \ L.Faddeev, A.Niemi, Aspects of electric and magnetic variables in

\ \ \ \ \ \ \ \ \ SU(2) Yang-Mills theory, arXiv: hep-th/0101078.

[26] \ \ \ A.Wereszczynski, Integrability and Hopf solitons in models with

\ \ \ \ \ \ \ \ \ explicitlly broken O(3) symmetry,

\ \ \ \ \ \ \ \ \ European Phys. Journal C 38 (2004) 261-265.

[27] \ \ \ R.Bartnik and J.McKinnon, Particle-like solutions of the

\ \ \ \ \ \ \ \ \ Einstein-Yang-Mills equations, Phys.Rev. Lett. 61 (1988)
141-144.

[28] \ \ \ M.Volkov, D.Galt'sov, Gravitating non Abelian solitonts and black

\ \ \ \ \ \ \ \ \ holes with Yang-Mills fields, Phys.Reports 319 (1999) 1-83.

[29] \ \ \ J.Smoller, A.Wasserman, S-T. Yau, J.McLeod, Smooth static

\ \ \ \ \ \ \ \ \ solutions of the Einstein-Yang-Mills equations,

\ \ \ \ \ \ \ \ \ Comm.Math.Phys. 143 (1991) 115-147.

[30] \ \ \ L.Witten, Static axially symmetric solutions of self-dual SU(2)
gauge

\ \ \ \ \ \ \ \ \ fields in Euclidean four-dimensional space,

\ \ \ \ \ \ \ \ \ Phys.Rev.D 19 (1979) 718-720.

[31] \ \ \ D.Korotkin, H.Nicolai, Isomonodromic quantization of

\ \ \ \ \ \ \ \ \ dimensionally reduced gravity, Nucl.Phys.B 475 (1996)
397-439.

[32] \ \ \ S.Bilson-Thomson, F.Macropoulou, L.Smolin, Quantum

\ \ \ \ \ \ \ \ \ gravity and the standard model, Class.Quant.Gravity 24
(2007)

\ \ \ \ \ \ \ \ \ 3975-3993.

[33] \ \ \ S.Bilson-Thompson, J.Hackett, L.Kauffman, L.Smolin,

\ \ \ \ \ \ \ \ \ Particle identification from symmetries of braided ribbon

\ \ \ \ \ \ \ \ \ network invariants, arXiv: 0804.0037.

[34] \ \ \ S.Bilson-Thomson, J.Hackett, L.Kauffman, Particle topology,

\ \ \ \ \ \ \ \ \ braids and braided belts, arXiv: 0903.137.

[35] \ \ \ B.List, Evolution of an extended Ricci flow system,

\ \ \ \ \ \ \ \ \ Communications in Analysis and Geometry 16 (2008)
1007-1048.

[36] \ \ \ B.List, Evolution of an extended Ricci flow system, PhD thesis,

\ \ \ \ \ \ \ \ \ Freie University of Berlin, 2005.

[37] \ \ \ A.Chamseddine, A.Connes, Why the Standard Model,

\ \ \ \ \ \ \ \ \ JGP 58 (2008) 38-47.

[38] \ \ \ H.Stephani, D.Kramer, M.MacCallum, C. Hoenselaers, E.Herlt,

\ \ \ \ \ \ \ \ \ Exact Solutions of Einstein's Field Equations,

\ \ \ \ \ \ \ \ \ 2nd Edition, Cambridge University Press, Cambridge, UK,
2006.

[39] \ \ \ R.Wald, General Relativity, The University of Chicago Press,

\ \ \ \ \ \ \ \ \ Chicago, IL,1984.

[40] \ \ \ B.O'Neill, Semi-Riemannian Geometry, Academic Press,

\ \ \ \ \ \ \ \ \ New York, 1983.

[41] \ \ \ C.Reina, A.Trevers, Axisymmetric gravitational fields,

\ \ \ \ \ \ \ \ \ Gen.Relativ.Gravit. 7 (1976) 817-838.

[42] \ \ \ F.Ernst, New formulation of the axially symmetric gravitational

\ \ \ \ \ \ \ \ \ field problem, \ Phys.Rev. 167 (1968) 1175-1178.

[43] \ \ \ F.Ernst, New formulation of the axially symmetric gravitational
field

\ \ \ \ \ \ \ \ \ problem. II, Phys.Rev. 168 (1968) 1415-1417.

[44] \ \ \ J.Isenberg, Parametrization of the space of solutions of
Einstein's

\ \ \ \ \ \ \ \ \ equations, Physical Rev.Lett. 59 (1987) 2389-2392.

[45] \ \ \ J.Isenberg, Constant mean curvature solutions of the Einstein's

\ \ \ \ \ \ \ \ \ constraint \ equations on closed manifolds,

\ \ \ \ \ \ \ \ \ Class.Quantum Grav.12 (1995) 2249-2274.

[46] \ \ \ C.N.Yang, \ Condition of self-duality for SU(2) gauge fields on

\ \ \ \ \ \ \ \ \ Euclidean four-dimensional space,

\ \ \ \ \ \ \ \ \ Phys.Rev.Lett. 38 (1977) 1477-1379.

[47] \ \ \ P.Forgacs, Z.Horvath, L.Palla, Generating the

\ \ \ \ \ \ \ \ \ Bogomolny-Prasad-Sommerfeld one monopole solution by a

\ \ \ \ \ \ \ \ \ B\"{a}cklund transformation, Phys.Rev.Lett. 45 (1980)
505-508.

[48] \ \ \ \ \ D.Singleton, Axially symmetric solutions for SU(2) Yang-Mills
theory,

\ \ \ \ \ \ \ \ \ J.Math.Phys. 37 (1996) 4574-4583.

[49] \ \ \ N.Manton, Complex structure of monoploles,

\ \ \ \ \ \ \ \ \ Nucl.Phys.B135 (1978) 319-332.

[50] \ \ \ \ A.Kholodenko, Towards physically motivated proofs of the
Poincare$^{\prime }$

\ \ \ \ \ \ \ \ \ and geometrization conjectures, J.Geom.Phys. 58 (2008)
259-290.

[51] \ \ \ A.Kholodenko, E. Ballard, From Ginzburg-Landau to

\ \ \ \ \ \ \ \ \ Hilbert-Einstein via Yamabe, Physica A 380 (2007) 115-162.

[52] \ \ \ D.Gal'tsov, Integrable systems in stringy gravity,

\ \ \ \ \ \ \ \ \ Phys.Rev.Lett. 74 (1995) 2863-2866.

[53] \ \ \ P.Breitenlohner,D.Maison, G.Gibbons, 4-dimensional black

\ \ \ \ \ \ \ \ \ holes from Kaluza-Klein theories,

\ \ \ \ \ \ \ \ \ Comm.Math.Phys.120 (1988) 295-333.

[54] \ \ \ E.Kiritsis, String Theory in a Nutshell, Princeton U.Press,

\ \ \ \ \ \ \ \ \ Princeton, 2007.

[55] \ \ \ A.Sen, Strong-weak coupling duality in three-dimensional

\ \ \ \ \ \ \ \ \ string theory, arXiv:hep-th/9408083.

[56] \ \ \ C.Reina, Internal symmetries of the axisymmetric gravitational

\ \ \ \ \ \ \ \ \ fields, J.Math.Phys. 20 (1979) 303-304.

[57] \ \ \ \ A.Kholodenko, Boundary conformal field theories,

\ \ \ \ \ \ \ \ \ limit sets of Kleinian groups and holography,

\ \ \ \ \ \ \ \ \ J.Geom.Phys.35 (2000) 193-238.

[58] \ \ \ K.Akutagawa, M.Ishida, C.Le Brun, Perelman's invariant,

\ \ \ \ \ \ \ \ \ Ricci flow and the Yamabe invariants of smooth manifolds,

\ \ \ \ \ \ \ \ \ Arch.Math. 88 (2007) 71-75.

[59] \ \ \ S.S. Chern, J.Simons, Characteristic forms and geometric

\ \ \ \ \ \ \ \ \ invariants, Ann.Math. 99 (1974) 48-69.

[60] \ \ \ S.S.Chern, On a conformal invariant of three-dimensional

\ \ \ \ \ \ \ \ \ manifolds, in Aspects of Mathematics and its Applications,

\ \ \ \ \ \ \ \ \ pp.245-252, Elsevier Science Publishers, \ Amsterdam, 1986.

[61] \ \ \ E.Witten, 2+1 dimensional gravity as an exactly solvable model,

\ \ \ \ \ \ \ \ \ Nucl.Phys. B311 (1988/89) 46-78.

[62] \ \ \ S.Gukov, Three-dimensional quantum gravity, Chern-Simons theory

\ \ \ \ \ \ \ \ \ and the A-polynomial, Comm.Math.Phys. 255 (2005) 577-627.

[63] \ \ \ J.Zinn-Justin, Quantum Field Theory and Critical Phenomena,

\ \ \ \ \ \ \ \ \ Clarendon Press, Oxford, UK, 1989.

[64] \ \ \ K.Huang, Quarks, Leptons and Gauge Fields, World Scientific,

\ \ \ \ \ \ \ \ \ Singapore, 1982.

[65] \ \ \ M-C.Hong, G.Tian, Global existence of m-equivariant

\ \ \ \ \ \ \ \ \ Yang-Mills flow in four dimensional spaces,

\ \ \ \ \ \ \ \ \ Comm.Analysis and Geometry 12 (2004) 183-211.

[66] \ \ \ \ P.Kronheimer, T.Mrowka, Monopoles and Three-Manifolds,

\ \ \ \ \ \ \ \ \ Cambridge U.Press, Cambridge, 2007.

[67] \ \ \ E.Frenkel, A.Losev, N.Nekrasov, Instantons beyond topological

\ \ \ \ \ \ \ \ \ theory II, arXiv:0803.3302.

[68] \ \ \ \ N.Manton, P.Sutcliffe, Topological Solitons,

\ \ \ \ \ \ \ \ \ Cambridge U.Press, Cambridge, 2004.

[69] \ \ \ \ D.Auckly, Topological methods to compute Chern-Simons
invariants,

\ \ \ \ \ \ \ \ \ Math.Proc.Camb.Phil. Soc.115 (1994) 220-251.

[70] \ \ \ P.Kirk, E.Klassen, Chern-Simons invariants of 3-manifolds and

\ \ \ \ \ \ \ \ \ representation spaces of knot groups, Math.Ann.287 (1990)
343-367.

[71] \ \ \ D.Auckly, L.Kapitanski, Analysis of S$^{2}$-valued maps and

\ \ \ \ \ \ \ \ \ Faddeev's model, Comm.Math.Phys. 256 (2005) 611-620.

[72] \ \ P.Forgacs, Z.Horvath, L.Palla, Soliton-theoretic framework

\ \ \ \ \ \ \ \ for generating multimonopoles, Ann.Phys. 136 (1981) 371-396.

[73] \ D.Harland, Hyperbolic calorons, monopoles and instantons,

\ \ \ \ \ \ \ Comm.Math.Phys. 280 (2008) 727-735.

[74] \ D.Harland, Large scale and large period limits of symmetric

\ \ \ \ \ \ \ calorons, arXiv: 0704.3695.

[75] \ M.Atiyah, Magnetic monopoles in hyperbolic space, in

\ \ \ \ \ \ \ Vector Bundles on Algebraic Varieties, pages 1-33, Tata
Institute,

\ \ \ \ \ \ \ Bombay, 1984.

[76] \ \ E.Witten, Some exact multipseudoparticle solutions of classical

\ \ \ \ \ \ \ Yang-Mills theory, PRL 38 (1977) 121-124.

[77] \ R.Radjaraman, Solitons and Instantons, North-Holland,

\ \ \ \ \ \ \ Amsterdam, 1982.

[78] \ N.Manton, Instantons on the line, Phys. Lett.B 76 (1978) 111-112.

[79] \ A.Kholodenko, Veneziano amplitudes, spin chains and

\ \ \ \ \ \ \ Abelian reduction of QCD, J.Geom.Phys.59 (2009) 600-619.

[80] \ A.Polyakov, Gauge Fields and Strings, Harwood Academic

\ \ \ \ \ \ \ Publishers, New York, 1987.

[81] \ A.Polyakov, Supermagnets and sigma models,

\ \ \ \ \ \ \ arXiv:hep-th/0512310.

[82] \ \ K.-I.Kondo, Magnetic monopoles and center vortices as

\ \ \ \ \ \ \ gauge-invariant topological defects simultaneously responsible

\ \ \ \ \ \ \ for confinement, J.Phys.G 35 (2008) 085001.

[83] \ \ A.Kholodenko, Heisenberg honeycombs solve Veneziano puzzle,

\ \ \ \ \ \ \ Int.Math.Forum 4 (2009) 441-509.

[84] \ A.Kholodenko, E.Ballard, Topological character of hydrodynamic

\ \ \ \ \ \ \ screening in suspensions of hard spheres: \ An example of the

\ \ \ \ \ \ \ universal phenomenon, Physica A 388 (2009) 3024-3062.

[85] \ \ A.Calini, Integrable dynamics of knotted vortex filaments, in

\ \ \ \ \ \ \ \ Geometry, Integrability and Quantization, pages 11--50,

\ \ \ \ \ \ \ \ Softex, Sofia, 2004.

[86] \ \ L.Faddeev, L.Takhtajan, Hamiltonian Methods in Theory of Solitons,

\ \ \ \ \ \ \ \ Springer-Verlag, Berlin, 1987.

[87] \ \ O.Pashaev, S.Sergeenkov, Nonlinear sigma model with noncompact

\ \ \ \ \ \ \ \ symmetry group in the theory of a nearly ideal Bose gas,

\ \ \ \ \ \ \ \ Physica A, 137 (1986) 282-294.

[88] \ \ P.Forgacs, N.Manton, Space-time symmetries in gauge theories,

\ \ \ \ \ \ \ \ Comm.Math.Phys. 72 (1980) 15-35.

[89] \ \ A.Jaffe, C.Taubes, Vortices and Monopoles, Birh\"{a}user, Boston,
1980.

[90] \ \ G. Landweber, Singular instantons with SO(3) symmetry,

\ \ \ \ \ \ \ \ arXiv:math/0503611.

[91] \ \ J.-H. Lee, O.Pashaev, Moving frames hierarchy and B-F theory,

\ \ \ \ \ \ \ \ J.Math.Phys. 39 (1998) 102-123.

[92] \ \ \ O.Pashaev, The Lax pair by dimensional reduction of Chern-Simons

\ \ \ \ \ \ \ \ gauge theory, J.Math.Phys. 37 (1996) 4368-4387.

[93] \ \ \ L.Faddeev, A.Niemi, U.Wiedner, Glueballs, closed fluxtubes and

\ \ \ \ \ \ \ \ \ $\eta (1440),$arXiv:hep-ph/0308240.

[94] \ \ \ Q.Ding, A note on the NLS and Schrodinger flow of maps,

\ \ \ \ \ \ \ \ \ Phys.Lett.A 248 (1998) 49-56.

[95] \ \ \ Q.Ding, J.-I. Inoguchi, Schrodinger flows, binormal motion

\ \ \ \ \ \ \ \ \ for curves and second AKNS hierarchies, Chaos, Solitons and

\ \ \ \ \ \ \ \ \ fractals, 21 (2004) 669-677.

[96] \ \ \ F.Dalfovo, S.Giorgi, L.Pitaevskii, S.Stringari, Theory of
Bose-Einstein

\ \ \ \ \ \ \ \ condensation in trapped gases, Rev.Mod.Phys. 71 (1999)
463-512.

[97] \ \ \ V.Zakharov, A.Shabat, Exact theory of two-dimensional
self-focusing

\ \ \ \ \ \ \ \ \ and one-dimensional self-modulation of waves in nonlinear
media,

\ \ \ \ \ \ \ \ \ Sov.Phys. JETP 34 (1972) 62-69.

[98] \ \ \ V.Zakharov, A.Shabat, Interaction between solitons in a stable

\ \ \ \ \ \ \ \ \ medium, Sov.Phys. JETP 37 (1973) 823-828.

[99] \ \ \ Y.Castin, C.Herzog, Bose-Einstein condensates in symmetry

\ \ \ \ \ \ \ \ \ breaking states, arXiv: cond-math/0012040

[100] \ \ J.Maki and T.Kodama, Phenomenological quantization scheme in a

\ \ \ \ \ \ \ \ \ nonlinear Schr\"{o}dinger equation, PRL 57 (1986)
2097-2100.

[101] \ \ E.Lieb, W.Linger, Exact analysis of an interacting Bose gas I.

\ \ \ \ \ \ \ \ \ The General solution and the ground state,

\ \ \ \ \ \ \ \ \ Phys.Rev.130 (1963) 1605-1616.

[102] \ \ M.Batchelor, X.Guan, J.McGuirre, Ground state of 1d bosons

\ \ \ \ \ \ \ \ \ with delta interaction: link to the BCS model, J.Phys. A
37 (2004)

\ \ \ \ \ \ \ \ \ L497-504.

[103] \ \ A.Ovchinnikov, On exactly solvable pairing models for bosons,

\ \ \ \ \ \ \ \ \ J.Stat.Mechanics: Theory and Experiment (2004) P07004.

[104] \ \ R.Richardson, Exactly solvable many-boson model,

\ \ \ \ \ \ \ \ \ J.Math.Physics 9 (1968) 1327-1343.

[105] \ \ R.Richardson, N. Sherman, Exact eigenstates of the

\ \ \ \ \ \ \ \ \ \ pairing-force hamiltonian, Nucl.Phys. 52 (1964) 221-238.

[106] \ \ \ A.Dhar, B.Shastry, Bloch walls and macroscopic string states

\ \ \ \ \ \ \ \ \ \ in Bethe solution of the Heisenberg ferromagnetic linear
chain,

\ \ \ \ \ \ \ \ \ \ PRL 85 (2000) 2813-2816.

[107] \ \ \ A.Dhar, B.Shastry, Solution of a generalized Stiltjes problem,

\ \ \ \ \ \ \ \ \ \ J.Phys. A 34 (2001) 6197-6208.

[108] \ \ \ J.Fuchs, A.Recati, W. Zwerger, An exactly solvable model of

\ \ \ \ \ \ \ \ \ \ the BCS-BEC crossover, PRL 93 (2004) 090408.

[109] \ \ \ J.Dukelsky,C.Esebbag, P.Schuck, Class of exactly solvable

\ \ \ \ \ \ \ \ \ \ pairing models, PRL 87 (2001) 066403.

[110] \ \ \ P.Ring, P.Schuck, The Nuclear Many-Body Problem,

\ \ \ \ \ \ \ \ \ \ Springer-Verlag, Berlin, 1980.

[111] \ \ \ J.Eisenberg, W.Greiner, Nuclear Theory, Vol.3,

\ \ \ \ \ \ \ \ \ \ Microscopic Theory of the Nucleus,

\ \ \ \ \ \ \ \ \ \ North-Holland, Amsterdam, 1972.

[112] \ \ \ D.Rowe, Nuclear Collective Motion,

\ \ \ \ \ \ \ \ \ \ Methhuen and Co.Ltd., London, 1970.

[113] \ \ \ A.Kerman, R.Lawson, M.Macfarlane, Accuracy of the

\ \ \ \ \ \ \ \ \ \ superconductivity approximation for pairing forces in

\ \ \ \ \ \ \ \ \ \ nuclei, Phys.Rev.124 (1961) 162-167.

[114] \ \ \ M.Goldhaber, E.Teller, On nuclear dipole vibrations,

\ \ \ \ \ \ \ \ \ \ Phys.Rev.74 (1948) 1046.

[115] \ \ \ A.Sykes, P.Drummond, M.Davis, Excitation spectrum of

\ \ \ \ \ \ \ \ \ \ bosons in a finite one-dimensional circular waveguide via

\ \ \ \ \ \ \ \ \ \ \ Bethe ansatz, Phys.Rev. A 76 (2007) 06320.

[116] \ \ \ \ D.Dean,M.Hjorth-Jensen, Pairing in nuclear systems:

\ \ \ \ \ \ \ \ \ \ \ from neutron stars to finite nuclei,

\ \ \ \ \ \ \ \ \ \ \ Rev.Mod.Phys. 75 (2003) 607-654.

[117] \ \ \ \ P.Braun-Munzinger, J.Wambach, \ Phase diagram of strongly

\ \ \ \ \ \ \ \ \ \ \ interacting \ matter, Rev.Mod.Phys. 81 (2009)
1031-1050.

[118] \ \ \ \ M.Barbaro, R.Cenni, A.Molinari, M.Quaglia, Analytic solution of

\ \ \ \ \ \ \ \ \ \ \ the pairing problem: one pair in many levels,

\ \ \ \ \ \ \ \ \ \ \ Phys.Rev.C 66 (2002) 03410.

[119] \ \ \ \ M.Barbaro, R.Cenni, A.Molinari, M.Quaglia, The many levels

\ \ \ \ \ \ \ \ \ \ \ pairing Hamiltonian for two pairs, The European Phys.J.

\ \ \ \ \ \ \ \ \ \ \ A22 (2004) 377-390.

[120] \ \ \ \ M.G\"{u}rses, B.Xanthopoulos, Axially symmetric, static
self-dual

\ \ \ \ \ \ \ \ \ \ \ SU(3) gauge fields and stationary Einstein-Maxwell
metrics,

\ \ \ \ \ \ \ \ \ \ \ Phys.Rev.D 26 (1982) 1912-1915.

[121] \ \ \ \ M.G\"{u}rses, Axially symmetric, static self-dual Yang-Mills
and

\ \ \ \ \ \ \ \ \ \ \ stationary Einstein-gauge field equations, Phys.Rev.D
30 (1984)

\ \ \ \ \ \ \ \ \ \ \ 486-488.

[122] \ \ \ \ P.Mazur, A relationship between the electrovacuum Ernst

\ \ \ \ \ \ \ \ \ \ \ equations and nonlinear $\sigma -$model, \ 

\ \ \ \ \ \ \ \ \ \ \ Acta Phys.Polonica B14 (1983) 219-234.

[123] \ \ \ \ W.Bonnor, Exact solutions of the Einstein-Maxwell equations,

\ \ \ \ \ \ \ \ \ \ \ Z.Phys.161 (1961) 439-444.

[124] \ \ \ \ I.Hauser,F.Ernst, SU(2,1) generation of electrovacs from

\ \ \ \ \ \ \ \ \ \ \ Minkowski space, J.Math.Phys. 20 (1979) 1041-1055.

[125] \ \ \ \ B.Ivanov, Purely electromagnetic spacetimes,

\ \ \ \ \ \ \ \ \ \ \ Phys.Rev.D 77 (2008) 044007.

[126] \ \ \ \ W.Cottingham, D.Greenwood, An Introduction to the

\ \ \ \ \ \ \ \ \ \ \ Standard Model of Particle Physics,

\ \ \ \ \ \ \ \ \ \ \ Cambridge U.Press, Cambridge, 2007.

[127] \ \ \ J.Charap, R.Jones, P.Williams, Unitary symmetry,

\ \ \ \ \ \ \ \ \ \ Rep.Progr.Phys. 30 (1967) 227-283.

[128] \ \ \ M.G\"{u}rses, Axially symmetric, static self-dual Yang-Mills and

\ \ \ \ \ \ \ \ \ \ stationary Einstein-gauge field equations,

\ \ \ \ \ \ \ \ \ \ Phys.Rev.D 30 (1984) 486-488.

[129] \ \ \ A.Herrera-Aguillar, O.Kechkin, String theory extensions of

\ \ \ \ \ \ \ \ \ \ Einstein-Maxwell fields: The stationary case,

\ \ \ \ \ \ \ \ \ \ J.Math.Physics 45 (2004) 216-229.

[130] \ \ \ L.Mason, E.Newman, A connection between the Einstein

\ \ \ \ \ \ \ \ \ \ \ and Yang-Mills equations, Comm.Math.Phys. 121 (1989)
659-668.

[131] \ \ \ A. Ashtekar, T.Jacobson, L.Smolin, A new characterization of

\ \ \ \ \ \ \ \ \ \ Half-flat solutions to Einstein's equation,

\ \ \ \ \ \ \ \ \ \ Comm.Math.Phys. 115 (1988) 631-648.

[132] \ \ \ T. Ivanova, Self-dual Yang-Mills connections and generalized

\ \ \ \ \ \ \ \ \ \ Nahm equations, in Tensor and Vector Analysis, pages
57-70,

\ \ \ \ \ \ \ \ \ \ Gordon and Breach, Amsterdam, 1998.

[133] \ \ \ R.Finkelstein, A.Cadavid, Masses and interactions of

\ \ \ \ \ \ \ \ \ \ q-fermionic knots, Int.J.Mod.Phys. A21 (2006) 4269-4302,

\ \ \ \ \ \ \ \ \ \ arXiv: hep-th/0507022.

[134] \ \ \ \ F.Lin, Y.Yang, Analysis on Faddeev knots and Skyrme solitons:

\ \ \ \ \ \ \ \ \ \ Recent progress and open problems, in Perspectives in
Nonlinear

\ \ \ \ \ \ \ \ \ \ Partial Differential Equations, pp 319-343, AMS
Publishers,

\ \ \ \ \ \ \ \ \ \ Providence, RI, 2007.

[135] \ \ \ N.Manton, S.Wood, Light nuclei as quantized Skyrmions:

\ \ \ \ \ \ \ \ \ \ Energy spectra and form factors, arXiv: 0809.350.

[136] \ \ \ J.Schechter, H.Weigel, The Skyrme model for barions,

\ \ \ \ \ \ \ \ \ \ arXiv: hep-ph/990755.

[137] \ \ \ F.Bonahon, Low-Dimensional Geometry.

\ \ \ \ \ \ \ \ \ \ From Euclidean Surfaces to Hyperbolic Knots.

\ \ \ \ \ \ \ \ \ \ AMS Publishers, Providence,RI. 2009.

[138] \ \ \ C.Reina, A.Treves, Gyromagnetic ratio of Einstein-Maxwell fields,

\ \ \ \ \ \ \ \ \ \ \ Phys.Rev.D 11 (1975) 3031-3032.

[139] \ \ \ \ F.Ernst, Charged version of Tomimatsu-Sato spinning mass field,

\ \ \ \ \ \ \ \ \ \ \ Phys.Rev.D 7 (1973) 2520-2521.\ \ \ 

[140] \ \ \ \ H.Pfister,M.King, The gyromagnetic factor in electrodynamics,

\ \ \ \ \ \ \ \ \ \ \ quantum theory and general relativity,

\ \ \ \ \ \ \ \ \ \ \ Class.Quant.Gravity 20 (2003) 205-213.

[141] \ \ \ \ T.Appelquist, A.Chodos, P.Freund, Modern Kaluza-Klein

\ \ \ \ \ \ \ \ \ \ \ Theories, Addison-Wesley Publ.Co., Reading, MA, 1987.

[142] \ \ \ \ C.Misner, J.Wheeler, Clsassical physics as geometry,

\ \ \ \ \ \ \ \ \ \ \ Ann.Phys. 2 (1957) 525-603.

[143] \ \ \ \ N. Sigbatulin, Oscillations and Waves in Strong Gravitational

\ \ \ \ \ \ \ \ \ \ \ and Electromagnetic Fields, Springer-Verlag, Berlin,
1991.

[144] \ \ \ \ J.Griffiths Colliding Plane waves in General Relativity,

\ \ \ \ \ \ \ \ \ \ \ Clarendon Press, Oxford, 1991.

[145] \ \ \ \ E.Herlt, Static and stationary axially symmetric gravitational

\ \ \ \ \ \ \ \ \ \ \ fields of bounded sourses. I Solutions obtainable from
the

\ \ \ \ \ \ \ \ \ \ \ van Stockum metric, Gen.Relativ.Gravit. 9 (1978)
711-719.

[146] \ \ \ \ Y.Nakamura, Symmetries of stationary axially symmetric

\ \ \ \ \ \ \ \ \ \ \ vacuum Einstein equations and the new family of exact

\ \ \ \ \ \ \ \ \ \ \ solutions, JMP 24 (1983) 606-609.

[147] \ \ \ \ T.Ortin, Gravity and Strings, Cambridge U. Press,

\ \ \ \ \ \ \ \ \ \ \ Cambridge, 2004.

[148] \ \ \ \ \ A.Besse, Einstein Manifolds, Springer-Verlag,

\ \ \ \ \ \ \ \ \ \ \ Berlin,1987.

[149] \ \ \ \ \ Y. Choquet-Bruhat, General Relativity and the

\ \ \ \ \ \ \ \ \ \ \ Einstein Equations, Oxford U.Press, Oxford, 2009.

[150] \ \ \ \ A.Kholodenko, Newtonian limit of Einsteinian gravity

\ \ \ \ \ \ \ \ \ \ \ and dynamics of Solar System, arXiv: 1006.4650.

\ \ \ \ \ \ \ \ \ \ \ 

\bigskip

\bigskip

\bigskip

\bigskip

\bigskip

\bigskip

\bigskip

\bigskip

\bigskip

\bigskip

\bigskip

\bigskip

\bigskip

\bigskip

\bigskip

\bigskip

\bigskip

\bigskip

\bigskip

\bigskip

\end{document}